\documentclass[11pt]{amsart}
\usepackage{url}
\usepackage{amsmath,amsxtra,amssymb,latexsym,epsfig,amscd,amsthm,fancybox,epsfig}
\usepackage{natbib}
\usepackage{amsfonts}
\usepackage{amssymb}
\usepackage{mathtools}
\usepackage{float}
\restylefloat{figure}
\usepackage{booktabs}
\usepackage{multirow}
\usepackage{color}
\usepackage{placeins}
\usepackage{lscape}
\usepackage{mathtools}
\usepackage{amsmath}

\usepackage[mathscr]{eucal}
\usepackage{graphicx}
\usepackage{subcaption}
\usepackage{babel,blindtext}
\usepackage{tikz}
\usepackage{caption}
\usepackage{multicol,xcolor}
\usepackage{epsfig} % for postscript graphics files
\usepackage{epstopdf}
\usepackage{cases}
\usepackage{color}
\usepackage{hyperref}
\usepackage{bm}  %JLK
\usepackage{bbm} %JLK
\usepackage[ruled,vlined]{algorithm2e}
\usepackage{algorithmic}
\usepackage{color, colortbl}
\usepackage{booktabs,xcolor}

\newtheorem{thm}{Theorem}[section]
\newtheorem{prop}{Proposition}[section]

\newtheorem{rem}{Remark}[section]
\newtheorem{remark}{Remark}[section]

\usepackage[top=0.8in, bottom=0.8in, left=0.9in, right=0.8in]{geometry}
\newtheorem {asp}{Assumption}[section]
\newtheorem{lm}{Lemma}[section]
\newtheorem{cor}{Corollary}[section]

\theoremstyle{definition}

\theoremstyle{remark}

\numberwithin{equation}{section}

\def\red#1{\textcolor{red}{#1}}

\newcommand{\E}{\mathbb{E}}

\newcommand{\PP}{\mathbb{P}}

\newcommand{\K}{\mathcal{K}}

\newcommand{\cC}{\mathcal{C}}

\numberwithin{equation}{section}

\def\wt{\widetilde}
\newcommand{\bbr}{\mathbb{R}}

\newcommand{\bed}{\begin{displaymath}}
\newcommand{\eed}{\end{displaymath}}
\newcommand{\bea}{\bed\begin{array}{rl}}
\newcommand{\eea}{\end{array}\eed}

\newcommand{\barray}{\begin{array}{ll}}
\newcommand{\earray}{\end{array}}

\def\a.s{\text{\;a.s.\;}}

\newcommand{\beq}[1]{\begin{equation} \label{#1}}
\newcommand{\eeq}{\end{equation}}
\title[Continuous-time optimal investment with portfolio constraints with reinforcement learning]{Continuous-time optimal investment with portfolio constraints: a reinforcement learning approach}
\author[H. Chau]{H. Chau }
\address{Department of Mathematics \\
University of Manchester\\
Manchester M13 9PL, United Kingdom}
\email{huy.chau@manchester.ac.uk}

\author[D. Nguyen]{D. Nguyen }
\address{Department of Mathematics\\
Marist College\\
3399 North Road\\
Poughkeepsie NY 12601\\
United States
}
\email{nducduy@gmail.com}
\author[T. Nguyen]{T. Nguyen}
\address{\'{E}cole d'Actuariat \\
Universit\'{e} Laval\\
 2325 Rue de l'Université, Québec, QC G1V 0A6, Canada}
\email{thai.nguyen@act.ulaval.ca}


\begin{document}
\maketitle
\begin{abstract}
In a reinforcement learning (RL) framework, we study the exploratory version of the continuous time expected utility (EU) maximization problem with a portfolio constraint that includes widely-used financial regulations such as short-selling constraints and borrowing prohibition. The optimal feedback policy of the exploratory unconstrained classical EU problem is shown to be Gaussian. In the case where the portfolio weight is constrained to a given interval, the corresponding constrained optimal exploratory policy follows a truncated Gaussian distribution. We verify that the closed form optimal solution obtained for logarithmic utility {and quadratic utility} for both unconstrained and constrained situations converge to the non-exploratory expected utility counterpart when the exploration weight goes to zero. Finally, we establish a policy improvement theorem and devise an implementable reinforcement learning algorithm by casting the optimal problem in a martingale framework. Our numerical examples {show that exploration leads to an optimal wealth process that is more dispersedly distributed with heavier tail compared to that of the case without exploration. This effect becomes less significant as the exploration parameter is smaller. Moreover, the numerical implementation also confirms} the intuitive
understanding that a broader domain of investment opportunities necessitates a higher exploration cost. Notably, when subjected to both short-selling and money borrowing constraints, the exploration cost becomes negligible compared to the unconstrained
case.
\end{abstract}
%{  \hypersetup{linkcolor=blue}
  %\tableofcontents
%}
%\tableofcontents

%\newpage

%\end{frontmatter}

\section{Introduction}
%\red{Add OR discussions}
Reinforcement learning (RL) is a dynamic and rapidly advancing subset within the field of machine learning. In recent times, RL has been applied across diverse disciplines to address substantial real-world challenges, extending its reach into domains such as game theory \citep{silver2016mastering,silver2017mastering}, control
theory \citep{bertsekas2019reinforcement},
information theory \citep{williams2017information}, and statistics \citep{kaelbling1996reinforcement}.
%Ones have also considered
%applying ideas of
%RL in quantitative finance.
{RL has been also applied to study many important issues of operations research (OR) such as manufacturing planning, control systems \citep{schneckenreither2019reinforcement} and inventory management \citep{bertsimas2006robust}. %Such OR problems are typically linked to a combinatorial optimization which are typically hard to solve and do not have an efficient polynomial-time solution. The reader is refereed to \citep{mazyavkina2021reinforcement} for a survey on RL approaches for combinatorial optimization}. 
In quantitative finance, RL has been used to study important
problems such
as algorithmic and high frequency trading,
portfolio management. 
%smart order routing
%using RL tools.
The availability of big data sets
has too facilitated
the rapid development
of using RL techniques in financial
engineering. As
a typical example, electronic markets
can provide a sufficient amount of microstructure data for training and adaptive learning,
much beyond what human traders and portfolio managers could handle
in old days. Numerous studies
have been done in this direction
including optimal order execution \citep{nevmyvaka2006reinforcement,schnaubelt2022deep},
optimal trading \citep{hendricks2014reinforcement}, portfolio allocation \citep{moody1998performance},
mean-variance  portfolio allocation \citep{wang2020continuous,wang2019large,dai2020learning,jia2022policy}.
Notably, 
the authors in \cite{wang2020continuous}
considered the continuous time mean-variance 
portfolio selection using a continuous RL framework.
Their algorithm outperforms
both traditional and deep neural network based
algorithms by a large margin. For recent reviews on applications of RL in OR, finance and economics, the 
interested reader is invited to delve into the review papers e.g. \cite{gosavi2009reinforcement,hambly2021recent,jaimungal2022reinforcement, charpentier2021reinforcement}.\\

Differing from conventional econometric or supervised learning approaches prevalent in quantitative 
finance research, which often necessitate the assumption of parametric models, an RL agent refrains from pre-specifying a structural model. Instead, she progressively learns optimal strategies through trial and error, engaging in interactions with a black-box environment (e.g.  the market).
%In contrast to econometric methods or
%supervised learning methods that are commonly used in
%quantitative finance research where
%one must assume some forms of
%parametric models,
%an RL agent {does not} pre-specify a structural model; she learns the
%best strategies based on {trial and error}, through interactions with the
%black-box environment (e.g. the market).
By repeatedly trying
a policy for actions,
receiving and evaluating
reward signals, and improving
the policy, the agent gradually improves her
performance.
The three key components
that essentially capture
the heart of RL are: i) exploration, ii) policy evaluation, and iii) 
policy improvement.
The agent first explores
the surrounding
environment by trying
a policy. She
then evaluates
her reward for the given policy.
Lastly, using the information
receiving from both  exploration
and current reward, she
devises a new policy
with larger reward.
The whole process is then repeated.\\

Despite the fast development and vast applications, very few existing RL studies are done for continuous settings
e.g. \cite{doya2000reinforcement,fremaux2013reinforcement,lee2021policy}, whereas a large body of RL investigations is limited to  discrete learning frameworks
such as discrete-time Markov decision processes (MPD) or deterministic settings, see e.g. \cite{sutton2018reinforcement, hambly2021recent,liu2020dynamic,lee2021multi}. Since discrete time dynamics
just provides an approximation for
real work systems,
it is important, from
the practical point of view, to consider
continuous time systems with both
continuous states and action spaces.
%For example, in a high-frequency stock trading system an agent can or actually needs to interact with environments at an ultra-high frequency trading rate.
As mentioned previously,
under RL settings,
an agent simultaneously interacts with
the surrounding environment (exploration)
and improves her overall performance (exploitation).
Since exploration is inherently costly in term of resources,
it is important to design
an active learning approach
which balances both exploration and exploitation. {Therefore, there is a critical necessity} to extend the RL techniques
to continuous settings where agent can find
best learning strategies that balance exploration and exploitation.\\

The studies of RL
under  a continuous time framework
in both time and space
have been initiated in a series of recent papers \citep{wang2020reinforcement,wang2020continuous,wang2019large,dai2020learning,jia2022policy}.
In \cite{wang2020reinforcement}
the authors
propose a theoretical framework, called \textit{exploratory formulation}, for studying
RL problems in continuous systems in both time and space
that captures repetitive learning 
under exploration in the continuous time limit.
\cite{wang2020continuous} adopts the RL setting of \cite{wang2020reinforcement}
to study the continuous-time mean variance
portfolio selection with a finite time horizon.
They show that in a learning framework incorporating both exploration and exploitation, 
the optimal feedback policy is Gaussian, with time-decaying variance.
As showed in \cite{jia2022a}, the framework in \cite{wang2020reinforcement} minimizes
the mean-square  temporal difference (TD) error \citep{barnard1993temporal,baird1995residual,sutton2018reinforcement}
when learning the value function, which later
turns out to be inconsistent for stochastic settings.  
\cite{dai2020learning} further considers the
equilibrium mean–variance strategies addressing the time-inconsistent issue of the problem. \cite{guo2022general, mou2021robust}
extend the formulation and results of \cite{wang2020continuous}  to mean-field games
and a mean-variance problem with drift
uncertainty, respectively. %Studies of convergence under the exploratory formulation have been considered in \cite{tang2022exploratorySIAM,huang2022convergence}.
In \cite{jia2022policy}, using a martingale approach,
the authors 
are able to present the gradient of the value
function with respect to 
a given parameterized stochastic policy  as the expected integration of an auxiliary running reward function that can be evaluated
using samples and the current value function. This representation effectively turns
policy gradient into a policy evaluation  problem.  Studies of convergence under the exploratory formulation can be found in \cite{tang2022exploratorySIAM,huang2022convergence}.\\

In this paper, we adopt the exploratory stochastic
control learning framework of \cite{wang2020reinforcement} 
to study the continuous-time optimal investment problem without and with a portfolio constraint which covers widely-used financial regulations such as short-selling constraints and borrowing prohibition \citep{cuoco1997optimal}. In both constrained and unconstrained settings with exploration, we manage to find the closed form solution of the exploratory problem for logarithmic utility {and quadratic} function. We show that the optimal feedback policy of the exploratory unconstrained expected utility (EU) problem is Gaussian, which is aligned with the result obtained in \cite{wang2020continuous} for the mean-variance setting. However, when the risky investment ratio is restricted on a given interval, the constrained optimal exploration policy now follows a truncated Gaussian distribution. \\

The explicit form of the solution to HJB equations enables us to obtain a policy improvement theorem which confirms that exploration can be performed within the class of (resp. truncated) Gaussian policies for the (resp. constrained) unconstrained EU problem. Moreover, by casting the optimal problem in a martingale framework as in \cite{jia2022policy}, we devise an implementable reinforcement learning 
algorithm. We observe that {compared to the classical case (without exploration), exploration procedure in the presence of a portfolio constraint leads to an optimal portfolio process that exhibits a more dispersed distribution with heavier tail. This effect becomes less significant as the exploration parameter is smaller}. Moreover, a decrease (resp. increase) in the lower (resp. upper)  bound of portfolio strategy leads to a corresponding increase in the exploration cost. This aligns with the intuitive understanding that a larger investment opportunity domain requires a higher exploration cost. Notably, when facing both short-selling and money borrowing constraints, the exploration cost becomes negligible compared to the unconstrained case. Our paper is the first attempt to explicitly analyze the classical continuous-time EU problem with possible portfolio constraints in an entropy-regularized RL framework. {Although this paper addresses specifically an optimal portfolio problem, {it is highly relevant to OR challenges from multiple perspectives.} The techniques and methodologies discussed here can be applied directly to various areas of OR that are related to decision making under model parameter uncertainty and the reliable data is used to learn the true model parameters.} {It is worth highlighting that our findings in Section 6 establish a direct connection between quadratic utility and the mean-variance (MV) framework, a cornerstone in OR for evaluating risk-return profiles under portfolio constraints. With reinforcement learning (RL) methods increasingly adopted in OR for solving dynamic stochastic problems, our work offers a versatile framework with clear applications in portfolio optimization, inventory management, supply chain systems, and other OR domains involving constrained stochastic decision-making.}\\

{While revising this paper, we acknowledge a recent work \cite{dai2023learning} where the authors consider the Merton optimal portfolio for a power utility function in a stochastic volatility setting with a recursive weighting scheme on exploration that endogenously discounts the current exploration reward by the past accumulative amount of exploration. By adopting a two-step optimization of the corresponding exploratory HBJ for an exploratory learning that incorporates the correlation between the risk asset and the volatility dynamics suggested in \cite{dai2021dynamic}, the authors in \cite{dai2023learning} are able to characterize the Gaussian optimal policy with biased structure due to incompleteness. Compared to \cite{dai2021dynamic, dai2023learning}, our paper studies the Merton problem for logarithmic and quadratic utility functions in complete market settings aiming at obtaining explicit solution, while accounting for possible portfolio constraints. We remark that the approach in \cite{dai2023learning} seems to be no longer applicable for the classical expected utility maximization when removing the endogenously recursive weighting scheme on exploration. In addition, it is not clear how the two-step optimization procedure in \cite{dai2023learning} is applied to our constrained setting.}\\
%By   makes the following
%contributions to the current literature:
%\begin{itemize}
%\item 
%\item
%\item 
%\end{itemize}

The rest of the paper is organized as follows: Section \ref{Formulation}
formulates the exploratory optimal
investment consumption
problem
incorporating both exploration and exploitation and %.Section \ref{HJB equation and optimal control distribution} studies
provides a general form of the optimal distributional control to the exploratory HJB equation. 
Section \ref{se:Log} studies the unconstrained optimal investment problem for logarithmic utility function. We elaborate the constrained EU problem with exploration and discuss the exploration cost and its impact in Section \ref{se:constrainedLog}. Section \ref{se: Learning} 
discuses an implementable reinforcement learning 
algorithm in a martingale framework and provides some numerical examples. {Section \ref{se:quadratiutility} studies the case with a quadratic utility function}. {An extension to random coefficient markets is presented in Section \ref{Sec: factor}.} Section \ref{Conclusion} concludes
the paper with future research perspectives. {Technical proofs and additional details can be found in the Appendix.}\\

{\noindent\textbf{Notation}. In this paper,
 we use $Y$ or $Y_t$
 to denote a stochastic process $Y:=\{Y_t\}_{t\in [0,T]}$;
 $Y_t$ also refers to the values of the 
 stochastic process when it is clear from the context.
 We use $\mathcal N(x|a,b^2)$
 to denote the normal distribution with mean $a$ and standard deviation $b>0$,
 $\pi_e\approx 3.14$ denotes the mathematical constant pi.
 Additionally, $f_t, f_x, f_{xx}$
 denote the partial first/second derivative with respect to the corresponding
 arguments. $\varphi(y)=\frac{1}{\sqrt{2\pi}}e^{-\frac{1}{2}y^2}$, $\Phi(y)=\int_{-\infty}^y\varphi(x)dx$
are the probability density and the probability cumulative density functions of the standard normal distribution, respectively.
 }

\section{Formulation} \label{Formulation} 
Consider a filtered probability space ($\Omega, \mathcal{F}^{{W}}, \{\mathcal{F}^{{W}}_t\}_{t \in [0,T]}, \PP^{{W}}$). Furthermore, let $\{ W_t \}_{t \in [0,T]}$ be a standard Brownian motion (restricted to $[0,T]$) under the real world measure $\PP^{{W}}$ and $\{\mathcal{F}^{{W}}_t\}_{t \in [0,T]}$ is the standard filtration of $\{ W_t \}_{t \in [0,T]}$.
For our analysis, we assume a Black-Scholes economy. To be more precise, there are two assets traded in the market, one risk-free asset $B_t$ earning a constant interest rate $r$ and one risky asset $S_t$ following a geometric Brownian motion with instantaneous rate of return $\mu > r$ and volatility $\sigma > 0$. Their dynamics are given by,
\begin{align}
\label{equ:Marketdynamics}
d B_t &= r B_t dt, \enspace B_0 = 1, \\
d S_t &= \mu S_t dt + \sigma S_t d W_t,\quad t\in [0,T], \enspace S_0 = s.
\end{align}
Consider an investor endowed with an initial wealth $x$, which will be used for investments.
We assume that the investor splits her initial wealth in the two assets given above. We use $\pi = \{\pi_t\}_{t\in[0,T]}$  to denote the fraction of wealth that the investor invests in the risky asset. The remaining money is invested in the risk-free asset. We assume that the processes $\{\pi_t\}_{t\in[0,T]}$ is  $\mathcal{F}^{{W}}$-adapted. The investor chooses an investment and consumption  strategy from the following admissible set
\begin{align*}
\mathcal{A}(x_0) :=&  \Bigg\{\pi \; \text{is progressively measurable}, \, X_t^{\pi}  \geq 0 \; \text{for all} \; t\in [0,T], \;  \int_0^{T} \pi_t^2 d t  < \infty\Bigg\}.
\end{align*}
Here, $X_t^{\pi} $ is the investor's wealth process which satisfies the following stochastic differential equation:
\begin{align} \label{eq_Xtnoc0}
d X_t^{\pi}  &= \left(r + \pi_t  (\mu-r) \right) X_t^{\pi} dt  + \sigma \pi_t  X_t^{\pi} d W_t \, , \quad X_0 = x_0>0. \end{align}
\subsection{The classic optimal investment problem}
The classical asset allocation problem can be written as
\begin{align}
& \max_{(\pi) \in \mathcal{A}(x_0) }  \mathbb{E}^{{\PP^W}} \Big[ %\int_{0}^{T}e^{-\rho t}{U}(c_tX_t^{\pi,c}) dt +
U(X_T^{\pi})\Big], \label{eq: classicaEU}%{eq: classicaEU} {eq: classicaEUnoc}
\end{align}
where $U$ is a concave utility function. %and $\rho>0$ is a discount rate parameter. 
The optimization problem \eqref{eq: classicaEU} is solely an exploitation problem,
which is a typical set up of classical stochastic control optimization problems. In a financial market with complete knowledge of the model parameters,
ones are readily to employ
the classical model-based stochastic control theory (see e.g. \cite{fleming2006controlled,pham2009continuous}), duality approach (see e.g. \cite{chen2017optimal,kamma2022near}) or to use the so-called martingale approach
(see e.g. \cite{karatzas1998methods}) to find the solution of Problem \eqref{eq: classicaEU}. When implementing the optimal strategy, one needs to estimate the
market parameters from historical time series of asset prices. However, in practice,  estimating
these parameters is notoriously
difficult with an acceptable accuracy, especially the
mean return $\mu$, also known as the mean-blur problem, see e.g. \cite{luenberger1997investment}.

\subsection{Optimal investment with exploration}

 In the RL setting where the underlying model is not known, dynamic learning is needed and the agent employs exploration to interact with and learn from the
unknown environment through trial and error. The key idea is to model exploration via a
distribution over the space of controls {$\cC\subseteq \bbr$ from which the trials are sampled. {Here we assume that the action space $\cC$ is continuous and randomization is restricted to those distributions
that have density functions.} In particular, at time $t\in[0,T]$, with the corresponding current wealth $X_t$, the agent considers a classical control $(\pi_t)$ sampled from a policy distribution $\lambda_t(\pi):=\lambda(\pi|t, X_t)$. Mathematically, such a policy $\lambda$ can be generated e.g. by introducing an additional random variable $Z$ that is uniformly distributed in $[0,1]$ and is independent of the filtration of the Brownian motion $W$, see e.g. \cite{jia2022a}. In this sense, the probability space is now expanded to ($\Omega, \mathcal{F}, \{\mathcal{F}_t\}_{t \in [0,T]}, \PP$), where $\mathcal{F}:=\mathcal{F}^{W}\vee \sigma(Z)$ and $\PP$ is now an extended probability measure on $\mathcal F$ whose projection on $\mathcal{F}^{W}$ is $\PP^{W}$.} %Therefore, the limiting setting induces a distribution control. 
This sampling is executed for $N$ rounds over the same time horizon. Intuitively, the utility of such a (feedback) policy becomes accurate
enough when N is large {by using the law of large numbers}.  This procedure, known as policy evaluation, is considered as a fundamental element of most RL algorithms in practice.  {For our continuous-time setting, we follow the exploratory setting suggested in \cite{wang2020reinforcement} and refer to e.g., \cite{wang2020reinforcement, wang2020continuous, jia2022a,jia2022policy} for motivations and additional details.}  
%For evaluating such a policy distribution in our continuous time setting, it is necessary to consider the limiting situation as $N\to\infty$.
%Let $W^i_t, i=1,\ldots, N$ be $N$ independent sample paths of the Brownian motion $W_t$ and $(\pi^i_t, c^i_t)\in \cC=\cC_\pi\times\cC_c$, $i=1,2,\ldots,N$ be $N$ samples obtained from {$\lambda_t=\lambda(\pi,c)$ at time $t$}, where {$\cC=\cC_\pi\times\cC_c$ is the agent's space of investment-consumption decisions.} Let $x_t^i,\;
 %i=1,\ldots, N$ be the copies of the wealth process respectively under the control $(\pi^i_t, c^i_t)$.
%Corresponding to each $(\pi^i_t, c^i_t)$,
%let
%{ 
%$$
%\Delta x^i_t\equiv x^i_{t+\Delta t}-x_t^i
%\approx A(\pi^i_t,c^i_t,x_t^i,)\Delta t+B(\pi^i_t,x_t^i)[W_{t+\Delta t}^i-W_t^i],
%$$ 
%}
%where
%{
%$$
%A(\pi,c,x):=(r + \pi  (\mu-r) ) x-cx;\quad B(\pi,x):=\sigma \pi x.
%$$ 
%}
  %By the law of large numbers, it holds that as $N\to\infty$,
%\begin{equation*}
%\frac{1}{N}\sum_{i=1}^N \Delta x^i_t\to \E\bigg[\int_\cC A(\pi_t,c_t,X_t) \lambda(\pi_t,c_t) d \pi_t d c_t \bigg]\Delta t,
%\label{eq:}
%\end{equation*}
%and
%\begin{equation*}
%\frac{1}{N}\sum_{i=1}^N (\Delta x^i_t)^2 \to \E\bigg[\int_\cC B^2(\pi_t,X_t) \lambda(\pi_t,c_t) d \pi_t d c_t \bigg]\Delta t,
%\label{eq:}
%\end{equation*}
%%where $\Lambda$ is a distribution (probability measure) on the space of controls  $\cC=\cC_\pi\times\cC_c$. 
%This means that both $\Delta x^i_t$ and $(\Delta x^i_t)^2$ are affected by repetitive learning under the given policy $\lambda(\pi_t,c_t)$. 
In particular,  we consider the exploratory version of the wealth dynamics \eqref{eq_Xtnoc0} given by
\begin{align} \label{eq_XtExplo}
d X_t^{\lambda} = {\hat{A}(t,X_t^{\lambda}; \lambda)} dt + {\hat{B}(t,  X_t^{\lambda}; \lambda) }d W_t \, , \quad X_0 = x_0,
\end{align}
where the exploration drift and exploration volatility are defined by
\begin{equation}
\hat{A}(t,x; \lambda):=\int_\cC A(\pi, x) \lambda (\pi|t,x) d \pi ; \quad
\hat{B}(t,\pi,x):=\sqrt{\int_\cC B^2(\pi,x)  \lambda (\pi|t,x) d \pi }.
\label{eq:hatAB}
\end{equation}
where {
$$
A(\pi,x):=(r + \pi  (\mu-r) ) x;\quad B(\pi,x):=\sigma \pi x.
$$ 
}
From \eqref{eq:hatAB} we observe that
\begin{equation}
\hat{A}{(t, x;\lambda)}=x\bigg(r+(\mu-r)\int_{\cC} \pi \lambda (\pi|t,x) d \pi%-\int_{\cC_c} c\lambda_t (c|t,x) d c
\bigg)
; \quad
\hat{B}^2{(t,x;\lambda)}=\sigma^2 x^2 \int_{{\cC} } \pi^2 \lambda(\pi|t,x) d \pi.
\end{equation}
%where with an abuse of notations $\lambda_t(\pi|t,x) $ and $ \lambda_t(c|t,x)$ are the marginal probability density of $\lambda_t(.|t,x)$ on $\cC_\pi$ and $\cC_c$ respectively. 
%In the same spirit, the exploration version of consumption at time $t$ is given by ${\E}\bigg[\int_{\cC }e^{-\rho t}{U}(cX_t^{\lambda})  \lambda_t (\pi,c|t,X_t^{\lambda}) d\pi dc\bigg]$. 
%$$
%%\frac{1}{N}\sum_{i=1}^N e^{-\rho t}{U}(c_t^i x_t^i)\Delta t\longrightarrow 
%{\E}\bigg[\int_{\cC }e^{-\rho t}{U}(c_tX_t)  \lambda_t (\pi,c) d\pi dc\bigg] %\Delta t.
%$$
{
As a result, the agent may want to maximize over all admissible policies $\lambda$ the exploratory expected investment utility $ \mathbb{E} [U(X_T^{\lambda})].$ %\int_{0}^{T}e^{-\rho t}\bigg(\int_\cC{U}(c_t X_t^{\lambda})  \lambda_t (\pi,c|t,X_t^{\lambda}) d\pi dc\bigg) dt +
%\begin{align*}
%%V(0,x;m):=& 
 %\mathbb{E} \Bigg[ %\int_{0}^{T}e^{-\rho t}\bigg(\int_\cC{U}(c_t X_t^{\lambda})  \lambda_t (\pi,c|t,X_t^{\lambda}) d\pi dc\bigg) dt +
%e^{-\rho T}U(X_T^{\lambda})\Bigg]. % \label{eq: explorationEU000}
%\end{align*}
Intuitively, the agent must account for exploration effect in her objective. As suggested in \cite{wang2020reinforcement}, for a given exploration distribution $\lambda$, the following Shannon's differential entropy penalty at time $t\in[0,T]$
\begin{equation*}
\K^\lambda(t,x):=-\int_\cC \lambda(\pi|t,x)  \ln\lambda(\pi|t,x) d \pi, 
\label{eq:}
\end{equation*}
can be used to measure the exploration impact. {Note that when the model is fully known, there would be no requirement for exploration and learning. In such a scenario, control distributions would collapse to Dirac measures, placing us in the domain of classical stochastic control \cite{fleming2006controlled,pham2009continuous}.}
%Note that when the model is fully known, exploration and learning would not
%be needed at all and the control distributions would all degenerate to the Dirac measures, and we would then be in the %realm of the classical stochastic control.
%Therefore, for the RL framework considered in this section, a regularization term that accounts for model uncertainty and exploration is needed.}  
}

{
Motivated by the above discussion, below we consider the following Shannon's entropy-regularized exploratory optimization problem}
\begin{align}
%V(0,x;m):=& 
\max_{\lambda \in \mathcal{H} }  \mathbb{E} \Bigg[ %
%\int_{0}^{T}e^{-\rho t}%\bigg(\int_\cC{U}(c_t X_t^{\lambda})  \lambda_t (\pi,c) d\pi dc
U(X_T^{\lambda})-m\int_{0}^{T}\int_\cC {\lambda(\pi|t,X_t^{\lambda})  \ln\lambda(\pi|t,X_t^{\lambda})} d \pi dt\Bigg], \label{eq: explorationEU0}
\end{align}
where $\mathcal{H}$ is the set of  admissible ({feedback}) exploration distributions (or more precisely density). {Mathematically, \eqref{eq: explorationEU0} is a relaxed stochastic control problem which has been widely studied in the control literature, see e.g. \cite{fleming1976generalized,fleming1984stochastic,nicole1987compactification}. Note that relaxed controls in general have a natural interpretation
that at each time $t$, the
agent chooses not a single action (strategy)  but rather a probability measure
over the action space $\cC$ from which a specific action is randomly sampled. {In more general settings of relaxed control e.g. \cite{fleming1976generalized,fleming1984stochastic,nicole1987compactification}, the admissible set may contain open-loop distributions that are measure-valued stochastic processes.} This is also in line with the notion of a mixed strategy in game theory, in which players choose probability measures over the action set, rather than
single actions (pure strategies).}

We note that the minus sign in front of the entropy term
accounts for the fact that exploration is inherently costly
in term of resources. Also, we can see 
that
the optimization problem \eqref{eq: explorationEU0} incorporates
both exploration (due to the entropy factor)
and exploitation.
The rate of exploration is determined by the exogenous temporal parameter
$m>0$ in the sense that a larger value of $m$ indicates that more explorations are needed 
and vice versus. Hence, {the agent can personalize her exploration rate by selecting an appropriate exogenous temporal parameter $m$. }% the agent can choose her own exploration rate by choosing a suitable exogenous temporal parameter $m$.

Next, we specify the set of admissible  controls as follows.

\begin{asp}
The set of admissible control $\lambda\in\mathcal{H}$ has the following
properties:
\begin{enumerate}
	\item For each $(t,x)\in[0,T]\times \bbr$, $\lambda (\cdot|t,x)$ is a density function on $\cC$.
	\item The mapping {$ [0,T]\times \bbr\times \cC \ni(t,x,\pi)\mapsto \lambda(\pi|t,x)$ is measurable}.
	%\item For each $E\in\cB(\cC)$, the process $\{\int_E\lambda_t(\pi)d\pi,\, t\in[0,T] \}$ is $\mathcal F$ progressively measurable;
	\item For each $\lambda\in\mathcal{H}$, the exploratory SDE \eqref{eq_XtExplo} admits a unique strong solution denoted by $X^\lambda$ {which is positive} and
	$$\mathbb{E} \Bigg[%e^{-\rho t}\bigg(\int_\cC{U}(c_tX_t^{\lambda})  \lambda_t (\pi,c) d\pi dc
	U(X_T^{\lambda})-m \int_{0}^{T}\int_\cC \lambda (\pi|t,X_t^{\lambda})\ln\lambda (\pi|t,X_t^{\lambda}) d \pi  dt\Bigg]<\infty.$$
\end{enumerate}
\end{asp}
%\thaicomment{\bf Discussion on these assumptions: TO BE ADDED}
\subsection{HJB equation and optimal distribution control}
%\label{HJB equation and optimal control distribution}
%\subsection{HJB equation}
For each $m\in\bbr$, we denote by $v(t,x;m)$ the optimal value function of Problem \eqref{eq: explorationEU0}, i.e.
\begin{align}
v(t,x;m):= \max_{\lambda \in \mathcal{H} } J(t,x,m,\lambda)\label{eq: explorationEU}:=  \mathbb{E} \Bigg[U(X_T^{\lambda})-m \int_{t}^{T}\int_\cC  \lambda (\pi|s,X_s^{\lambda})\ln\lambda (\pi|s,X_s^{\lambda})d \pi ds |X^\lambda_t=x\Bigg]\notag.
\end{align}
By standard arguments of Bellman's principle of optimality, 
%\begin{align*}
%v(t,x;m)=\sup_{\lambda \in\mathcal{H}}\E\bigg[e^{-\rho (s-t)}v(s,X_s^\lambda; m)+ \int_{t}^{s}e^{-\rho (u-t)}\bigg(\int_\cC ({U}(c_{{u}} X_{{u}}^{\lambda}) -m\ln\lambda_u(\pi,c))\lambda_u (\pi,c) d \pi d c \bigg) du|X^\lambda_t=x\bigg].
%\end{align*}
 the optimal value function $v$ satisfies the following HJB equation
\begin{equation}
v_t(t,x;m)+\sup_{\lambda}\bigg\{\hat{A}(t,x;\lambda)v_x(t,x;m)+\frac{1}{2}\hat{B}^2(t,x;\lambda)v_{xx}(t,x;m)-m\int_\cC\ln\lambda(\pi|t,x)\lambda(\pi|t,x) d \pi \bigg\}=0,\label{eq:HJBnoc}
\end{equation}
with terminal condition $v(T,x;m)=U(x)$. We observe that the formula in the bracket of \eqref{eq:HJBnoc} can be expressed as
$$
\int_\cC \bigg((r+\pi(\mu-r)))xv_x(t,x;m)+\frac{1}{2}\sigma^2 x^2 \pi^2 v_{xx}(t,x;m)-m\ln \lambda(\pi|t,x)\bigg)\lambda(\pi|t,x) d \pi.
$$ The optimal distribution $\lambda^*$ can be obtained by Donsker and Varadhan's variational formula (see e.g. \cite{donsker2006large})
\begin{equation}
\lambda^*(\pi|t, x;m){\propto \exp\bigg\{\frac{1}{m} 
\bigg((r+\pi(\mu-r)))xv_x(t,x;m)+\frac{1}{2}\sigma^2 x^2 \pi^2 v_{xx}(t,x;m)\bigg)\bigg\}},
%\frac{\exp\bigg\{\frac{1}{m} 
%\bigg((r+\pi(\mu-r)))xv_x(t,x;m)+\frac{1}{2}\sigma^2 x^2 \pi^2 v_{xx}(t,x;m)\bigg)\bigg\}}
%{\int_\cC \exp\bigg\{\frac{1}{m} \bigg((r+\pi(\mu-r)))xv_x(t,x;m)+\frac{1}{2}\sigma^2 x^2 \pi^2 v_{xx}(t,x;m)\bigg) \bigg\} d \pi},
\label{eq:lambdaopt} 
\end{equation}
which is a feedback form and can be seen as a Boltzmann distribution { (see e.g., \cite{tang2022exploratorySIAM} for a similar result)}. {Here the notation $\lambda^*(\pi|t, x;m)$ is introduced to indicate that $\lambda^*$ depends on the exploration parameter $m$.}
Note that we can write $\lambda^*(\pi|t,x;m)$ as
\begin{equation*}
\lambda^*(\pi| t, x;m)\propto\exp\left(\frac{1}{2m}\sigma^2x^2v_{xx}\big(\pi-\frac{(r-\mu)xv_x}{\sigma^2 x^2v_{xx}}\big)^2 \right).
%\exp\left(\frac{1}{m}(U(cx)-cxv_{x}) \right).
\label{eq:}
\end{equation*}
{%It appears that the exploratory problem concerning both consumption and investment remains unresolved. It would be interesting to understand the distribution that follows from equation \eqref{eq:lambdaopt}. However, in this paper, our focus is solely on the investment problem with portfolio constraints. 
which is Gaussian with mean $\alpha$ and variance $\beta^2$ ({assuming that $v_{xx}<0$}) defined by
\begin{equation}
\alpha=-\frac{(\mu-r)xv_x}{\sigma^2 x^2 v_{xx}};\quad \beta^2=-\frac{m}{\sigma^2 x^2 v_{xx}}.
\label{eq:alphabeta}
\end{equation}
Substituting \eqref{eq:lambdaopt}  back to the HJB \eqref{eq:HJBnoc} we obtain the following non-linear PDE
\begin{align}
 v_t(t,x;m)+rxv_x(t,x;m)-\frac{1}{2}\frac{(\mu-r)^2v_x^2(t,x;m)}{\sigma^2 v_{xx}(t,x;m)}+\frac{m}{2}\ln\bigg(-\frac{2\pi_e m}{\sigma^2 x^2 v_{xx}(t,x;m)}\bigg)=0, \label{eq:HJBnocsimp}
\end{align}
with terminal condition $\quad v(T,x;m)=U(x).$
%{\bf TODO}
%\begin{enumerate}
%	\item Solve \eqref{eq:HJBnocsimp} for each case: $U(x)=\ln x$, $U(x)=-e^{-\gamma x}$, $U(x)=x^{\gamma}$
%	\item In case closed-form solution does not exist, find expansion when $m\to 0$.
%\end{enumerate}

\vspace{2mm}
\section{Unconstrained optimal investment problem with exploration}\label{se:Log}
\subsection{Unconstrained optimal policy}
Below we show that the optimal solution to \eqref{eq:HJBnocsimp} can be given in explicit form for the case of logarithmic utility. {Logarithmic utility is widely recognized as the preference function for rational investors, as highlighted in \cite{rubinstein1977strong}. The study in \cite{pulley1983mean} supports this assertion by showcasing that the logarithmic utility serves as an excellent approximation in the context of the Markowitz mean-variance setting and remarkably, the optimal portfolio under logarithmic utility is almost identical to that derived from the mean-variance strategy.
It is crucial to emphasize that while our optimal strategy with logarithmic utility maintains time consistency, the mean-variance strategy is only precommitted, meaning it is only optimal at time 0. This stands out as one of the most critical drawbacks within the mean-variance setting. Logarithmic utility's versatility extends to diverse applications, including long-term investor preferences \citep{gerrard2023optimal}, optimal hedging theory and within semimartingale market models \citep{merton1975optimum}, categorizing it under the hyperbolic absolute risk aversion (HARA) class. This class has been explored in portfolio optimization under finite-horizon economies \citep{cuoco1997optimal}, infinite-horizon consumption-portfolio problems \citep{el1998optimization}, and scenarios involving allowed terminal debt \citep{chen2017optimal}.}

%We finally remark that logarithmic utility is widely used in various contexts, such as long-term investor preferences \citep{gerrard2023optimal}), optimal hedging theory \citep{merton1975optimum} and in general semimartingale market models \citep{goll2000optimal}. It is also part of the so-called hyperbolic absolute risk aversion (HARA) class of utility functions, explored in portfolio optimization under finite-horizon economies \citep{cuoco1997optimal}, infinite-horizon consumption-portfolio problems \citep{el1998optimization}, allowed terminal debt \citep{chen2017optimal}.  

We summarize the results {for the unconstrained case i.e. $\cC=\bbr$} in the following theorem . %Recall that $\pi_e=3.14\ldots$ is the regular constant pi.
\begin{thm}\label{Thm:logx}
 The optimal value function of the entropy-regularized exploratory optimal investment problem with logarithmic utility $U(x)=\ln x$ is given by \footnote{{Recall that $\pi_e=3.14\ldots$ is the regular constant pi.}}
\begin{align}
v(t,x;m)=&\ln x+
 \bigg(r+\frac{1}{2}\frac{(\mu-r)^2}{\sigma^2} \bigg)(T-t)+\frac{m}{2} \ln(\sigma^{-2}{2\pi_e m})(T-t)
\label{eq:valueopt}
\end{align}
for $(t,x)\in[0,T]\times \bbr_+$. Moreover, the optimal feedback distribution control $\lambda^*$ (which is independent of $(t,x)$) is a Gaussian with mean $\frac{(\mu-r) }{\sigma^2}$ and variance $\frac{m}{\sigma^2 }$, i.e.
\begin{equation}
\lambda^*(\pi)=\mathcal N\left(\pi\bigg|\frac{(\mu-r) }{\sigma^2}, \frac{m}{\sigma^2 }\right)
\label{eq:lambdaopt}
\end{equation}
and the associated optimal wealth under $\lambda^*$ is given by the following SDE
\begin{align} 
d X_t^{\lambda*} = X_t^{\lambda^*} \bigg(r+\frac{(\mu-r)^2 }{\sigma^2}\bigg) dt + \sqrt{\bigg(m +\frac{(\mu-r)^2 }{\sigma^2}\bigg)} X_t^{\lambda^*}  d W_t \, , \quad X_0 = x_0 \,.\label{eq:Wealthoptlamb}
\end{align}
\end{thm}
From Theorem \ref{Thm:logx} we can see that the {best} control distribution to balance exploration and exploitation
is {Gaussian}. This demonstrates
the popularity of Gaussian distribution in RL studies.  The {mean} of the optimal distribution control is
given by the {Merton} strategy 
\begin{equation}
\pi^{Merton}:=\frac{\mu-r }{\sigma^2},
\label{eq:piMerton}
\end{equation}
whereas the {variance} of the optimal Gaussian policy is controlled by the degree of exploration $m$. We also observe that at any $t\in [0,T]$, the exploration variance {decreases} as $\sigma$ increases, which means that exploration is less necessary in more random market environment.

\begin{rem}\label{Rem:1}
The optimal value function of the regularized exploration problem can be expressed as
\begin{align}
v(t,x;m)= & v(t,x;0)+ \psi(t,x;m),
\label{eq:valuerela}
\end{align}
where 
$$
 \psi(t,x;m):= \frac{m}{2}\ln(\sigma^{-2}{2\pi_e m}) (T-t)
$$
Intuitively, $ \psi(t,x;m)$ measures the exploration effect. It is easy to see that  $ \psi(t,x;m)\to 0$ as $m\to 0$, hence, $v(t,x;m)\to v^{Merton}(t,x)=v(t,x;0)=\ln x+(r+\frac{1}{2}\frac{(\mu-r)^2}{\sigma^2})(T-t)$ which is the optimal value function in the absence of exploration. %Clearly, this is equivalent to $k(t)\to 1$ and $l(t)\to (r+\frac{1}{2}\frac{(\mu-r)^2}{\sigma^2}) $. 
\end{rem}

The following shows solvability equivalence between the classical and exploratory EU problem, { meaning that the solution to one directly provides the solution to the other, without requiring a separate resolution.} 
%\subsection{ }This can be done easily
\begin{thm}\label{Thm:equivalence}
The following statements are equivalent %{hold true}
\begin{enumerate}
	\item [(a)]	The function \begin{align*}
v(t,x;m)=& \ln x+(r+\frac{1}{2}\frac{(\mu-r)^2}{\sigma^2}) (T-t)+ \frac{m}{2}\ln(\sigma^{-2}{2\pi_e m}) (T-t)
\end{align*}
is the optimal value function of the exploratory problem. 
%\begin{align}
%v(0,x;m):=& \max_{\lambda \in \mathcal{H} }  \mathbb{E} \Bigg[ \ln(X_T^{\lambda})-m\int_\cC \lambda_t \ln\lambda_t(\pi) d \pi  dt \Bigg], \label{eq: explorationLogEU}
%\end{align}
%\begin{align} \label{eq_XtnocExplo}
%d X_t^{\lambda} = & \bigg((r+(\mu-r)\int_\cC \pi_t \lambda_t (\pi_t) d \pi )  dt + \sigma \int_\cC \pi_t  \lambda_t (\pi) d \pi_t d W_t \bigg) X_t^{\lambda}\, , \quad X_0 = x_0 \,.
%\end{align}
The optimal distribution control $\lambda^*(\pi| t,x;m)$ is given by \eqref{eq:lambdaopt} 
%can be expressed as a Gaussian density
 %\begin{equation}
%\lambda^*(\pi; t,x,m):=\mathcal N\bigg(\pi\bigg|
 %\frac{(\mu-r) }{\sigma^2},\frac{m}{\sigma^2 }\bigg).
%\end{equation}
and the exploratory wealth process is given by \eqref{eq:Wealthoptlamb}.
\item [(b)] 	The function 
$$v^{Merton}({t},x):=v(t,x;0)=\ln x+ (r+\frac{1}{2}\frac{(\mu-r)^2}{\sigma^2}) (T-t)$$
is the optimal value function of the classic (without exploration) EU problem 
%\begin{align}
%& \max_{\lambda \in \mathcal{H} }  \mathbb{E} [ \ln(X_T^{\lambda})], \label{eq: LogEU}
%\end{align}
where the optimal investment strategy is given by the Merton fraction $\pi_t^{Merton}=\frac{(\mu-r) }{\sigma^2}$.

\end{enumerate}
\end{thm}
\proof
It is easy to see that the $v(t,x;m)$ and $v(t,x;0)$  solve the HJB equation \eqref{eq:HJBnocsimp} with and without exploration respectively. The admissibility of these two optimal strategies are straightforward to see. 
\endproof
%\subsection{Exploration EU with exponential utility}
\subsection{Exploration cost and exploration effect }\label{se:Explorationcost}
The exploration cost for a general RL problem is defined by the difference between the expected utility following the corresponding optimal control under the classical objective and the exploratory objective, net of the value of
the entropy. Remark that the exploration cost is well defined only if both the
classical and the exploratory problems are solvable.
Let $v(t,x;0)$ be the value function of the classical EU problem. {Below, we often write for shorthand $\lambda_s(\pi)=\lambda(\pi|s,X^{\lambda}_s)$ when there is no confusion.} The exploration cost at time $t=0$ is defined by
\begin{equation}
v(0,x;0)-\bigg(v(0,x;m)+m \E\bigg[\int_{0}^{T}\int_\cC \lambda_s^* (\pi)\ln\lambda_s^*(\pi) d \pi  ds |X^{\lambda^*}_0=x\bigg]\bigg). 
\label{eq:exploration cost0}
\end{equation}
The first term is the classical value function at time $t=0$. The second term is the expected utility (before being regularized) of the exploration problem at time $t=0$. 
\begin{prop}
\label{eq:ExploratoryTheorem}
Assume that one the statements in Theorem \ref{Thm:equivalence} holds. Then, the exploration cost is given by $\frac{mT}{2}$.
\end{prop}
\proof By Remark \ref{Rem:1}, the optimal value function of the regularized exploration problem can be expressed as
\begin{align}
v(0,x;m)= & v^{Merton}(0,x)+ \frac{m}{2}T\ln(\sigma^{-2}{2\pi_e m}). 
\label{eq:}
\end{align}
On the other hand, the entropy term can be computed using the Gaussian form of $\lambda^*$ as follows
\begin{align}
\int_{0}^{T}\E\bigg[\int_{-\infty}^{+\infty} \lambda_s^* (\pi) \ln\lambda_s^*(\pi) d \pi   |X^{\lambda^*}_0=&x\bigg]ds=T\int_{-\infty}^{+\infty} \big(-\frac{1}{2\beta^2}(\pi-\alpha)^2-\ln(\sqrt{2\pi_e \beta^2})\big)\mathcal N (\pi|\alpha,\beta^2) d \pi\notag\\
& =-T(\ln\sqrt{\frac{2m\pi_e}{\sigma^2 }}+\frac{1}{2}). \label{eq:gaussianentropy}
\end{align}
Substituting this into \eqref{eq:exploration cost0} leads to the desired conclusion.\endproof
%Intuitively, $ \psi(t,x;m)$ measures the exploration effect. It is easy to see that  $ \psi(t,x;m)\to 0$ as $m\to 0$, hence, $v(t,x;m)\to v(t,x;0)=\ln x+(r+\frac{1}{2}\frac{(\mu-r)^2}{\sigma^2})$ which is the optimal value function in the absence of exploration. Clearly, this is equivalent to $k(t)\to 1$ and $l(t)\to (r+\frac{1}{2}\frac{(\mu-r)^2}{\sigma^2}) $. 

Observe that the exploration cost converges to zero as $m\to 0$. 
Since the exploration weight as been taken as an exogenous parameter $m$ which reflects the level of exploration desired by the learning agent, it is intuitive to expect that the smaller $m$ is, the more
emphasis is placed on exploitation. Moreover, when $m$ is sufficiently close to zero, the exploratory formulation is getting close to the problem without exploration. The following theorem confirms a
desirable result that the entropy-regularized EU problem converges to its classical EU counterpart when the exploration weight $m$ goes to zero.
\begin{thm}\label{m-consistency}
Assume that one the statements in Theorem \ref{Thm:equivalence} holds. Then, for each $x\in\mathbb R_+$, it holds that
 \begin{equation}
\lambda^*(\cdot|t,x;m)\to \delta_{\pi^{Merton}}(\cdot)\quad\mbox{when}\quad m\to 0.
\end{equation} 
where $\delta_{\pi^{Merton}}$ is the Dirac measure at the Merton strategy. 
Furthermore, for each $(t,x)\in[0,T]\times \bbr_+$,
$$
\lim_{m\to0}|v(t,x;m)-v^{Merton}(t,x)|=0.
$$
\end{thm}
\proof The convergence of the optimal distribution control $\lambda^*$ to the Dirac measure $\delta_{\pi^{Merton}}$ is straightforward because it is Gaussian density with mean identical to $\pi^{Merton}=\frac{(\mu-r) }{\sigma^2}$ and variance $\frac{m}{\sigma^2}\to 0$ as $m\to0$. The convergence of the value function follows directly from the relation \eqref{eq:valuerela} in Remark \ref{Rem:1}. \endproof
\subsection{Policy improvement}
The following policy improvement theorem is crucial for interpretable RL algorithms as it ensures that the iterated value functions is non-decreasing and  converges to the optimal value function. 
Below, for some given policy $\lambda$ (not necessarily Gaussian), we denote the corresponding value function by
\begin{align}
&v^{\lambda}(t,x;m):=\mathbb{E} \Bigg[U(X_T^{\lambda})-m\int_t^T \int_{-\infty }^{+\infty} \lambda_s (\pi)  \ln\lambda_s(\pi)  d \pi|X^{\lambda}_t=x\Bigg].
\label{eq:valuefunction}
\end{align}
%{where $U(x)$ =\ln x$.}
\begin{thm}\label{Thm-update}
%\coprod
For some given admissible policy $\lambda$ (not necessarily Gaussian), we assume that
the corresponding value function
%\begin{align}
%&v^{\lambda}(t,x;m):=\mathbb{E} \Bigg[U(X_T^{\lambda})-m\int_t^T \int_\cC \lambda_s (\pi_s)  \ln\lambda_s(\pi_s)  d \pi_s|X^{\lambda}_t=x\Bigg].
%\end{align}
 $v^\lambda(\cdot,\cdot; m) \in C^{1,2}([0,T)\times \bbr_+)\cap C([0,T]\times \bbr_+)$ and satisfies $v^\lambda_{xx}(t,x,; m)<0$ for any $(t,x)\in[0,T)\times \bbr_+.$ Suppose furthermore that the feedback policy $\wt{\lambda}$ defined by
\begin{equation}\label{eq:pischeme}
\wt{\lambda}(\pi| t,x; m)= \mathcal N\bigg(\pi\bigg|
-\frac{(\mu-r)xv_x^\lambda}{\sigma^2 x^2 v_{xx}^{\lambda}},-\frac{m}{\sigma^2 x^2 v_{xx}^\lambda}\bigg)
\end{equation}
is admissible. Let $v^{\wt{\lambda}}(t,x; m)$ be the value function corresponding to this new (Gaussian) policy $\wt{\lambda}$. Then, 
\begin{equation}
v^{\wt{\lambda}}(t,x; m)\ge v^{\lambda}(t,x; m), \quad  (t,x)\in[0,T)\times \bbr_+.
\label{eq:}
\end{equation}
\end{thm}

{Note that Theorem \ref{Thm-update} holds true for general utility function $U$.}  One important implication of Theorem \ref{Thm-update} is that 
for any given (not necessarily
Gaussian) policy $\lambda$, there are always policies in the Gaussian
family that improve the value function of $\lambda$ (i.e. providing higher expected utility values). Therefore, it is intuitive to focus
on the Gaussian policies when choosing an initial exploration distribution. Note that the
optimal Gaussian policy given by \eqref{eq:lambdaopt} also suggests that a candidate of the initial
feedback policy may take the form ${\lambda}^0(\pi| t,x, m)= \mathcal N\big(\pi\big|
a, b^2\big)$ {for some real number $a,b$}. As shown below, such a choice leads to a {fast} convergence of both
value functions and policies in a finite number of iterations.
\begin{thm}\label{Th:policyupdatelog}
Let ${\lambda}^0(\pi|t,x, m)= \mathcal N\big(\pi\big|
 a, b^2 \big)$ with some real number $a,b >0$ {and consider the logarithmic utility function $U(x)=\ln x$}. Define the sequence of feedback policies $\{\lambda^n  (\cdot| t,x; m)\}_{n\ge 1}$ updated by the policy improvement scheme \eqref{eq:pischeme}, 
i.e.,
\begin{equation}
\lambda^{n}(\pi| t,x; m)= \mathcal N\bigg(\pi\bigg|
-\frac{(\mu-r)xv_x^{\lambda^{n-1}}(t,x; m)}{\sigma^2 x^2 v^{\lambda^{n-1}}_{xx}(t,x; m)},-\frac{m}{\sigma^2 x^2 v^{\lambda^{n-1}}_{xx}(t,x; m)}\bigg),\quad n=1,2,\ldots,
\end{equation}
where $v^{\lambda^{n-1}}$ is the value function corresponding to the policy $\lambda^{n-1}$ defined by
\begin{align}
v^{\lambda^{n-1}}(t,x;m):=\mathbb{E} \Bigg[U(X_T^{\lambda^{n-1}})-m\int_t^T \int_{-\infty }^{+\infty}  \lambda^{n-1}  (\pi|s,X^{\lambda^{n-1}}_s) \ln\lambda^{n-1} (\pi|s,X^{\lambda^{n-1}}_s)  d \pi|X^{\lambda^{n-1}}_t=x\Bigg].
\end{align}
Then, 
\begin{equation}
\lim_{n\to \infty}\lambda^{n}(\cdot| t,x; m)=\lambda^*(\cdot| t,x; m),\quad \mbox{weakly}
\label{eq:}
\end{equation}
and 
\begin{equation}
\lim_{n\to \infty}v^{\lambda^n}( t,x; m)=v^{\lambda^*}( t,x; m)=v(t,x;m),
\label{eq:}
\end{equation}
which is the optimal value function given by \eqref{eq:valueopt}.
\end{thm}

As shown {in Appendix \ref{Ap: update}}, starting with a Gaussian policy  the optimal solution of the exploratory EU with logarithmic utility is attained rapidly (after {one update}). 
{\begin{remark}It is valuable to extend Theorem \ref{Th:policyupdatelog} to contexts with general utility functions. However, obtaining closed-form solutions for highly nonlinear average PDEs (see Appendix \ref{Ap: update}) is not feasible. Note that even the solution existence of the next-step average PDE is not obvious and would require a thorough analysis, including asymptotic expansions. %We leave this as an interesting direction for future research.
\end{remark}}
\subsection{Policy evaluation and algorithm}\label{se:PE}
%\subsection{Policy evaluation with martingale condition}
Following \cite{jia2022a,jia2022policy} we consider in this section the policy evaluation that takes the martingale property into account. Recall with an abusive notation the value function
\begin{equation}
J(t,x;{\lambda}):=\E\bigg[U(X_T^\lambda) -m\int_t^T \int_{-\infty }^{+\infty} \lambda (\pi|s,X^{\lambda}_s)  \ln\lambda (\pi|s,X^{\lambda}_s)  d \pi ds\bigg|X_t^\lambda=x\bigg],
\label{eq:}
\end{equation}
{where $\lambda_t$ is a given learning (feedback) policy.} By Feymann-Kac's theorem, it can be seen that $J(t,x;{\lambda})$ solves (e.g. in viscosity sense) the following average PDE
\begin{equation}
\int_{-\infty }^{+\infty} \bigg(\mathcal L J(t,x;\lambda)-m  \ln\lambda (\pi|t,x)\bigg) \lambda (\pi|x,t)  d \pi =0,
\label{eq:averagePDE}
\end{equation}
where $\mathcal L$ is the infinitesimal operator associated to $X_t^\lambda$, i.e.
\begin{equation}
\mathcal L J(t,x;\lambda):= J_t(t,x;\lambda)+\hat{A}({t,x;}\lambda)J_x(t,x;\lambda)+\frac{1}{2}\hat{B}^2({t,x;}\lambda)J_{xx}(t,x;\lambda).
\label{eq:}
\end{equation}
As in \cite{jia2022a,jia2022policy}, the value function of a policy $\lambda$ can be characterized by the following martingale property. 
\begin{thm}\label{Thm:martingale}
A function $J(\cdot,\cdot;{\lambda})$ is the value function associated with the policy $\lambda$ if and only if it satisfies the terminal condition $J(T,x;{\lambda})=U(x)$, and
\begin{align}
Y_s:=J(t,X^{\lambda}_s;{\lambda})-m\int_t^s \int_{-\infty }^{+\infty}  \lambda (\pi|u,X^{\lambda}_u)  \ln\lambda (\pi|u,X^{\lambda}_u)  d \pi du, \quad {s\in[t,T]}
\end{align}
is a martingale on $[t,T]$.
\end{thm}
%\proof The proof is similar to that in \cite{jia2022a,jia2022b} and can be done as follows. First, by the Markov property of the process $X^\lambda_t$, it can be observed directly that
%\begin{align}
%Y_s&=J(t,X^{\lambda}_s;{\lambda})-m\int_t^s \int_\cC \lambda (\pi|X^{\lambda}_u,u)  \ln\lambda (\pi|X^{\lambda}_u,u)  d \pi du\notag\\&=
%\E\bigg[U(X_T^\lambda) -m\int_s^T \int_\cC \lambda (\pi|X^{\lambda}_u,u)  \ln\lambda (\pi|X^{\lambda}_u,u)  d \pi du\bigg|X_s^\lambda\bigg]-m\int_t^s \int_\cC \lambda (\pi|X^{\lambda}_u,u)  \ln\lambda (\pi|X^{\lambda}_u,u)  d \pi du\notag\\
%&=
%\E\bigg[U(X_T^\lambda) -m\int_t^T \int_\cC \lambda (\pi|X^{\lambda}_u,u)  \ln\lambda (\pi|X^{\lambda}_u,u)  d \pi du\bigg|X_s^\lambda\bigg]=\E\big[Y_T|X_s^\lambda\big]\notag
%.
%\end{align} 
%\endproof
It follows from Itô's Lemma and Theorem \ref{Thm:martingale} that $\E[\int_0^T h_s d Y_s]=0$ for all adapted square-integrable processes $h$ satisfying $\E[\int_0^T h^2_s d \langle Y\rangle_s]<\infty $. Equivalently, 
\begin{align}
\E\bigg[\int_0^T h_t \bigg(d J(t,X^{\lambda}_t;{\lambda})-m\int_{-\infty }^{+\infty} \lambda (\pi|t,X^{\lambda}_t)  \ln\lambda (\pi|t,X^{\lambda}_t)  d \pi\bigg) dt\bigg]=0.
\end{align}
Such a process $h$ is called test function. Let $J^\theta(t,X^{\lambda}_t;{\lambda})$ be a parameterized family that is used to approximate $J$, where $\theta\in\Theta\subset \bbr^n, n\ge 1$. { Here $J^\theta$ satisfies Assumption \ref{Ass:1} below and our goal is to find the best parameter $\theta^*$ such that the martingale property in Theorem \ref{Thm:martingale} holds. In this sense,}  the process $$Y_t^{\theta^*}=J^{\theta^{*}}(t,X^{\lambda}_t;{\lambda})-m\int_0^t \int_{-\infty }^{+\infty}  \lambda (\pi|u,X^{\lambda}_u)  \ln\lambda (\pi|u,X^{\lambda}_u)  d \pi du$$ {should be} a martingale in $[0,T]$ with terminal value $$Y_T^{{\theta^*}}=U(X^\lambda_T)-m\int_0^T \int_{-\infty }^{+\infty} \lambda (\pi|u,X^{\lambda}_u)  \ln\lambda (\pi|u,X^{\lambda}_u)  d \pi du.$$ {As discussed in \cite[page 18]{jia2022a}, the martingale condition of $Y^{\theta^*}$ is further equivalent to a requirement that the process at any given time $t<T$ is the
expectation of the terminal value conditional on all the information available at that time. A fundamental property of the conditional expectation then yields that $Y^{\theta_t^*}$ minimizes the $L^2$-error between $Y^\theta_T$ and any ${\mathcal F}_t$-measurable random variables. Therefore,} {our objective is to minimize the martingale cost function defined by} $\E[\int_0^T (Y_T^ {\theta}-Y_t^{\theta})^2 dt]$. In other words, for the policy evaluation (PE) procedure, we look at the following (offline) minimization problem
\begin{equation}
\min_{\theta\in\Theta} \E\bigg[\int_0^T \bigg( U(X^{\lambda}_T)- J^\theta(t,X^{\lambda}_u;{\lambda})-m\int_t^T \int_{-\infty }^{+\infty}  \lambda (\pi|u,X^{\lambda}_u)  \ln\lambda (\pi|u,X^{\lambda}_u)  d \pi d u\bigg)^2 dt\bigg]=0
\label{eq:}
\end{equation}
and check the martingale orthogonal property
\begin{align}\label{eq:martJtheta}
\E\bigg[\int_0^T h_t \bigg(d J^\theta(t,X^{\lambda}_t;{\lambda})-m\int_{-\infty }^{+\infty} \lambda (\pi|t,X^{\lambda}_t)  \ln\lambda (\pi|t,X^{\lambda}_t)  d \pi\bigg) dt\bigg]=0.
\end{align}
The orthogonality condition \eqref{eq:martJtheta} can be assured by solving the following minimization problem {(which is a quadratic form of \eqref{eq:martJtheta})} \footnote{{Our procedure is inspired from \cite[page 22]{jia2022a}. In particular, we are
optimizing with respect to the parameter $\theta$ which is a vector in general. In fact, we have to make
sure that the parameterization satisfies the martingale orthogonality condition \eqref{eq:martJtheta}) by choosing a suitable test function $h$. For numerical
approximation methods involving a parametric family $\{J^\theta, \theta\in\bbr^n\}$, in principle, we need at least $n$
equations as our martingale orthogonality conditions in order to fully determine $\theta$. Consequently, $h$
may be vector-valued, making \eqref{eq:martJtheta} a vector-valued equation or, equivalently, a system of equations.
}}
\begin{align}
\label{eq:changeinJtheta}
\min_{\theta\in\Theta} & \E\bigg[\int_0^T h_t \bigg(d J^\theta(t,X^{\lambda}_t;{\lambda})-m\int_{-\infty }^{+\infty} \lambda (\pi|X^{\lambda}_t,t)  \ln\lambda (\pi|t,X^{\lambda}_t)  d \pi\bigg) dt\bigg]^\intercal \times {\Sigma} \times\notag\\
&\E\bigg[\int_0^T h_t \bigg(d J^\theta(t,X^{\lambda}_t;{\lambda})-m\int_{-\infty }^{+\infty}  \lambda (\pi|t,X^{\lambda}_t)  \ln\lambda (\pi|t,X^{\lambda}_t)  d \pi\bigg) dt\bigg],
\end{align}
where $\Sigma$ is a positively definite matrix with suitable size (e.g. $\Sigma=I$ or $\Sigma= \E[\int_0^T h_t^\intercal h_t dt]$) for any test process $h$. A common choice of test process is to take $h_t=\frac{\partial}{\partial \theta} J^\theta(t,X^{\lambda}_t;\lambda)$. The following assumption\footnote{There is a rich theory on solution existence and regularity of general parabolic PDEs, see e.g. \cite{friedman2008partial}. However, the average PDE \eqref{eq:averagePDE} appears to be a new type of PDEs for which further studies on well-posedness (existence and
uniqueness) and regularity of solution are needed. See \cite{tang2022exploratorySIAM} for related discussions on a similar exploratory elliptic PDEs. } is needed in such an approximation procedure, {see e.g. \cite{jia2022policy}.} 

\begin{asp}\label{Ass:1} For all  $\theta\in\Theta\subset \bbr^n$, $J^\theta$ is a $C^{1,2}([0,T)\times \bbr_+)\cap C([0,T]\times \bbr_+)$ with polynomial growth condition in $x$. {Moreover, $J^\theta$ is smooth in $\theta$ and its two first derivatives are $C^{1,2}([0,T)\times \bbr_+)\cap C([0,T]\times \bbr_+)$ and satisfy the polynomial growth condition in $x$.}
\end{asp}

%\subsection{Policy gradient algorithm}
Suppose that a PE step can be proceeded to obtain an estimate of the value function for an given admissible policy $\lambda$. If the policy $\lambda^\phi$ can be parameterized by $\phi\in\Phi\subset \bbr^d, {d\ge1}$, it is possible to learn the corresponding value function by following a policy gradient (PG) step. Recall first that the value function  $J(t,x,\lambda^\phi)$ satisfies the average PDE
\begin{equation}
\int_{-\infty }^{+\infty} \bigg(\mathcal L J(t,x;\lambda^\phi)-m \ln\lambda^\phi (\pi|t,x)\bigg) \lambda^\phi (\pi|t,x)   d \pi=0.
\label{eq:Jphi}
\end{equation}
By differentiating \eqref{eq:Jphi} w.r.t. $\phi$ we obtain the following average PDE for the policy gradient $g(t,x;\phi):=\frac{\partial}{\partial\phi} J(t,x,\lambda^\phi)$
\begin{equation}
\int_{-\infty }^{+\infty} \bigg( \psi (t,x,\pi;\phi)+\mathcal L g(t,x;\phi)\bigg) \lambda^\phi (\pi|t,x)    d \pi=0,
\label{eq:Jphi2}
\end{equation}
with terminal condition $ g(T,x;\phi)=\frac{\partial}{\partial\phi} J(T,x;\lambda^\phi)=0$, 
where 
\begin{equation}
\psi (t,x,\pi;\phi):=\bigg(\mathcal L  J(t,x;\lambda^\phi)-m  \ln\lambda^\phi(\pi|t,x)\bigg)\frac{\frac{\partial}{\partial\phi} \lambda^\phi(\pi|t,x)}{\lambda^\phi(\pi|t,x)}-m \frac{\partial}{\partial\phi}\ln \lambda^\phi(\pi|t,x).
\label{eq:}
\end{equation}
By Feynman-Kac's formula, $g(t,x;\phi)$ admits the following representation
\begin{equation}
g(t,x;\phi)=\E\Big[\int_t^T \int_{-\infty }^{+\infty} \psi (s,X^{\lambda^\phi}_s,\pi;\phi) \lambda^\phi (\pi|s,X^{\lambda^\phi}_s)  d \pi ds|X^{\lambda^\phi}_t=x\Big].
\label{eq:}
\end{equation}
Since $\psi $ cannot be observed nor computed without knowing the environment, it is important to replace $\psi (s,X^{\lambda^\phi}_s,\pi;\phi) ds$ by a suitable approximation. To this end, using Itô's formula, the term $\mathcal L J(s,X^{\lambda^\phi}_s;\lambda^\phi) ds $ can be replaced by $d J(s,X^{\lambda^\phi}_s;\lambda^\phi)-\frac{\partial}{\partial {x}}J(s,X^{\lambda^\phi}_s;\lambda^\phi) \hat{B}(\pi,\lambda_s^\phi)dW_s$. Taking into account the fact that the term $\frac{\partial}{\partial {x}}J(s,X^{\lambda^\phi}_s;\lambda^\phi) \hat{B}(\pi,\lambda_s^\phi)dW_s$ is a martingale, we obtain the following.

\begin{thm}
Let $\lambda^\phi$ be an admissible policy. The policy gradient $g(t,x;\phi)=\frac{\partial}{\partial\phi} J(t,x,\lambda^\phi)$ admits the following representation
\begin{align}
g(t,x;\phi)&=\E\Bigg[\int_t^T \int_{-\infty }^{+\infty}  \bigg(\frac{\partial}{\partial\phi} \ln \lambda^\phi(\pi|s,X^{\lambda^\phi}_s)\big(
d J(s,X^{\lambda^\phi}_s;\lambda^\phi)-m \ln \lambda^\phi(\pi|s,X^{\lambda^\phi}_s)\big)\notag\\
&-m \frac{\partial}{\partial\phi}\ln \lambda^\phi(\pi|s,X^{\lambda^\phi}_s)\bigg)\lambda^\phi(\pi|s,X^{\lambda^\phi}_s)d\pi ds \bigg | X^{\lambda^\phi}_t=x\Bigg].
\label{eq:derivationwrtphi}
\end{align}
\end{thm}
\proof The proof is similar to that of Theorem 5 in \cite{jia2022policy}, and hence is omitted. \endproof

Note that for any {arbitrary} policy, the gradient of the corresponding value function given by the above expectation \eqref{eq:derivationwrtphi} is not zero in general. As mentioned in \cite{jia2022a,jia2022policy}, all the terms inside the expectation above are all only computable when the action trajectories and the corresponding state trajectories on $[t, T]$ are available. The sample is also needed to estimate the value function $J$ obtained in the previous PE step.

\section{Constrained optimal investment problem with exploration}\label{se:constrainedLog}
In this section, we extend the results {obtained in Section \ref{se:Log}} to settings with a convex portfolio constraint see e.g. \cite{cuoco1997optimal}. In particular, we assume that the agent does not consume over her investment period and in addition, due to regulatory reasons, her risky investment ratio $\pi$ is set in a given interval $[a,b]$, where $a<b$ are real numbers. Observe that the well-known short-selling constraint can be included by taking $a=0$, $b>0$. If, in addition $b=1$, then both short-selling and money borrowing are prohibited. The case of buying constraint can be also covered by setting $b=0$. Clearly, we are back to the unconstrained setting when sending $a\to-\infty$ and $b\to +\infty$.  In such a constrained framework, the set of admissible investment strategies is now defined by
\begin{align*}
\mathcal{A}_{[a,b]}(x_0) :=&  \Bigg\{ (\pi)\,\, \bigg \vert \,\; a\le\pi \le b\; \text{is progressively measurable}, \, X_t^{\pi} \geq 0 \; \text{for all} \; t\geq 0, \;  \int_0^{T} \pi_t^2 d t  < \infty\Bigg\},
\end{align*}
where $X_t^{\pi}$ is the non-exploration wealth process starting with  $x$ for strategy $\pi$. The objective is to maximize the terminal expected utility
\begin{align}
& \max_{\pi \in \mathcal{A}_{[a,b]}(x_0) }  \mathbb{E} \Big[U(X_T^{\pi})\Big], \label{eq: constrainedEUnoc}
\end{align}
where $U$ is a utility function. 

As in the unconstrained case, we look at the exploratory version of the wealth dynamics given by \eqref{eq_XtExplo} with exploration drift {$\hat{A}(t,x; \lambda)$} and exploration volatility {$\hat{B}(t,x;\lambda)$} are defined by \eqref{eq:hatAB}. Intuitively, the corresponding exploration setting is slightly adjusted to taking the constraint $\pi_t\in[a,b]$ into account. In particular, let $\mathcal{H}_{[a,b]}$ be the set of admissible exploration distributions $\lambda$ in $\cC =[a,b]$ that satisfy the following
properties:
%{ADJUST THE BELOW ASSUMPTION}
\begin{enumerate}
	\item  For each $(t,x)\in[0,T]\times \bbr$, $\lambda (\cdot|t,x)$ is a density function on $[a,b]$.%For each $t\in[0,T]$, $\lambda_t$ is a density function on $[a,b]$ a.s.;
	\item The mapping {$ [0,T]\times \bbr\times [a,b] \ni(t,x,\pi)\mapsto \lambda(\pi|t,x)$ is measurable}.%For each $E\in\cB([a,b])$, the process $\{\int_E\lambda_t(\pi)d\pi,\, t\in[0,T] \}$ is $\mathcal F_t$ progressively measurable;
	\item For each $\lambda\in\mathcal{H}_{[a,b]}$, the exploration SDE \eqref{eq_XtExplo} admits a unique strong solution denoted by $X^{\lambda^{{,[a,b]}}}$ {which is positive} and
	$\mathbb{E} \bigg[ U(X_T^{\lambda^{{,[a,b]}}})-m\int_0^T\int_a^b \lambda(\pi|t,X^{\lambda^{{,[a,b]}}}_t) \ln\lambda(\pi|t,X^{\lambda^{{,[a,b]}}}_t) d \pi dt \bigg]<\infty.$
\end{enumerate}
The exploration optimization is now stated by 
\begin{align} \max_{\lambda\in\mathcal{H}_{[a,b]  }}  \mathbb{E} \bigg[ U(X_T^{\lambda^{{,[a,b]}}})-m\int_0^T {\int_a^b} \lambda(\pi|t,X^{\lambda^{{,[a,b]}}}_t) \ln\lambda(\pi|t,X^{\lambda^{{,[a,b]}}}_t) d \pi dt\bigg].
\end{align}

As before, the optimal value function satisfies the following HJB equation
\begin{align}
 v^{[a,b]}_t(t,x;m)+\sup_{\lambda\in\mathcal{H}_{[a,b]}}\bigg\{{\hat{A}(t,x;\lambda)}v^{[a,b]}_x(t,x;m)&+\frac{1}{2} {\hat{B}^2(t,x;\lambda)}v^{[a,b]}_{xx}(t,x;m)\notag\\
&-m{\int_a^b}  \lambda(\pi|t,x) \ln\lambda(\pi|t,x) d \pi \bigg\}=0,\label{eq:constrainedHJBnoc}
\end{align}
with terminal condition $v(T,x;m)=U(x)$.
%First, observe that the formula in the bracket of \eqref{eq:HJB} can be expressed as
%$$
%\int_\cC \bigg((r+\pi(\mu-r))xv_x(t,x;m)+\frac{1}{2}\sigma^2 x^2 \pi^2 v_{xx}(t,x;m) -m\ln\lambda_t\bigg)\lambda_t (\pi) d \pi.
%$$
Using again the standard argument of {DPP} we observe that under the portfolio constraint $\pi\in[a,b]$, the optimal feedback policy now follows a truncated Gaussian distribution. 
\begin{lm}
In the exploratory constrained EU setting, {if $v^{[a,b]}_{xx}<0$} then the optimal distribution $\lambda^{*,[a,b]}$ is a Gaussian distribution with mean $\alpha$ and variance $\beta^2$, truncated on interval $[a,b]$, where 
\begin{equation}
\alpha=-\frac{(\mu-r)xv_x}{\sigma^2 x^2 v^{[a,b]}_{xx}};\quad \beta^2=-\frac{m}{\sigma^2 x^2 v^{[a,b]}_{xx}}.
\label{eq:alphabeta}
\end{equation}
The density of the optimal policy $\lambda^{*,[a,b]}$ is given by
\begin{align}
 \lambda^{*,[a,b]}(\pi|t, x;m)&:=\frac{1}{\beta}\frac{\varphi(\frac{\pi-\alpha}{\beta})}{\Phi(\frac{b-\alpha}{\beta})-\Phi(\frac{a-\alpha}{\beta})},
\label{eq:constrainedoptpolicy}
\end{align}
where $\varphi$ and $\Phi$ are the PDF and CDF functions of the standard normal distribution, respectively.
\end{lm}
Note that the first two moments and entropy of a truncated Gaussian distribution can be computed explicitly, see e.g. \cite{kotz2004continuous}. Substituting \eqref{eq:constrainedoptpolicy} back to the HJB \eqref{eq:constrainedHJBnoc} we obtain the following highly non-linear PDE
\begin{align}
 v^{[a,b]}_t(t,x;m)+rxv^{[a,b]}_x(t,x;m)&-\frac{1}{2}\frac{(\mu-r)^2(v^{[a,b]}_x)^2(t,x;m)}{\sigma^2 v^{[a,b]}_{xx}(t,x;m)}\notag\\&+\frac{m}{2}\ln\bigg(-\frac{2\pi_e m}{\sigma^2 x^2 v^{[a,b]}_{xx}(t,x;m)}\bigg)+{m\ln Z(m)}=0, \label{eq:constrainedHJBnocsimp}
\end{align}
with terminal condition $v(T,x;m)=U(x)$, where by abusing notations, 
\begin{equation}
Z(m):={\Phi(B(m))-\Phi(A(m))}, %\frac{\phi(A(m))-\phi(B(m))}
\label{eq:}
\end{equation}
with
\begin{equation}
A(m):=\bigg(a+\frac{(\mu-r)}{\sigma^2}\frac{ v^{[a,b]}_{x}}{ xv^{[a,b]}_{xx}}\bigg)\sqrt{\frac{-\sigma^2 x^2 v^{[a,b]}_{xx}}{m}};\quad B(m):=\bigg(b+\frac{(\mu-r)}{\sigma^2}\frac{ v^{[a,b]}_{x}}{ xv^{[a,b]}_{xx}}\bigg)\sqrt{\frac{-\sigma^2 x^2 v^{[a,b]}_{xx}}{m}}.
\label{eq:}
\end{equation}
%{\bf TODO}
%\begin{enumerate}
%	\item Solve \eqref{eq:HJBnocsimp} for each case: $U(x)=\ln x$, $U(x)=-e^{-\gamma x}$, $U(x)=x^{\gamma}$
%	\item In case closed-form solution does not exist, find expansion when $m\to 0$.
%\end{enumerate}

\vspace{2mm}
\subsection{Optimal policy with portfolio constraints}
Below we show that the optimal solution to \eqref{eq:constrainedHJBnocsimp} can be given in explicit form for the case of logarithmic utility. We summarize the results in the following theorem.
\begin{thm}[Exploratory optimal investment under portfolio constraint]\label{Thm:constrainedlogx}
For logarithmic utility $U(x)=\ln x$, the optimal value function of the entropy-regularized exploratory constrained optimal investment problem is given by
\begin{align}
v^{[a,b]}(t,x;m)=&\ln x+
 \bigg(r+\frac{1}{2}\frac{(\mu-r)^2}{\sigma^2}\bigg)(T-t)+\frac{m}{2} \ln(\sigma^{-2}{2\pi_e m})(T-t){+m\ln Z_{a,b}(m) (T-t)}
\label{eq:constrainedvalueopt}
\end{align}
for $(t,x)\in[0,T]\times \bbr_+$, where
\begin{equation}
Z_{a,b}(m):=\Phi\bigg((b-\pi^{Merton})\sigma m^{-1/2}\bigg)-\Phi\bigg((a-\pi^{Merton})\sigma m^{-1/2}\bigg).
\label{eq:Zab}
\end{equation} Moreover, the optimal feedback distribution control $\lambda^{*,[a,b]}$ is a Gaussian distribution with parameter $\pi^{Merton}=\frac{(\mu-r) }{\sigma^2}$ and  $\frac{m}{\sigma^2 }$, truncated  on the interval $[a,b]$ i.e.
\begin{equation}
\lambda^{*,[a,b]}(\pi)=\mathcal N\left(\pi\bigg|\frac{(\mu-r) }{\sigma^2}, \frac{m}{\sigma^2 }\right)\bigg|_{[a,b]}
\label{eq:constrainedlambdaopt}
\end{equation}
and the associated optimal wealth under $\lambda^{*,[a,b]}$ is given by the following SDE
\begin{align} 
d X_t^{\lambda^{*,[a,b]}} = X_t^{\lambda^{*,[a,b]}}  \bigg(r+\pi_t^{*,[a,b]} (\mu-r) \bigg) dt + \sigma {q_t^{*,[a,b]}}  X_t^{\lambda^{*,[a,b]}}  d W_t \, , \quad X_0 = x_0 \,,
\label{eq:constrainedwealth}
\end{align}
%\thaicomment{The second moment of truncated normal distribution is more complicated: TO BE CORRECTED IN THE CODE}
{where
\begin{align} 
(q_t^{*,[a,b]})^2=  (\pi_t ^{*,[a,b]})^2+
\frac{m}{\sigma^2}\bigg(1+
\frac{A(m) \varphi(A(m))-B(m)\varphi(B(m))}{Z_{a,b}(m)}-
\bigg(\frac{\varphi(A(m))-\varphi(B(m))}{Z_{a,b}(m)}\bigg)^2\bigg)
\end{align}
with, by abusing notations, $A(m):=(a-\pi^{Merton})\sigma m^{-1/2}$, $B(m):= (b-\pi^{Merton})\sigma m^{-1/2}$ and
} 
\begin{equation}
\pi^{*,[a,b]}_t=\pi^{Merton} +\frac{\varphi\bigg((a-\pi^{Merton})\sigma m^{-1/2}\bigg)-\varphi\bigg((b-\pi^{Merton})\sigma m^{-1/2}\bigg)}{\sigma  m^{-1/2}Z_{a,b}(m)}.
\label{eq:constrainedpi}
\end{equation}
\end{thm}
\begin{remark} At any $t\in [0,T]$, the exploration variance {decreases} as $\sigma$ increases, which means that exploration is less necessary in more random market environment. However, in contrast to the unconstrained section, both the {mean} and variance of the optimal distribution $\lambda^{*,[a,b]}$ are controlled by the degree of exploration $m$. 
\end{remark}
From Theorem \ref{Thm:constrainedlogx} we can see that in such a setting with portfolio constraints, the {best} control distribution to balance exploration and exploitation is truncated {Gaussian}, demonstrating once again
the popularity of Gaussian distribution in RL studies, even in constrained optimization settings. 

%Theorem \ref{Thm:constrainedlogx} suggests that instead of taking a generic (possibly non Gaussian) policy, we
%only need to learn among the class of Gaussian policies generated by 
%$\lambda^*(\pi)=\mathcal N\left(\pi\big|\frac{(\mu-r) }{\sigma^2}, \frac{m}{\sigma^2}\right)$, truncated on the interval $[a,b]$, which can considerably increase the learning speed. Also, 
\begin{lm}\label{lm:constrained1}
The optimal value function of the regularized exploration constrained problem can be expressed as
\begin{align}
v^{[a,b]}(t,x;m)= & v^{[a,b]}(t,x;0)+ \psi^{[a,b]}(t,x;m),
\label{eq:constrainedvaluerela}
\end{align}
where 
\begin{align}
v^{[a,b]}(t,x;0):=&\ln x+ \bigg(r+(\mu-r)\pi^{0,[a,b]}-\frac{1}{2}(\pi^{0,[a,b]})^2 \sigma^2\bigg)(T-t),%\notag\\=& \ln x+\bigg(r+\frac{1}{2}\frac{(\mu-r)^2}{\sigma^2}+f(a) {\bf 1}_{\pi^{Merton}<a}+f(b){\bf 1}_{\pi^{Merton}>b}\bigg)(T-t),
\label{eq:mconstrainedvaluerela}
\end{align}
which is the constrained optimal value function without exploration  using the corresponding constrained optimal investment strategy 
\begin{equation}
\pi^{0,[a,b]}:=\pi^{Merton}{\bf 1}_{\pi^{Merton}\in[a,b]}+a {\bf 1}_{\pi^{Merton}<a}+b{\bf 1}_{\pi^{Merton}>b},
\label{eq:constrainedMerton}
\end{equation} 
where
$$
\psi^{[a,b]}(t,x;m)= \frac{m}{2}\ln(\sigma^{-2}{2\pi_e m}) (T-t){+m\ln Z_{a,b}(m)} (T-t)+(f(a) {\bf 1}_{\pi^{Merton}<a}+f(b){\bf 1}_{\pi^{Merton}>b})(T-t)
$$
and
$$
f(y):=\frac{1}{2}y^2 \sigma^2-y (\mu-r)+\frac{1}{2}\frac{(\mu-r)^2}{\sigma^2}=\frac{1}{2} \sigma^2 (y-\pi^{Merton})^2.
$$

\end{lm}
Note that $f(\pi^{Merton})=0$. The following lemma confirms the intuition that $\psi^{[a,b]}(t,x;m)$, which measures the exploration effect, converges to 0 as $m\to 0$. 
\begin{lm}\label{Le:Zab}
Consider $Z_{a,b}(m)$ defined by \eqref{eq:Zab} for each $m>0$. We have
\begin{equation}
\lim_{m\to 0} {m\ln Z_{a,b}(m)}=\begin{cases}
0 &{\pi^{Merton}\in[a,b]}\\
-f(a)&{\pi^{Merton}<a}\\
-f(b)& {\pi^{Merton}>b}.
\end{cases}
\label{eq:}
\end{equation}
\end{lm}
\proof It can be seen directly using e.g. the classical L'Hôpital's rule. \endproof %Clearly, this is equivalent to $k(t)\to 1$ and $l(t)\to (r+\frac{1}{2}\frac{(\mu-r)^2}{\sigma^2}) $. 

Similar to Theorem \ref{Thm:equivalence} for the unconstrained case, it is straightforward to obtain the solvability equivalence between classical and exploratory constrained EU problem.
\subsection{Exploration cost and exploration effect under portfolio constraints }\label{se:constrainedExplorationcost}
We now turn our attention to the exploration cost. As before, the exploration cost at time $t=0$ is defined by
\begin{equation}
\begin{split}
L^{[a,b]}(T,x;m)&:=v^{[a,b]}(0,x;0)-v^{[a,b]}(0,x;m)\\
&-m \E\bigg[\int_{0}^{T}\int_a^b \lambda_s^{*,[a,b]}(\pi) \ln\lambda_s^{*,[a,b]}(\pi) d \pi  ds |X^{\lambda^{*,[a,b]}}_0=x\bigg],
\end{split}
\label{eq:exploration cost}
\end{equation}
where $v^{[a,b]}(0,x;0)$ and $v^{[a,b]}(0,x;m)$ are the optimal value function at time $t=0$ of the constrained EU problem without and with exploration respectively (defined by \eqref{eq:constrainedvaluerela} and \eqref{eq:mconstrainedvaluerela}). The integral term is the expected utility (before being regularized) of the exploration problem at time $t=0$. 
\begin{prop}
\label{eq:ExploratoryTheorem2}
In the constrained problem with exploration, the exploration cost is given by 
$$
L^{[a,b]}(T,x;m)=\frac{mT}{2}+mT \frac{A(m)\varphi(A(m))-B(m)\varphi(B(m))}{2Z_{a,b}(m)},
$$
where, by abusing notations,
$$
A(m):=(a-\pi^{Merton})\sigma m^{-1/2};\quad B(m):=(b-\pi^{Merton})\sigma m^{-1/2}.
$$
Moreover, $\lim_{m\to 0} L^{[a,b]}(T,x;m)=0$.
\end{prop}
\proof Recall first that the optimal distribution $\lambda_t^{*,[a,b]}$ is a Gaussian distribution with mean $\alpha=\pi^{Merton}$ and variance 
$\beta^2=\frac{m}{\sigma^2} $ truncated on the interval $[a,b]$. It is known e.g. \cite{kotz2004continuous} that for any Gaussian distribution $\mathcal N(\alpha, \beta^2)$ truncated on $[a,b]$, its entropy is given by
\begin{equation}
\ln\sqrt{{2\pi_e\beta^2}}+\frac{1}{2}+\ln\bigg(\Phi(l(b))-\Phi(l(a))\bigg) +\frac{l(a)\varphi(l(a))-l(b)\varphi(l(b))}{2 (\Phi(l(b))-)\Phi(l(a)))},
\label{eq:}
\end{equation}
where
$
l(y):=\frac{y-\alpha}{\beta}
$. The explicit form of the exploration cost now follows directly by Lemma \ref{lm:constrained1}. Finally, it is straightforward to verify that $\lim_{m\to 0} L^{[a,b]}(T,x;m)=0$. 
\endproof
%Intuitively, $ \psi(t,x;m)$ measures the exploration effect. It is easy to see that  $ \psi(t,x;m)\to 0$ as $m\to 0$, hence, $v(t,x;m)\to v(t,x;0)=\ln x+(r+\frac{1}{2}\frac{(\mu-r)^2}{\sigma^2})$ which is the optimal value function in the absence of exploration. Clearly, this is equivalent to $k(t)\to 1$ and $l(t)\to (r+\frac{1}{2}\frac{(\mu-r)^2}{\sigma^2}) $. 

As shown above, the exploration cost converges to zero as $m\to 0$. Since the exploration weight as been taken as an exogenous parameter $m$ which reflects the level of exploration desired by the learning agent, it is intuitive to expect that the smaller $m$ is, the more
emphasis is placed on exploitation; and when $m$ is sufficiently close to zero, the exploratory formulation is getting close to the problem without exploration. The following lemma confirms the intuition that exploration cost under a portfolio constraint is smaller than that without constraint when the investment strategy is restricted in an interval. % that contains the classical optimal unconstrained Merton strategy. 

\begin{lm}\label{Le:compare}
If the investment strategy is restricted in $[a,b]$, where $a<\pi^{Merton}< b$, then $L^{[a,b]}(T,x;m)\leq L(T,x;m)$, for all $x>0$. The same conclusion also holds for the case where $\pi^{Merton}\le a<b\le \pi^{Merton}+\sqrt{m}\sigma^{-1}$ or $\pi^{Merton}-\sqrt{m}\sigma^{-1}\le a<b\le \pi^{Merton}$. 
\end{lm}
%\proof Observe first that the function $x\varphi(x)$ is increasing in $(-1,1)$ and takes negative (resp. positive) values when $x<0$ (resp. $x>0$). Therefore, $A(m)\varphi(A(m))-B(m)\varphi(B(m))\le 0$
%when 
%\begin{itemize}	
%\item $A(m)< 0< B(m)$, which implies that $a<\pi^{Merton}<b$;
	%\item $0\le A(m)\le B(m)\le 1$, which implies that $\pi^{Merton}\le a<b\le \pi^{Merton}+\sqrt{m}\sigma^{-1}$;
	%\item $-1\le A(m)\le B(m)\le 0$,  which implies that $\pi^{Merton}-\sqrt{m}\sigma^{-1}\le a<b<\pi^{Merton}$.
%\end{itemize}
%For each of these cases, by compararing Propositions \ref{eq:ExploratoryTheorem2} and \ref{eq:ExploratoryTheorem} we obtain
%$$
%L^{[a,b]}(T,x;m)=\frac{mT}{2}+mT \frac{A(m)\varphi(A(m))-B(m)\varphi(B(m))}{2Z_{a,b}(m)}
%\leq \frac{mT}{2}=L(T,x;m)
%$$ 
%and the Lemma is proved. \endproof
The following theorem confirms a
desirable result that the entropy-regularized constrained EU problem converges to its non-exploratory constrained EU counterpart when the exploration weight $m$ goes to zero.
\begin{thm}\label{m-consistency}
For each $x\in\mathbb R +$, 
 \begin{equation}
\lambda^{*,[a,b]}(\cdot| t,x,m)\to \delta_{\pi^{0,[a,b]}}{(\cdot)}\quad\mbox{when}\quad m\to 0,
\end{equation} 
{where the (without exploration) constrained Merton strategy $\pi^{0,[a,b]}$ is defined by \eqref{eq:constrainedMerton}.}
Furthermore, for each $(t,x)\in[0,T]\times \bbr+$,
$$
\lim_{m\to0}|v^{*,[a,b]}(t,x;m)-v^{*,[a,b]}(t,x;0)|=0.
$$
\end{thm}
\proof The convergence of the optimal distribution control $\lambda^{*,[a,b]}$ to the Dirac measure $\delta_{{\pi^{0,[a,b]}}}$ is straightforward. The convergence of the value function follows directly from fact that $\psi^{[a,b]}(t,x;m)$ given by \eqref{eq:constrainedpi} converges to 0 as $m\to 0$ by Lemma \ref{lm:constrained1}. \endproof
\subsection{Policy improvement}
%The following policy improvement theorem is crucial for interpretable RL algorithms as it ensures that the iterated value functions in non-decreasing and  converges to the optimal value function. 
Below, for some given {admissible} (feedback)} policy $\lambda ^{[a,b]}$ (not necessarily truncated Gaussian) on $[a,b]$, we denote the corresponding value function by
\begin{align}
&v^{\lambda^{[a,b]}}(t,x;m):=\mathbb{E} \Bigg[U(X_T^{\lambda^{[a,b]}})-m\int_t^T {\int_a^b}\lambda_s^{[a,b]} (\pi)  \ln\lambda_s^{[a,b]}(\pi)  d \pi|X^{\lambda^{[a,b]}}_t=x\Bigg].
\end{align}
\begin{thm}\label{Thm-updateconstrained}
%\coprod
For some given policy $\lambda^{[a,b]}$ (not necessarily Gaussian) on the interval $[a,b]$, we assume that
the corresponding value function
%\begin{align}
%&v^{\lambda}(t,x;m):=\mathbb{E} \Bigg[U(X_T^{\lambda})-m\int_t^T \int_\cC \lambda_s (\pi_s)  \ln\lambda_s(\pi_s)  d \pi_s|X^{\lambda}_t=x\Bigg].
%\end{align}
 $v^{\lambda^{[a,b]}}(\cdot,\cdot; m) \in C^{1,2}([0,T)\times \bbr_+)\cap C([0,T]\times \bbr_+)$ and satisfies $v^{{\lambda^{[a,b]}}}_{xx}(t,x,; m)<0$ for any $(t,x)\in[0,T)\times \bbr_+.$ Suppose furthermore that the feedback policy $\wt{\lambda}^{[a,b]}$ defined by
\begin{equation}\label{eq:pischemeconstrained}
\wt{\lambda}^{[a,b]}(\pi| t,x; m)= \mathcal N\bigg(\pi\bigg|
-\frac{(\mu-r)xv^{{\lambda^{[a,b]}}}_x}{\sigma^2 x^2 v^{{\lambda^{[a,b]}}}_{xx}},-\frac{m}{\sigma^2 x^2 v^{{\lambda^{[a,b]}}}_{xx}}\bigg)\bigg|_{[a,b]}
\end{equation}
(Gaussian truncated on $[a,b]$) is admissible. Let $v^{\wt{\lambda}^{[a,b]}}(t,x; m)$ be the value function corresponding to this new truncated (Gaussian) policy $\wt{\lambda}^{[a,b]}$. Then, 
\begin{equation}
v^{\wt{\lambda}^{[a,b]}}(t,x; m)\ge v^{\lambda^{[a,b]}}(t,x; m), \quad  (t,x)\in[0,T)\times \bbr_+.
\label{eq:}
\end{equation}
\end{thm}

\proof The proof is similar to that of the unconstrained case hence, omitted. 
\endproof
Theorem \ref{Thm-updateconstrained}, %it is intuitive to focus on the truncated Gaussian policies when choosing an initial exploration distribution. %it can be observed that for any given (not necessarily Gaussian) policy $\lambda$ on $[a,b]$, there are always policies in the truncated Gaussian family that improve the value function of $\lambda$. 
also suggests that a candidate of the initial
feedback policy may take the form ${\lambda}^{0,[a,b]}(\pi| t,x; m)= \mathcal N\big(\pi\big|
\alpha, \beta^2\big)|_{[a,b]}$ for some parameters $\alpha, \beta^2$. {Similar to the unconstrained case}, such a choice {of policy} leads to the convergence of both
the value functions and the policies in a finite number of iterations.
\begin{thm}\label{Thm-convergenceconstrained}
Let ${\lambda}^{0,[a,b]}(\pi; t,x, m)= \mathcal N\big(\pi\big|
 \alpha, \beta^2\big)|_{[a,b]}$ with $\alpha,\beta >0$ {and assume that $U(x)=\ln x$.} Define the sequence of feedback policies $(\lambda^{n,{[a,b]}}  (\pi; t,x, m))$ updated by the policy improvement scheme \eqref{eq:pischemeconstrained}, 
i.e.,
\begin{equation}
\lambda^{n,{[a,b]}}(\pi| t,x; m)= \mathcal N\bigg(\pi\bigg|
-\frac{(\mu-r)xv_x^{\lambda^{n-1,{[a,b]}}}(t,x; m)}{\sigma^2 x^2 v^{\lambda^{n-1,{[a,b]}}}_{xx}(t,x; m)},-\frac{m}{\sigma^2 x^2 v^{\lambda^{n-1,{[a,b]}}}_{xx}(t,x; m)}\bigg)\bigg|_{[a,b]},\quad n=1,2,\cdots,
\end{equation}
where $v^{\lambda^{n-1,{[a,b]}}}$ is the value function corresponding to the policy $\lambda^{n-1,{[a,b]}}$ defined by

\begin{align}
&v^{\lambda^{n-1,{[a,b]}}}(t,x;m):=\mathbb{E} \Bigg[U(X_T^{\lambda^{n-1,{[a,b]}}})-m\int_t^T \int_a^b\lambda_s^{n-1,{[a,b]}} (\pi)  \ln\lambda_s^{n-1,{[a,b]}} (\pi)  d \pi|X^{\lambda^{n-1,{[a,b]}}}_t=x\Bigg].
\end{align}
Then, 
\begin{equation}
\lim_{n\to \infty}\lambda^{n,{[a,b]}}(\cdot| t,x; m)=\lambda^{*,{[a,b]}}(\cdot| t,x; m),\quad \mbox{weakly}
\label{eq:}
\end{equation}
and 
\begin{equation}
\lim_{n\to \infty}v^{\lambda^{n,{[a,b]}}}( t,x; m)=v^{\lambda^{*,{[a,b]}}}( t,x; m),
\label{eq:}
\end{equation}
which is the optimal value function given in Lemma \ref{lm:constrained1}.
\end{thm}
\proof The proof can be done using Feymann-Kac representation and the update result of Theorem \ref{Thm-updateconstrained}, similarly to the unconstrained problem.  
\endproof
The above improvement theorem allows to establish the policy evaluation and algorithm taking the martingale property into account. Since the steps are similar to the unconstrained problem in Section \ref{se:PE}, we skip the detail. 

	\section{Learning implementation and numerical example}\label{se: Learning}
	We are ready now in the position
to discuss a RL algorithm
for this problem. Note that a common practice
in RL literature is that
one can represent the value function $J^\theta$ and
the policy $\lambda^\phi$ using (deep) neural networks.
In this section, we don't follow this path.  Inspired by the (offline) actor-critic approach in \cite{jia2022policy}, 
%\thaicomment{and stochastic gradient Langevin dynamics (see \cite{welling2011bayesian,chau2021stochastic})}, 
we instead take advantages
of the explicit parametric form
of the optimal value function $v$ and the improvement policy
which are given in Theorem \ref{Thm:equivalence}.
This will in turn facilitate the learning
process and will lead to faster learning and convergence.

Below, to implement a workable algorithm,
one can approximate a generic expectation % $E(\theta)$
in the following manner: 
Let $0=t_0<t_1<\ldots<t_l<t_{l+1}=T$
be a partition for the finite interval $[0,T]$.
We then collect a sample $\mathcal D=\{(t_i,x_i): i=0,1,\ldots,l+1\}$ as follows:
For $i=0$, the initial sample point is $(0,x_0)$. Next, at each $t_i, i=0,1,\ldots,l$,
$\lambda_{t_{i}}(\pi)$ is sampled
to obtain an allocation $\pi$
for the risky asset. The wealth $x_{t_i+1}$ at the next time instant $t_{i+1}$ is computed by equation \eqref{eq_XtExplo}. As a result, a generic expectation can be approximated by
the following finite sum
\begin{equation}
\label{eq:BellmanError2}
E(\theta)\approx \sum_{(t_i,x_i)\in\mathcal D}
h_{t_i}\bigg(J^\theta(t_{i+1},X^{\lambda}_{t_{i+1}};{\lambda})-J^\theta(t_{i},X^{\lambda}_{t_{i}};{\lambda})-m\int_{-\infty }^{+\infty} \lambda (\pi|t_i,X^{\lambda}_{t_i})  \ln\lambda (\pi|t_i,X^{\lambda}_{t_i})  d \pi\bigg)\Delta t.
\end{equation}

%Note that we have from the body of the proof of Theorem \ref{eq:ExploratoryTheorem} that
%\begin{equation}
%\label{eq:EntropyIntegral}
%\int_\cC \lambda_{u}^{*} (\pi) \ln\lambda_{u}^{*}(\pi) d \pi=
%-\frac{1}{2}-\frac{1}{2}\ln\left(\frac{2\pi em}{\sigma^2[m(T-t)+1]} \right)
%=-\frac{1}{2}-\frac{1}{2}\ln(2\pi em)+\frac{1}{2}\ln(\sigma^2[m(T-t)+1])
%\end{equation}

\subsection{Implementation for the unconstrained problem}
Recall first that the exploratory optimal value function is given by
\begin{align}
v(t,x;m)%=%&\ln x+  \bigg(r+\frac{1}{2}\frac{(\mu-r)^2}{\sigma^2}\bigg)(T-t)+\frac{m}{2} \ln(\sigma^{-2}{2\pi_e m})(T-t)\notag\\
=&\ln x+\bigg(r+\frac{1}{2}\frac{(\mu-r)^2}{\sigma^2}\bigg)(T-t)+\frac{m}{2} \ln({2\pi_e m})(T-t)- m\ln(\sigma)(T-t).
\label{eq:Vtheta}
\end{align}
We learn it from the following parametric form
\begin{align}
\label{eq:Vtheta11}
J^{\theta}(t,x;m)=\ln x+ \theta(T- t)+\frac{m}{2} \ln({2\pi_e m})(T-t).
\end{align}
which clearly satisfies Assumption \ref{Ass:1}. It follows from \eqref{eq:Vtheta} that
\begin{align}
\label{eq:thetav}
\theta=r+\frac{1}{2}\frac{(\mu-r)^2}{\sigma^2}- m\ln \sigma.
 \end{align}
Recall that for a Gaussian distribution $\mathcal N(.|a,b^2)$
its entropy is given by $\frac{1}{2}+\ln\left(b\sqrt{2\pi_e} \right)$.
% Hence,
%\begin{align}
%\mathcal H(\lambda):=\int_\cC \lambda_{u}^{\phi} (\pi) \ln\lambda_{u}^{\phi}(\pi) d \pi=
%-\frac{1}{2}-\frac{1}{2}\ln(2\pi_e m \sigma^{-2}) .
%\end{align}
Since we do not know the model's parameters
$\mu$ and $\sigma$, we will learn them using Theorem \ref{Thm:equivalence} with the parametric policy 
$
\lambda^\phi(\pi)=\mathcal N(\pi|\phi_1,e^{\phi_2}m).
$
%Recall that (see e.g. \eqref{eq:gaussianentropy}) for a Gaussian distribution $\mathcal N(.|\mu,\sigma^2)$ its entropy is given by $\frac{1}{2}+\ln\left(\sigma\sqrt{2\pi_e} \right)$.  
It follows that
\begin{align}
\hat{p}(\phi_1,\phi_2):= -\int_{-\infty}^{+\infty} \lambda^{\phi} (\pi) \ln\lambda^{\phi}(\pi) d \pi=
\frac{1}{2}+\frac{1}{2}(\phi_2+\ln(2\pi_e m)).\label{eq:entrophat}
\end{align}
Also,
$$
\ln \lambda^\phi(\pi)=-\frac{1}{2}\phi_2-\frac{1}{2m}(\pi-\phi_1)^2 e^{-\phi_2}-\frac{1}{2}\ln(2\pi_e m).
$$
Therefore
$$
\frac{\partial}{\phi_1}\ln \lambda^\phi(\pi)=(\pi-\phi_1)\frac{e^{-\phi_2}}{m},\quad
\frac{\partial}{\phi_2}\ln \lambda^\phi(\pi)=-\frac{1}{2}+\frac{1}{2}(\pi-\phi_1)^2\frac{e^{-\phi_2}}{m}.
$$
Moreover, it can be seen from \eqref{eq:entrophat} that $
\frac{\partial \hat p}{\partial \phi_1}=0$ and $\frac{\partial \hat p}{\partial \phi_2}=\frac{1}{2}.$ 
Finally, in such a learning framework, we choose the test function $h_t=\frac{\partial J^{\theta}}{\partial \theta}=(T-t)$.

For learning step, we adopt the (offline) actor-critic approach in \cite{jia2022policy} which is summarized in Algorithm \ref{Martingale: Algorithm2} below.

\begin{algorithm}[H]
        \caption{PE/PG: Actor-Critic Algorithm: unconstrained problem}\label{Martingale: Algorithm2}
        \begin{algorithmic}[1]
            \REQUIRE Market simulator \textit{Market}
            \REQUIRE Learning  rates: $\eta_\theta,\eta_\phi$, exploration rate $m$, number of iterations $M$%, %Langevin parameters $L_L,L_T$
            \FOR {$k=1,2,\ldots, M$}
            \FOR {$i=1,2,\ldots, \frac{T}{\Delta t}$}
            \STATE Sample $(t_i^k,x_i^k)$ from \textit{Market} under $\lambda^\phi$% from \eqref{eq_XtExplo00}
            \STATE Obtain collected samples $\mathcal D=\{(t_i^k,x_i^k):1\leq i\leq \frac{T}{\Delta t}\}$
            \STATE Compute $\Delta_\theta J$ using \eqref{eq:changeinJtheta}
            \STATE Compute  $\Delta_\phi J$ using \eqref{eq:derivationwrtphi}
            \STATE Update $\theta\leftarrow\theta+l(i)\eta_{\theta}\Delta_\theta J$%+\sqrt{2L_L L_T}\epsilon_t$ with $\epsilon_t\sim N(0,1)$
            \STATE Update $\phi\leftarrow\phi-l(i)\eta_{\phi}\Delta_\phi J$%+\sqrt{2L_L L_T}\epsilon_t$ with $\epsilon_t\sim N(0,1)$
            \ENDFOR
            \STATE Update the policy $\lambda^{\phi}\leftarrow \mathcal N\left(\phi_1,e^{\phi_2}m\right)$
            \ENDFOR
            
            \end{algorithmic}
\end{algorithm}
\subsection{Implementation for the constrained problem}

First, recall from \eqref{eq:constrainedvalueopt} that the exploratory optimal value function is given by

\begin{align*}
v^{[a,b]}(t,x;m)=&\ln x+
 (r+\frac{1}{2}\frac{(\mu-r)^2}{\sigma^2})(T-t)+\frac{m}{2} \ln(\sigma^{-2}{2\pi_e m})(T-t){+}m\ln Z_{a,b}(m) (T-t)
\end{align*}
for $(t,x)\in[0,T]\times \bbr_+$, where $Z_{a,b}(m)$ is given by \eqref{eq:Zab}. 
%\begin{equation}
%Z_{a,b}(m):=\Phi\bigg((b-\pi^{Merton})\sigma m^{-1/2}\bigg)-\Phi\bigg((a-\pi^{Merton})\sigma m^{-1/2}\bigg).
%\label{eq:Zab}
%\end{equation} 
%\begin{align}
%v(t,x;m)=&\ln x+
 %\bigg(r+\frac{1}{2}\frac{(\mu-r)^2}{\sigma^2}\bigg)(T-t)+\frac{m}{2} \ln(\sigma^{-2}{2\pi_e m})(T-t){+}m\ln Z_{a,b}(m) (T-t)\notag\\
%=&\ln x+\bigg(r+\frac{1}{2}\frac{(\mu-r)^2}{\sigma^2}\bigg)(T-t)+\frac{m}{2} \ln({2\pi_e m})(T-t)- m\ln(\sigma)(T-t){+m}\ln Z_{a,b}(m) (T-t).
%\label{eq:Vtheta}
%\end{align}
We learn it from the following parametric form
\begin{align}
\label{eq:Vtheta11}
J^{\theta}(t,x;m)=\ln x+ \theta_1(T- t)+\theta_2(T- t)+\frac{m}{2} \ln({2\pi_e m})(T-t),
\end{align}
which clearly satisfies Assumption \ref{Ass:1}. It follows from \eqref{eq:Vtheta11} that
\begin{align}
\label{eq:thetav}
\theta_1=r+\frac{1}{2}\frac{(\mu-r)^2}{\sigma^2} -m\ln(\sigma), \quad \theta_2=m\ln Z_{a,b}(m).
 \end{align}
%and
 %\begin{align}
 %\label{eq:theta1}
%\theta_1=m\ln Z_{a,b}(m).
 %\end{align}
% Hence,
%\begin{align}
%\mathcal H(\lambda):=\int_\cC \lambda_{u}^{\phi} (\pi) \ln\lambda_{u}^{\phi}(\pi) d \pi=
%-\frac{1}{2}-\frac{1}{2}\ln(2\pi_e m \sigma^{-2}) .
%\end{align}
Again, we  parametrize the policy from Theorem \ref{Thm:constrainedlogx} as the following
$$
\lambda^\phi(\pi)=\mathcal N(\pi|\phi_1,e^{\phi_2}m)\vert_{[a,b]}=\frac{1}{\sqrt{e^{\phi_2}m}}\frac{\varphi(\frac{\pi-\phi_1}{\sqrt{e^{\phi_2}m}})}{ \left(\Phi(\frac{b-\phi_1}{\sqrt{e^{\phi_2}m}})-\Phi(\frac{a-\phi_1}{\sqrt{e^{\phi_2}m}}) \right)}.
$$
Also recall that for $ \lambda^\phi$, a Gaussian distribution truncated on $[a,b]$, its entropy is given by
\begin{align*}
\hat{p}(\phi_1,\phi_2)=&\ln\sqrt{{2\pi m}}+\frac{1}{2}\phi_2+\frac{1}{2}+\ln\left(\Phi(\frac{b-\phi_1}{\sqrt{e^{\phi_2}m}})-\Phi(\frac{a-\phi_1}{\sqrt{e^{\phi_2}m}}) \right) +\frac{\frac{a-\phi_1}{\sqrt{e^{\phi_2}m}}\varphi(\frac{a-\phi_1}{\sqrt{e^{\phi_2}m}})-\frac{b-\phi_1}{\sqrt{e^{\phi_2}m}}\varphi(\frac{b-\phi_1}{\sqrt{e^{\phi_2}m}})}{2 \left(\Phi(\frac{b-\phi_1}{\sqrt{e^{\phi_2}m}})-\Phi(\frac{a-\phi_1}{\sqrt{e^{\phi_2}m}}) \right)}.
\end{align*}

\noindent Also,
$$
\ln \lambda^\phi(\pi)=-\frac{1}{2}\phi_2-\frac{1}{2m}(\pi-\phi_1)^2 e^{-\phi_2}-\frac{1}{2}\ln(2\pi_e m) -\ln\left(\Phi(\frac{b-\phi_1}{\sqrt{e^{\phi_2}m}})-\Phi(\frac{a-\phi_1}{\sqrt{e^{\phi_2}m}}) \right).
$$
Therefore
$$
\frac{\partial}{\phi_1}\ln \lambda^\phi(\pi)=(\pi-\phi_1)\frac{e^{-\phi_2}}{m}+\frac{\left({\varphi}(\frac{b-\phi_1}{\sqrt{e^\phi_2m}})-{\varphi}(\frac{a-\phi_1}{\sqrt{e^{\phi_2}m}}) \right)}{\left(\Phi(\frac{b-\phi_1}{\sqrt{e^{\phi_2}m}})-\Phi(\frac{a-\phi_1}{\sqrt{e^{\phi_2}m}}) \right)}
\frac{1}{\sqrt{e^{\phi_2}m}},
$$
$$
\frac{\partial}{\phi_2}\ln \lambda^\phi(\pi)=-\frac{1}{2}+\frac{1}{2}(\pi-\phi_1)^2\frac{e^{-\phi_2}}{m}+
\frac{\left({\varphi}(\frac{b-\phi_1}{\sqrt{e^\phi_2m}})\frac{b-\phi_1}{\sqrt{m}}-{\varphi}(\frac{a-\phi_1}{\sqrt{e^{\phi_2}m}})\frac{a-\phi_1}{\sqrt{m}} \right)}{\left(\Phi(\frac{b-\phi_1}{\sqrt{e^{\phi_2}m}})-\Phi(\frac{a-\phi_1}{\sqrt{e^{\phi_2}m}}) \right)}\frac{1}{2}e^{-\phi_2/2}.
$$
{The partial derivatives of $\hat{p}(\phi_1,\phi_2)$ can be obtained explicitly. %, see Appendix \ref{Ap:learning}.}

Similarly to the unconstrained case, the learning step can be done using the Actor-Critic Algorithm adopted for the constrained problem in Algorithm \ref{Martingale: Algorithm3} below.
\begin{algorithm}[h]
        \caption{PE/PG Actor-Critic Algorithm: constrained problem}\label{Martingale: Algorithm3}
        \begin{algorithmic}[1]
            \REQUIRE Market simulator \textit{Market}
            \REQUIRE Learning  rates: $\eta_\theta,\eta_\phi$, exploration rate $m$, number of iterations $M$, portfolio bounds $a,b$%, Langevin parameters $L_L,L_T$
            \FOR {$k=1,2,\ldots, M$}
            \FOR {$i=1,2,\ldots, \frac{T}{\Delta t}$}
            \STATE Sample $(t_i^k,x_i^k)$ from \textit{Market} under $\lambda^\phi$% from \eqref{eq_XtExplo00}
            \STATE Obtain collected samples $\mathcal D=\{(t_i^k,x_i^k):1\leq i\leq \frac{T}{\Delta t}\}$
            \STATE Compute $\Delta_\theta J$ using \eqref{eq:changeinJtheta}
            \STATE Compute  $\Delta_\phi J$ using \eqref{eq:derivationwrtphi}
            \STATE Update $\theta\leftarrow\theta+l(i)\eta_{\theta}\Delta_\theta J$%+\sqrt{2L_L L_T}\epsilon_t$ with $\epsilon_t\sim N(0,1)$
            \STATE Update $\phi\leftarrow\phi-l(i)\eta_{\phi}\Delta_\phi J$%+\sqrt{2L_L L_T}\epsilon_t$ with $\epsilon_t\sim N(0,1)$
            \ENDFOR
            \STATE Update the policy $\lambda^{\phi}\leftarrow \mathcal N\left(\phi_1,e^{\phi_2}m\right)|_{[a,b]}$
            \ENDFOR
            
            \end{algorithmic}
\end{algorithm}

\subsection{Numerical demonstration}\label{Numerical demonstration}
We provide an
an example to demonstrate our results
under a portfolio constraint in a setting with $T=1$ and $\Delta t= 1/250$. The annualized interest rate is $r=3\%$.
We choose $\mu=0.08,\sigma=0.3$
to simulate sample paths of the
diffusion wealth process \eqref{eq_Xtnoc}
based on the the update policy
after each  episode. First,
for $m\in [0.001,2]$ we plot the corresponding exploration cost function for
  both constrained and unconstrained cases.
Figure \ref{fig: Cost function versus exploration rate1}
clearly confirms the theoretical results obtained
in Proposition \ref{eq:ExploratoryTheorem}. Interestingly, 
the exploration cost of unconstrained case ($L(T,x;m)$) 
is much larger than that of constrained case ($L^{[a,b]}(T,x;m)$). Consistent to Lemma \ref{Le:compare}, this could be explained that 
for the constrained case,
one just needs to search for the optimal strategy
in a finite domain as opposed to an infinite domain for the unconstrained case.

\begin{figure}[!ht] % <---
\hfill % <---
   \begin{subfigure}{0.48\textwidth}
       \includegraphics[width=90mm,scale=0.7]{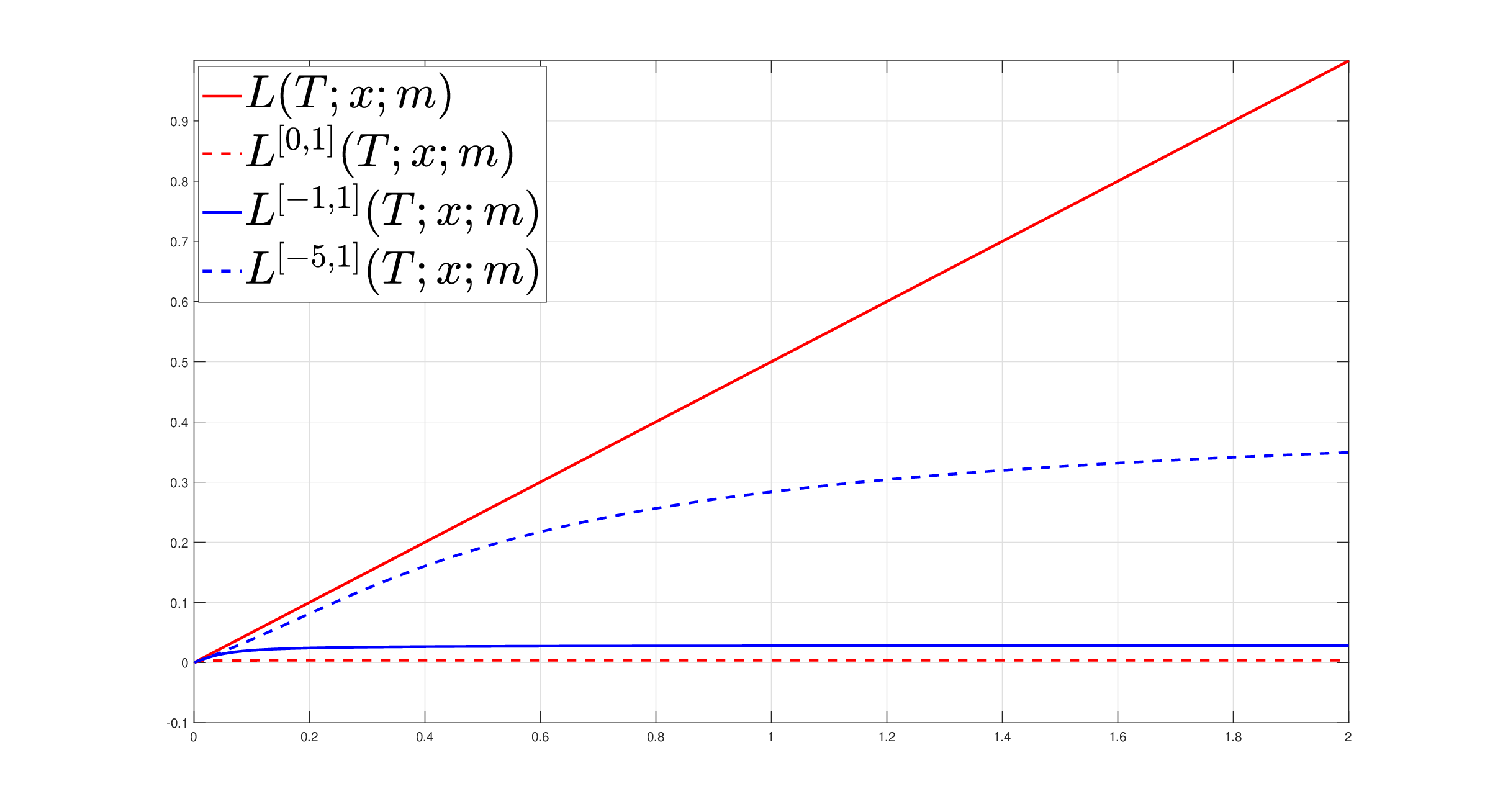}
       \caption{Impact of lower bound $a$}
       \label{fig:subim3}
   \end{subfigure}
   \hfill % <--- 
%   \begin{subfigure}{0.32\textwidth}
%       \includegraphics[width=\linewidth]{image/VminusVclass_lambda001eps.eps}
%       \caption{$m=0.01$}
%       \label{fig:subim1}
%   \end{subfigure}
%\hfill % <--- 
   \begin{subfigure}{0.48\textwidth}
       \includegraphics[width=90mm,scale=0.7]{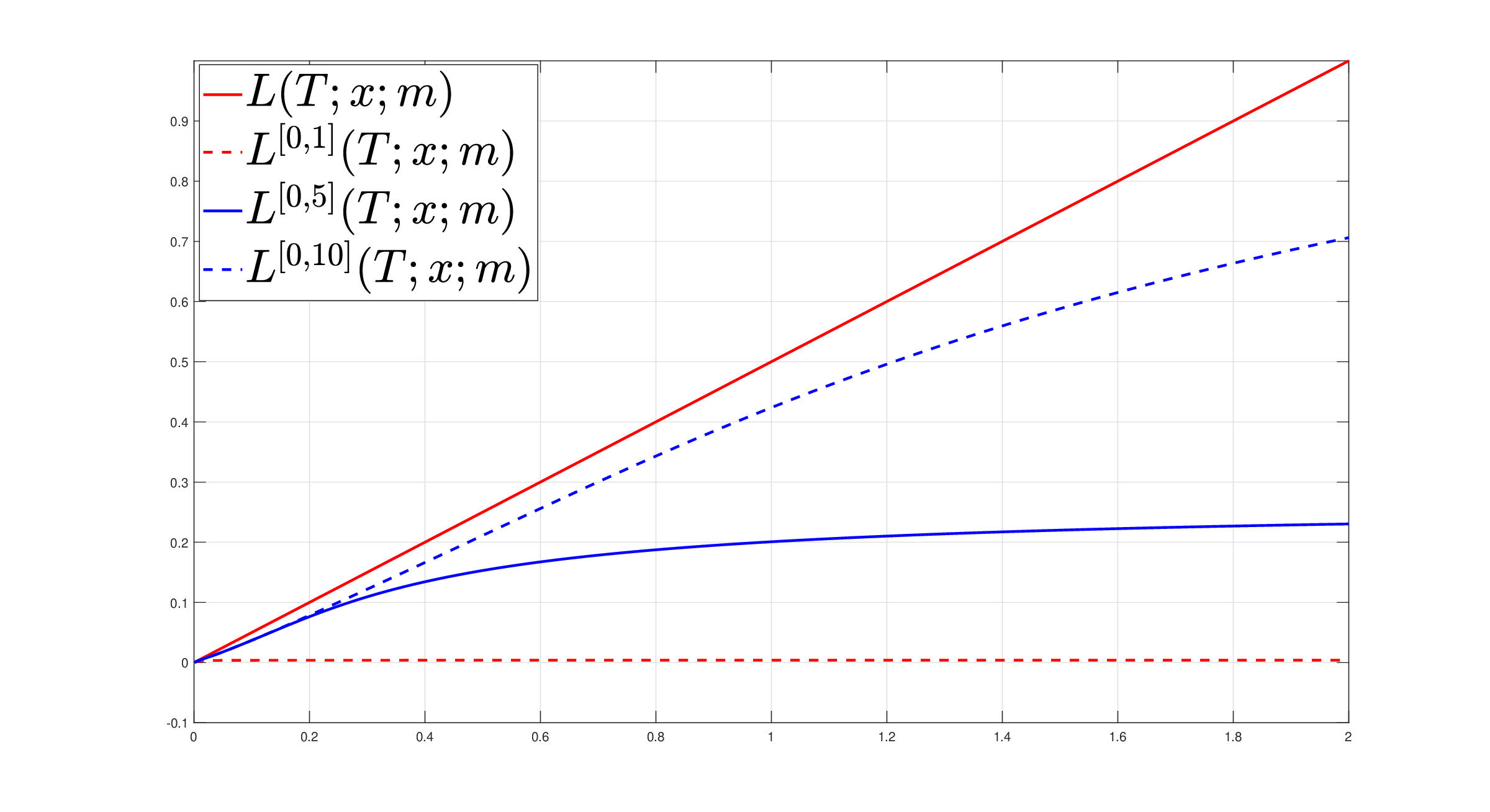}
       \caption{Impact of upper bound $b$}
       \label{fig:subim2}
   \end{subfigure}
    \caption{Exploration cost versus exploration rate}
   \label{fig: Cost function versus exploration rate1}
\end{figure}
The left (resp. right) panel of Figure 1 demonstrates the impact of the lower (resp. upper) portfolio bound on the exploration cost. Clearly, as the lower (resp. upper) bound of the portfolio decreases (resp. increases), the exploration cost increases. This finding aligns with the intuitive understanding that a broader domain of investment opportunities necessitates higher exploration expenses. Notably, when subjected to both short-selling (i.e., $a = 0$) and money borrowing constraints (i.e., $b = 1$), the exploration cost becomes negligible compared to the unconstrained case.  

%\begin{figure}[h] 	% <---
%          \includegraphics[width=110mm,height=7cm]{image/Plot_cost_funtion_vs_m_constrainedandUnconstrained.eps}
%       			   \caption{\small Exploration cost versus exploration rate}
%				\label{fig: Cost function versus exploration rate}
% \end{figure}

Recall from Theorem \ref{eq:ExploratoryTheorem} that the difference between the constrained value function $v^{*,[a,b]}(t,x;0)$
and the exploratory constrained value function $v^{*,[a,b]}(t,x;m)$
is determined by the temporal exploration rate $m$. 
In Figure \ref{fig:Convergence of value function}, we plot
the difference $|v^{*,[a,b]}(t,x;m)-v^{*,[a,b]}(t,x;0)|$ for $m=2$ (left panel) and
$m=0.001$ (right panel). It is clear
from Figure \ref{fig:Convergence of value function} the difference
is substantial for a large exploration rate $m$
but decreases greatly for a small exploration rate.

\begin{figure}[H] % <---
\hfill % <---
   \begin{subfigure}{0.48\textwidth}
       \includegraphics[width=90mm,scale=0.5]{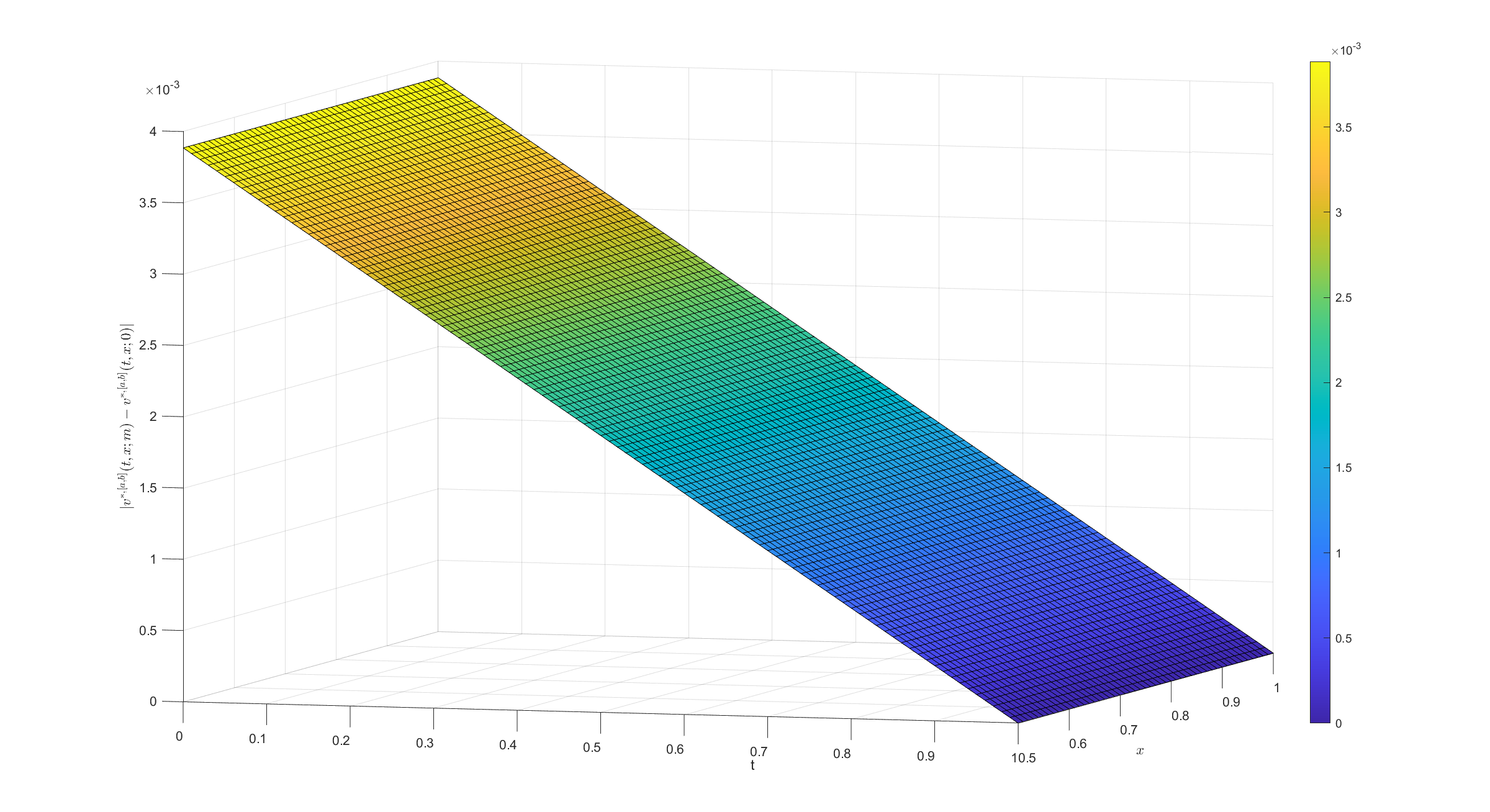}
       \caption{$m=2$}
       \label{fig:subim3}
   \end{subfigure}
   \hfill % <--- 
%   \begin{subfigure}{0.32\textwidth}
%       \includegraphics[width=\linewidth]{image/VminusVclass_lambda001eps.eps}
%       \caption{$m=0.01$}
%       \label{fig:subim1}
%   \end{subfigure}
%\hfill % <--- 
   \begin{subfigure}{0.48\textwidth}
       \includegraphics[width=90mm,scale=0.5]{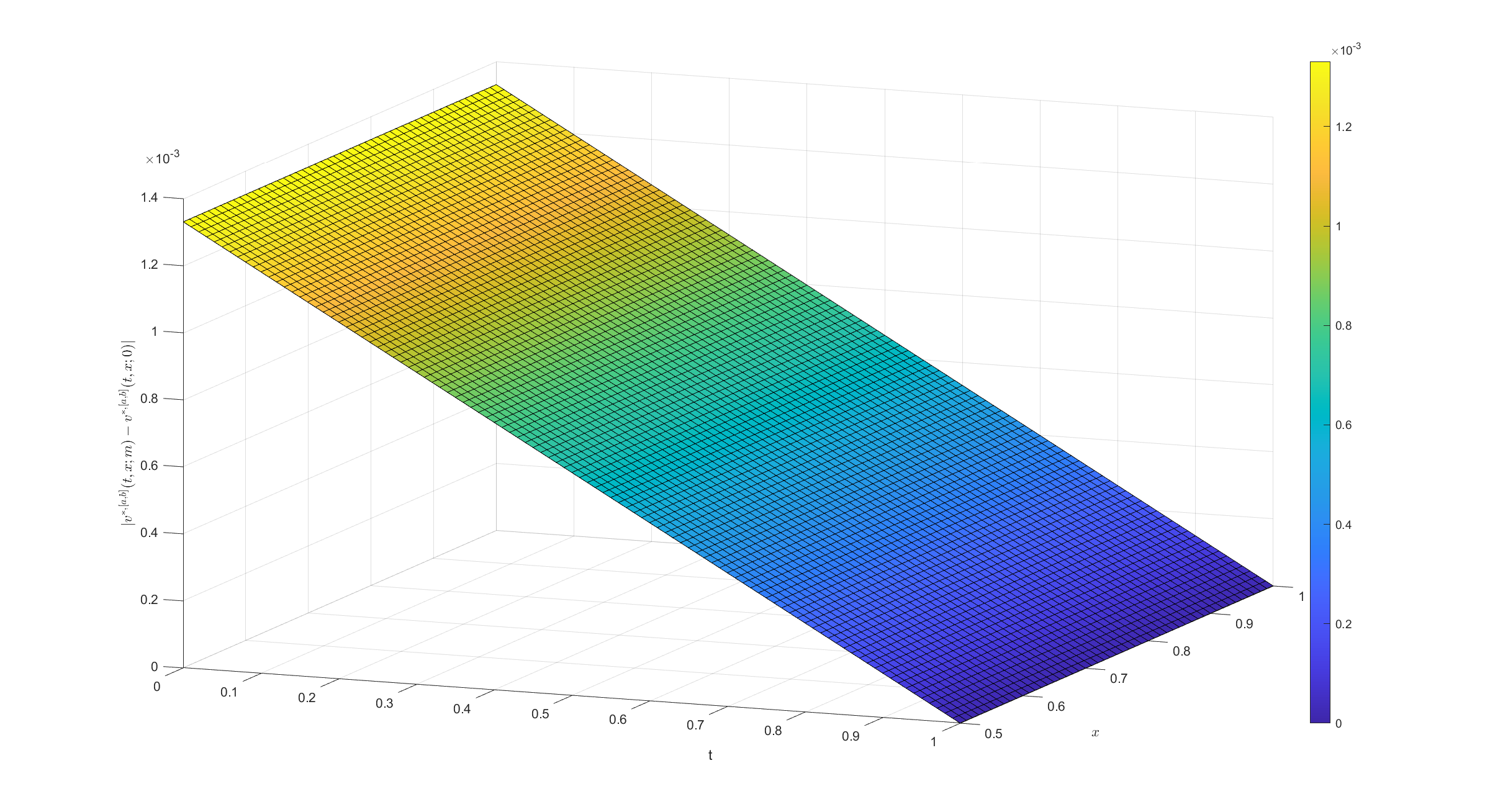}
       \caption{$m=0.001$}
       \label{fig:subim2}
   \end{subfigure}

   \caption{The effect of the exploration rate $m$ on $|v^{*,[a,b]}(t,x;m)-v^{*,[a,b]}(t,x;0)|$.}
   \label{fig:Convergence of value function}
\end{figure}

{
Next, we choose the learning rates which are used to update $\theta$ and $\phi$ to be $\eta_\theta=0.01$,
$\eta_\phi=0.001$, respectively with the decay rate $l(i)=1/i^{0.51}$. From the closed-from formula, the true value of $(\theta_1,\theta_2)$ is given by $\theta^*=( 0.055929, -0.001497)$.  For each learning episode, we start with an initial guess $$\theta=(\theta_1, \theta_2)=\theta^*+(\epsilon_1,\epsilon_2)$$ where $\epsilon_i, i=1,2$ are independent and generated from a (Gaussian noise) normal distribution. %$N(0,1)$.  
For each iteration, we randomly sample $10000$
1-year trajectories of the wealth process  $\mathcal D=\{(t_i^k,x_i^k):1\leq i\leq \frac{T}{\Delta t}\}$. The gradients of the value function $\Delta_\theta J$ are calculated by the above approximate discrete sum \eqref{eq:BellmanError2}. The model is trained for $N=10000$ iterations.
The results reported in this section is 
the average of $20$ training episodes. The portfolio strategy is limited by the lower bound $a=0$ (i.e. short-selling is not possible) and the upper bound $b=1$ (i.e. money borrowing is also prohibited).
Finally, we choose the test function $h_t=\frac{\partial J^{\theta}}{\partial \theta}=[(T-t);(T-t)]^T$.
}

%\begin{figure}[h] 	% <---
          %\includegraphics[width=110mm,height=7cm]{image/Convergence phi1Interation}
       			   %\caption{\small Learned  policy mean vs iteration steps for $m=0.01$. \thaicomment{TO BE REDONE}}
				%\label{fig:constrainedphi1}
 %\end{figure}
			%
				%Figure \ref{fig:constrainedphi1} shows the convergence of the learned mean which is the parameter $\phi_1$ in $\mathcal N(\phi_1,e^{\phi_2}m)$ truncated on $[0,1]$, as a function of the number of iterations temperature for $m=0.01$. 

\begin{figure}[H] 	% <---
   \begin{subfigure}{0.32\textwidth}
       \includegraphics[width=55mm,height=7cm]{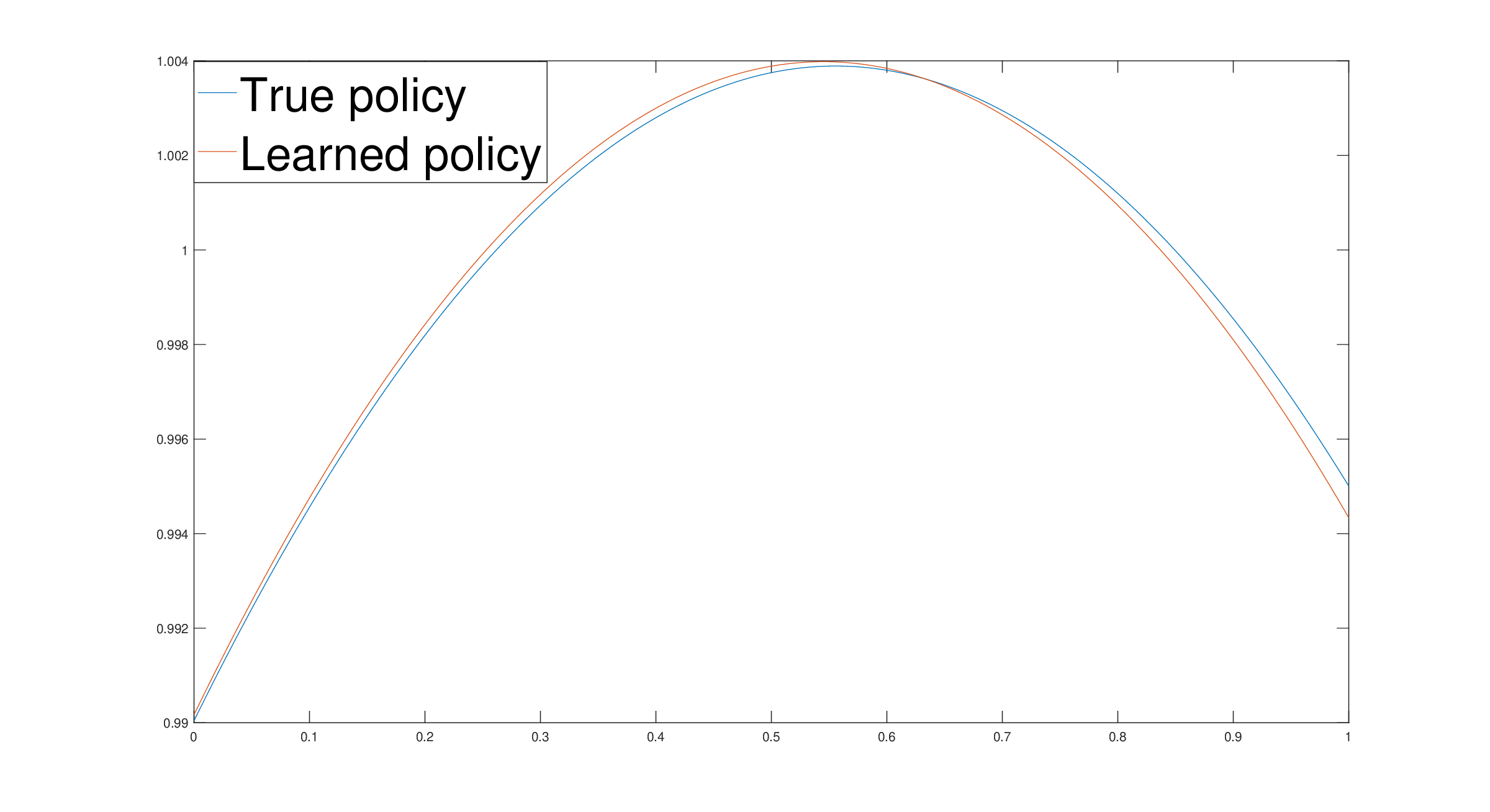}
       \caption{$m=1$}
       \label{fig:constdistrm1}
   \end{subfigure}
  \begin{subfigure}{0.32\textwidth}
       \includegraphics[width=55mm,height=7cm]{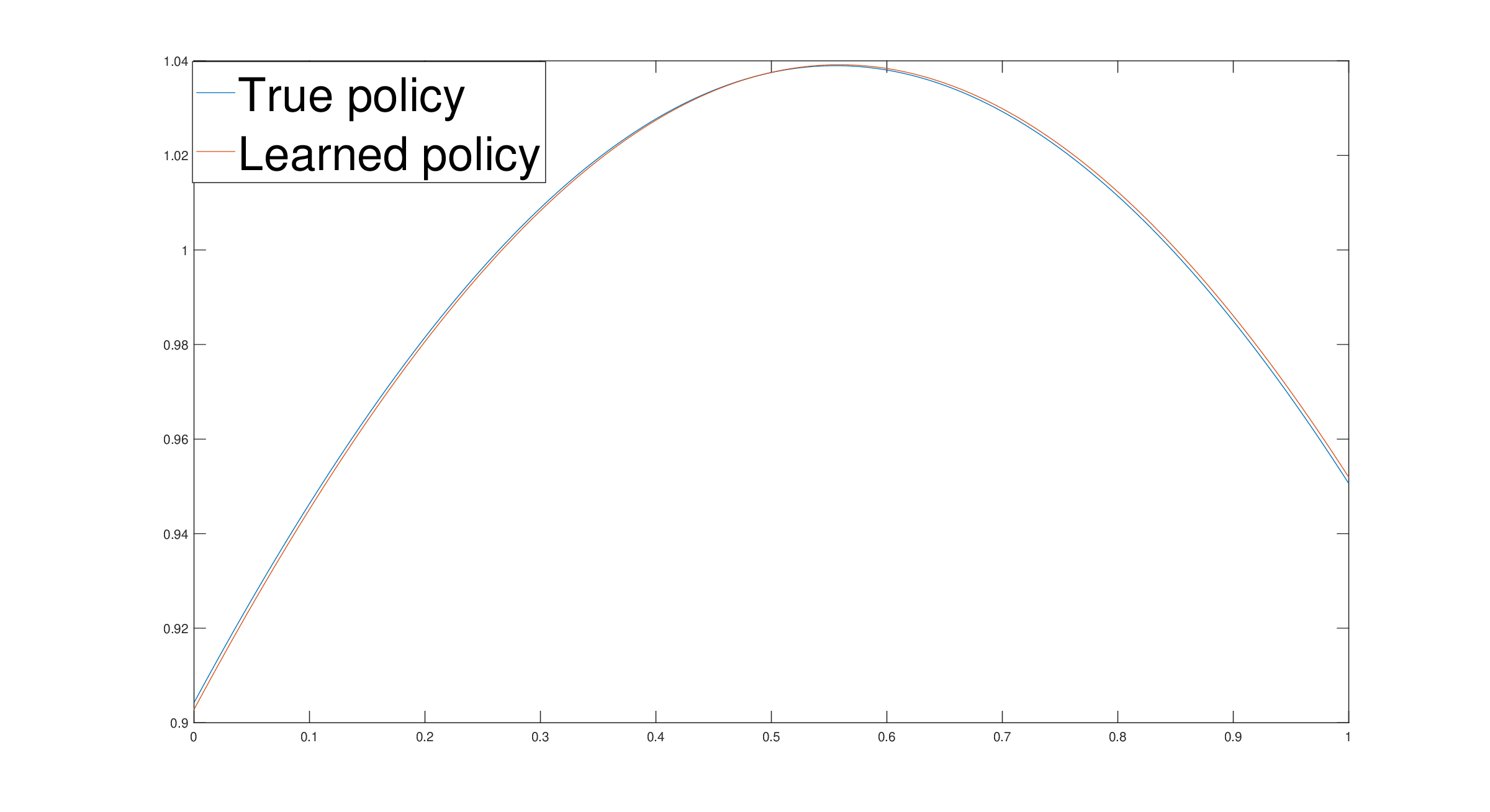}
       \caption{$m=0.1$}
       \label{fig:constdistrm01}
   \end{subfigure}
	  \begin{subfigure}{0.32\textwidth}
       \includegraphics[width=55mm,height=7cm]{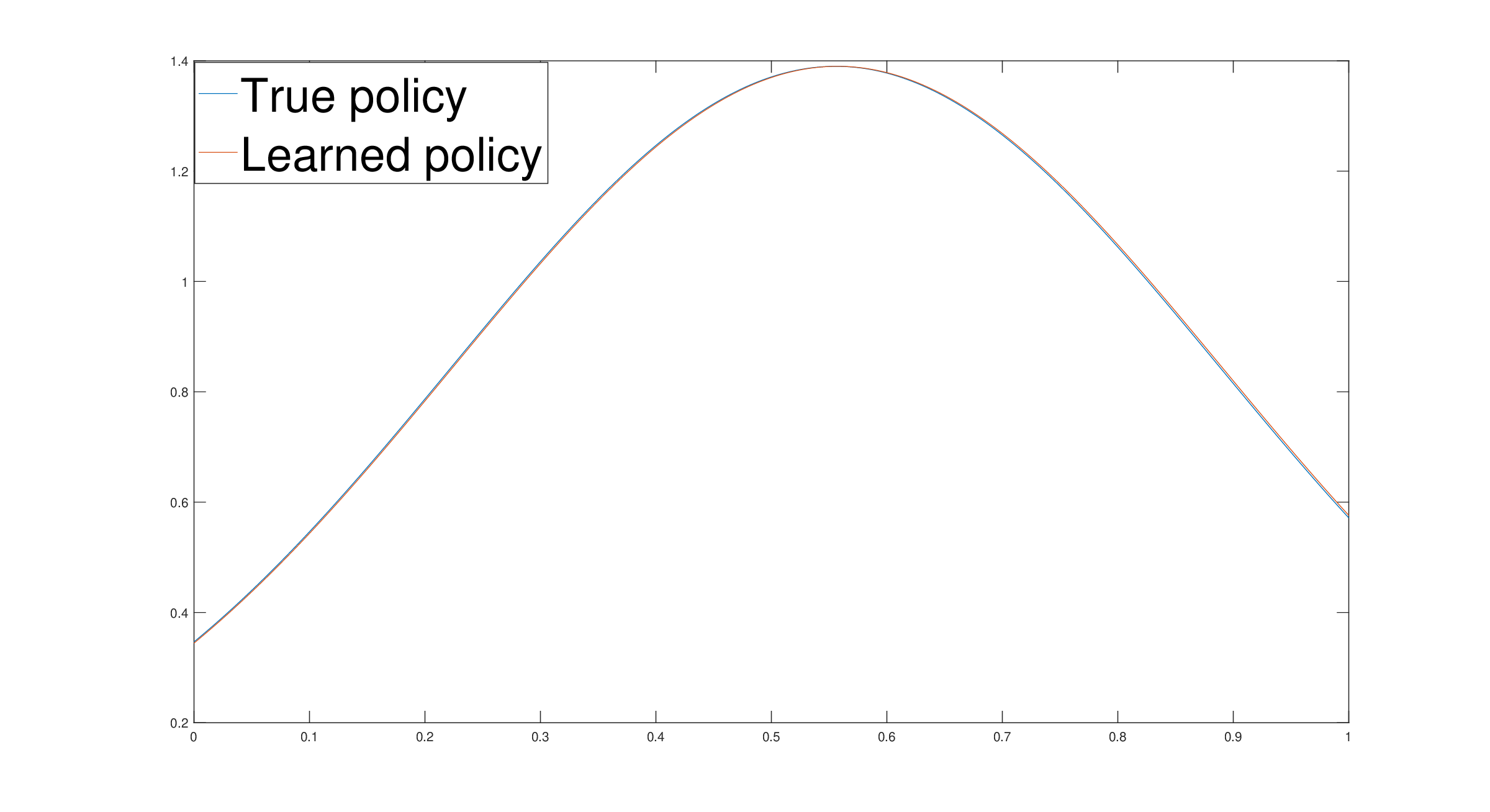}
       \caption{$m=0.01$}
       \label{fig:constdistm001}
			  \end{subfigure}
			   \caption{\small Learned policy for different exploration rates.}
				\label{fig:constrainedpolicy}
 \end{figure}

Figure \ref{fig:constrainedpolicy} reports the learned policy in comparison with the true optimal policy which is a Gaussian distribution truncated on $[a,b]$.  It can be observed that the smaller the value of the exploration rate is, the more closely matched the learned optimal strategy is to that of the ground true optimal policy. As shown in Figure \ref{fig:constrainedpolicyconvergence}, when $m$ is small, the (truncated Gaussian) optimal
policy gets closer to the Dirac distribution at the ``constrained'' Merton strategy $\pi^{0,[a,b]}$ defined in \eqref{eq:constrainedMerton}, which is consistent to the result obtained in Theorem \ref{Thm-convergenceconstrained}.
\begin{figure}[H] 	% <---
   \begin{subfigure}{0.48\textwidth}
       \includegraphics[width=95mm,height=7cm]{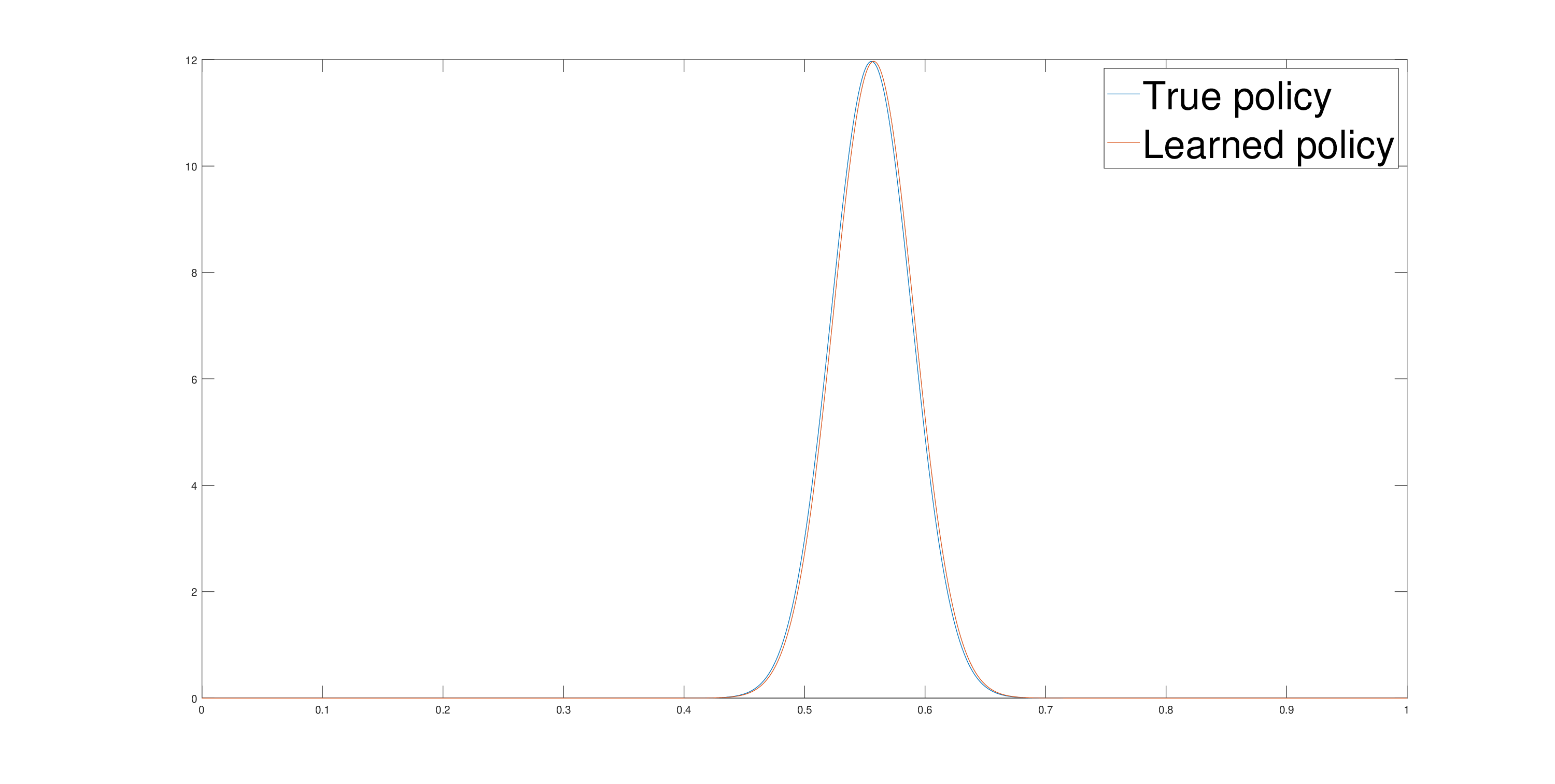}
       \caption{$m=0.0001$}
       \label{fig:constdistrm1}
   \end{subfigure}
  \begin{subfigure}{0.48\textwidth}
       \includegraphics[width=95mm,height=7cm]{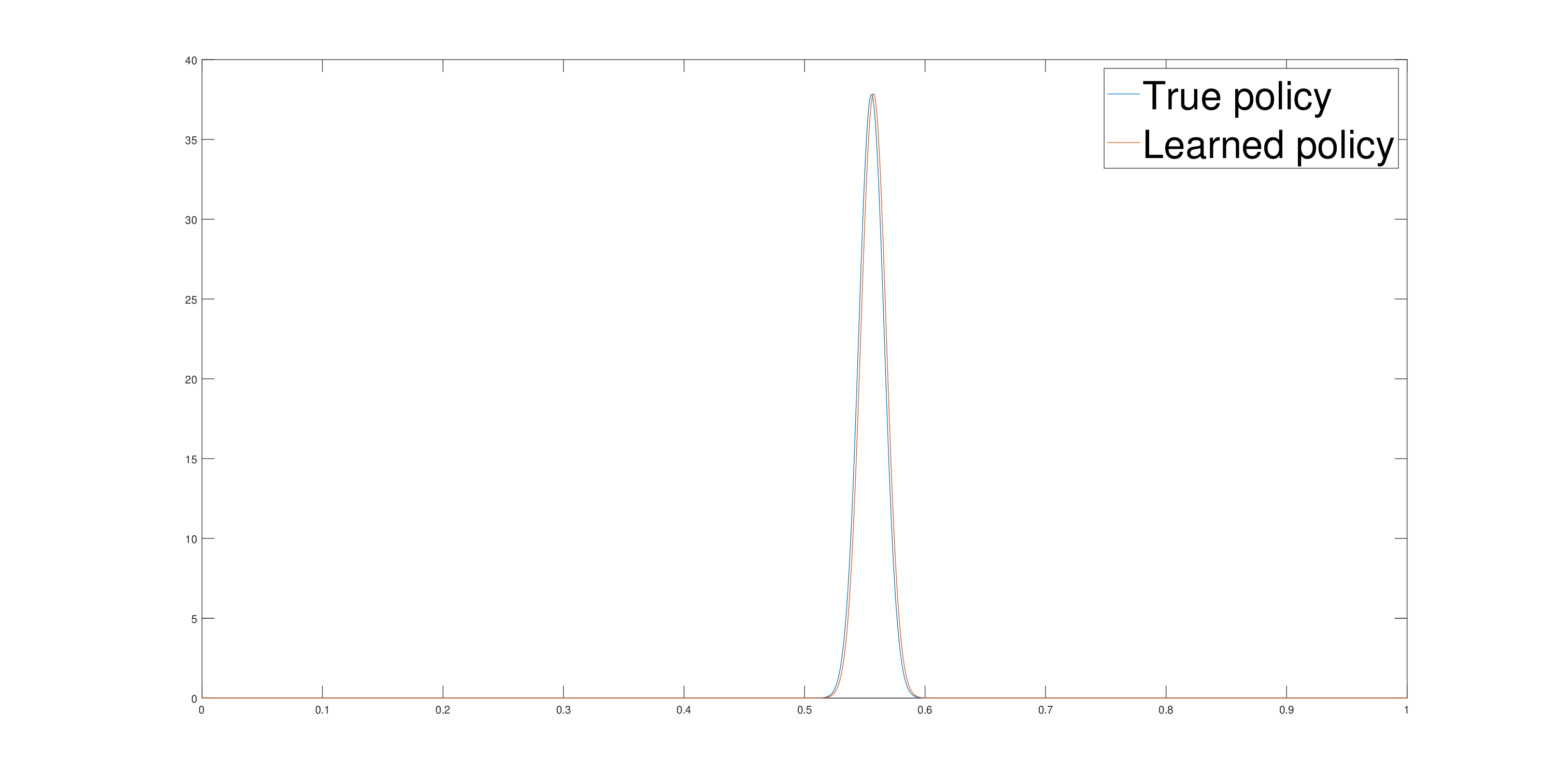}
       \caption{$m=0.00001$}
       \label{fig:constdistrm01}
   \end{subfigure}
			   \caption{\small Convergence of the constrained policy when the exploration rate $m\to 0$.}
				\label{fig:constrainedpolicyconvergence}
 \end{figure}

To have a more concrete comparison
between the true optimal optimal policy
and the learned optimal policy, in Table \ref{tab:truthTables},
we compare the true mean, the learned mean,
and the standard deviation of the learned mean. In addition,
we report the (empirical) Kullback–Leibler divergence (see e.g. \cite{csiszar1975divergence})
from the learned policy to the true policy for different $m$, see $D_{KL}$ column. 

\begin{table}[h]
\begin{center}\begin{tabular}{ |c|c|c|c|c| } 
 \hline
 $m$ & True mean & Learned mean & std  &$D_{KL}$ \\ 
% \hline
% 0.001 &2.555556  & 2.556511 &0.000004 \\ 
 \hline
 0.01 & 0.555556 & 0.555927 &0.000005 & 
0.000000\\ \hline
  0.1 &0.555556  &0.554679 &0.000007&0.000000\\
 \hline
  1 & 0.555556&0.555750  &0.000189&0.000052\\
 \hline
\end{tabular}
\caption{\small True mean versus Learned mean under portfolio constraint for different $m$}
    \label{tab:truthTables}  
\end{center}\end{table}

{Next we consider the dependence of $|v^{[a,b]}(t,x,m)-J^\theta(t,x,m)|$
on the number of iterations used. 
To easy the comparison, we consider
the difference $|v^{[a,b]}(0.5,0.5,0.01)-J^\theta(0.5,0.5,0.01)|$
as a function of iterations.
More specifically, we run Algorithm 1 to learn
the coefficients $\theta_1,\theta_2$ from data.
We then calculate the difference $|v^{[a,b]}(0.5,0.5,0.01)-J^\theta(0.5,0.5,0.01)|$.
In Figure \ref{Figure: Error versus iterations}, we report the difference
$|v^{[a,b]}(0.5,0.5,0.01)-J^\theta(0.5,0.5,0.01)|$ by averaging 20 runs.}

\begin{figure}[h]
	\centering
	\includegraphics[scale=0.4]{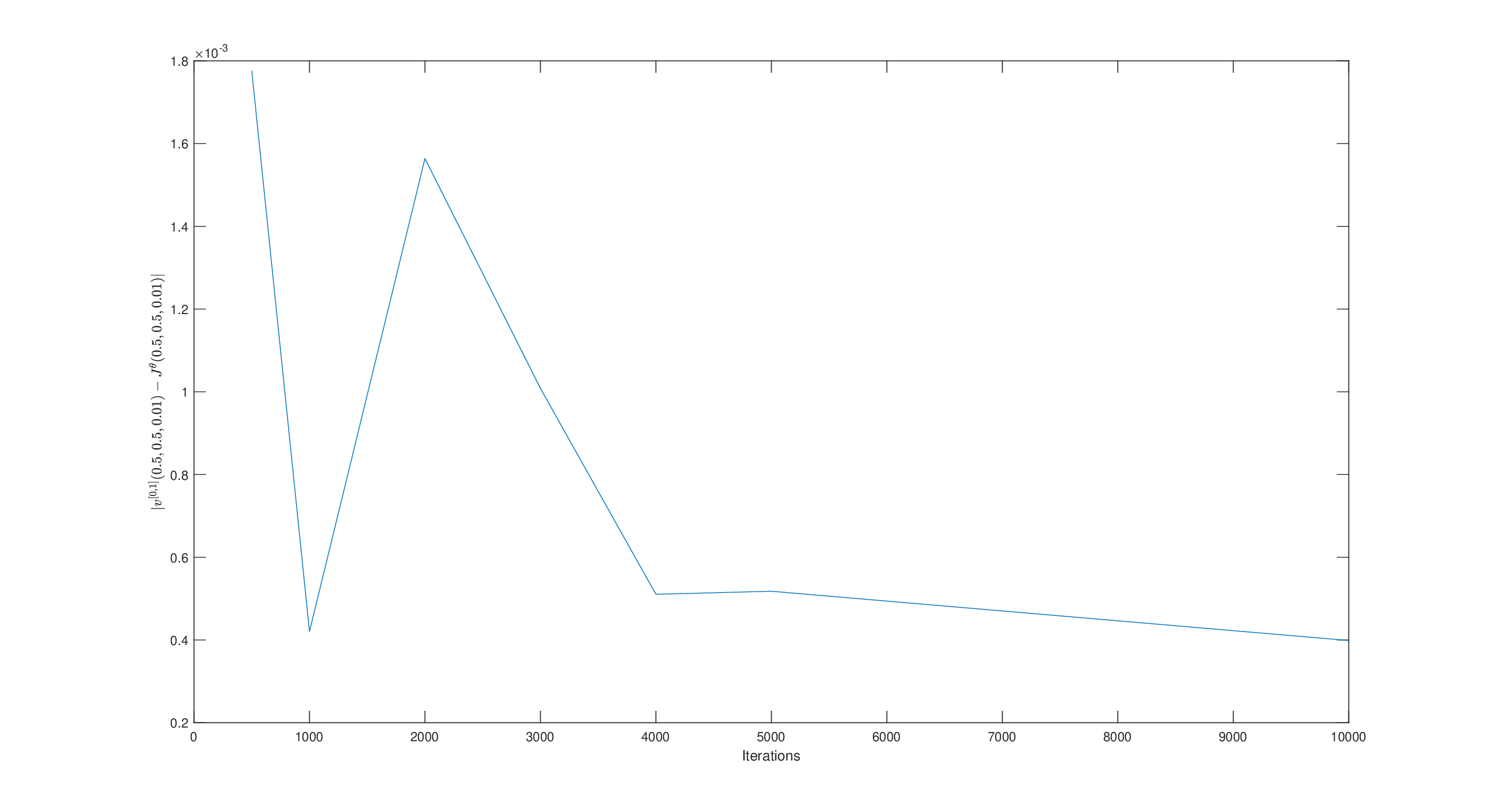}
\caption{{Error versus iterations}}\label{Figure: Error versus iterations}
\end{figure}
{It is clear from Figure \ref{Figure: Error versus iterations}
the error between the optimal value function
and the learned value function is decreasing
as the number of iterations increases.}

{To have a more concrete comparison,
we report
the true parameters  $(\theta_1,\theta_2)$, $(\phi_1,\phi_2)$
versus
the corresponding learned $(\theta_1,\theta_2)$, $(\phi_1,\phi_2)$
in Table \ref{tab:truthTables2}. As shown in Table \ref{tab:truthTables2}, the quality of learning is improving
as the number of iterations increases.}

\begin{table}[h]{
\begin{center}\begin{tabular}{|c |c|c||c|c|} 
 \hline
Iteration& True $(\theta_1,\theta_2)$ & Learned $(\theta_1,\theta_2)$&True $(\phi_1,\phi_2)$ & Learned $(\phi_1,\phi_2)$\\  \hline
500&(0.055929,-0.001497) &(0.052141,-0.001261) &(0.555556,2.407946) &(0.552537,2.407276)\\ \hline
1000&(0.055929,-0.001497) &(0.057109
,-0.001836) &(0.555556,2.407946) &(0.554388,2.413809
)\\ \hline
2000&(0.055929,-0.001497) &(0.054673,-0.003369
) &(0.555556,2.407946) &(0.555207,2.410082)\\ \hline
3000&(0.055929,-0.001497) &(0.057058,-0.004642
) &(0.555556,2.407946) &(0.555872,2.408604)\\ \hline
4000&(0.055929,-0.001497) &(0.054119
,0.001334) &(0.555556,2.407946) &(0.554001,2.409800)\\ \hline
5000&(0.055929,-0.001497) &(0.055721,-0.000253) &(0.555556,2.407946) & (0.557341,2.404103)\\ \hline
10000&(0.055929,-0.001497) &(0.056456,-0.002821) &(0.555556,2.407946) &(0.554809,2.405365)\\ \hline
\end{tabular}  
\caption{\small True parameters versus learned parameters} \label{tab:truthTables2}
   \end{center}}\end{table}

{Lastly in this section,
we would like to compare
the effect of the exploration
parameter $m$ on the 
distribution of the optimal wealth process.
In Figure \ref{fig:HistogramOptimalWealth}
we plot
the distribution of
the optimal wealth processes
at time $t=T/2=0.5$ using the dynamics in \eqref{eq:constrainedwealth} and  a sample size of $10000$.
From Figure \ref{fig:HistogramOptimalWealth}, 
it can be seen that
the exploration parameter $m$ plays a central role
on the distribution of the constrained optimal wealth process.
We can clearly distinguish the difference
between the case with exploration (i.e. $m\neq 0$) and the case without exploration (i.e. $m=0$). In particular, exploration leads to a more dispersed distribution with heavier tail. Again, as confirmed
from the theory in the previous section, this effect becomes less significant as the exploration parameter $m$ is smaller.}

\begin{figure}[H] 	% <---
   \begin{subfigure}{0.32\textwidth}
       \includegraphics[width=55mm,height=7cm]{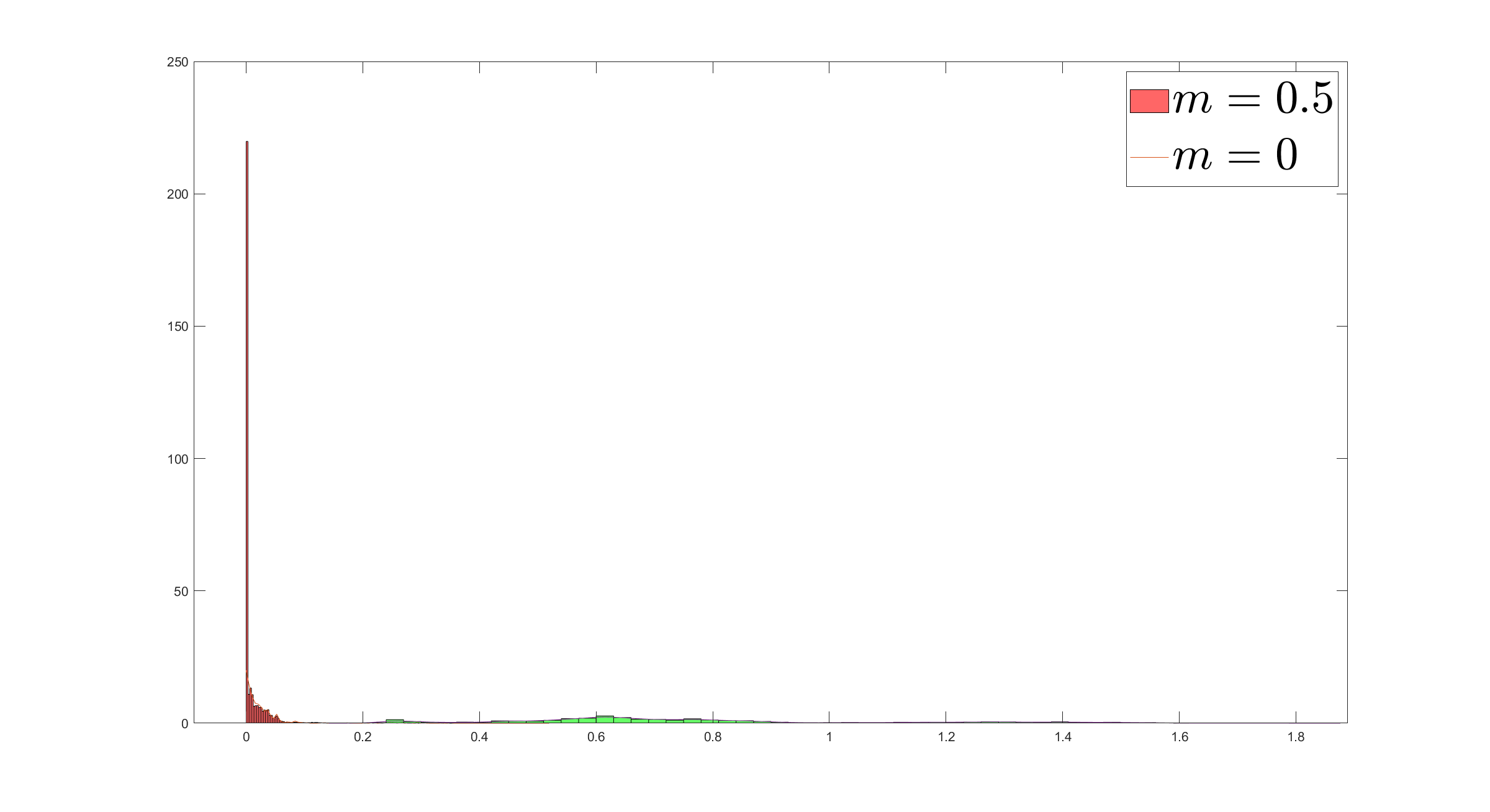}
       \caption{$m=0.5$}
       \label{fig:constdistrm1}
   \end{subfigure}
  \begin{subfigure}{0.32\textwidth}
       \includegraphics[width=55mm,height=7cm]{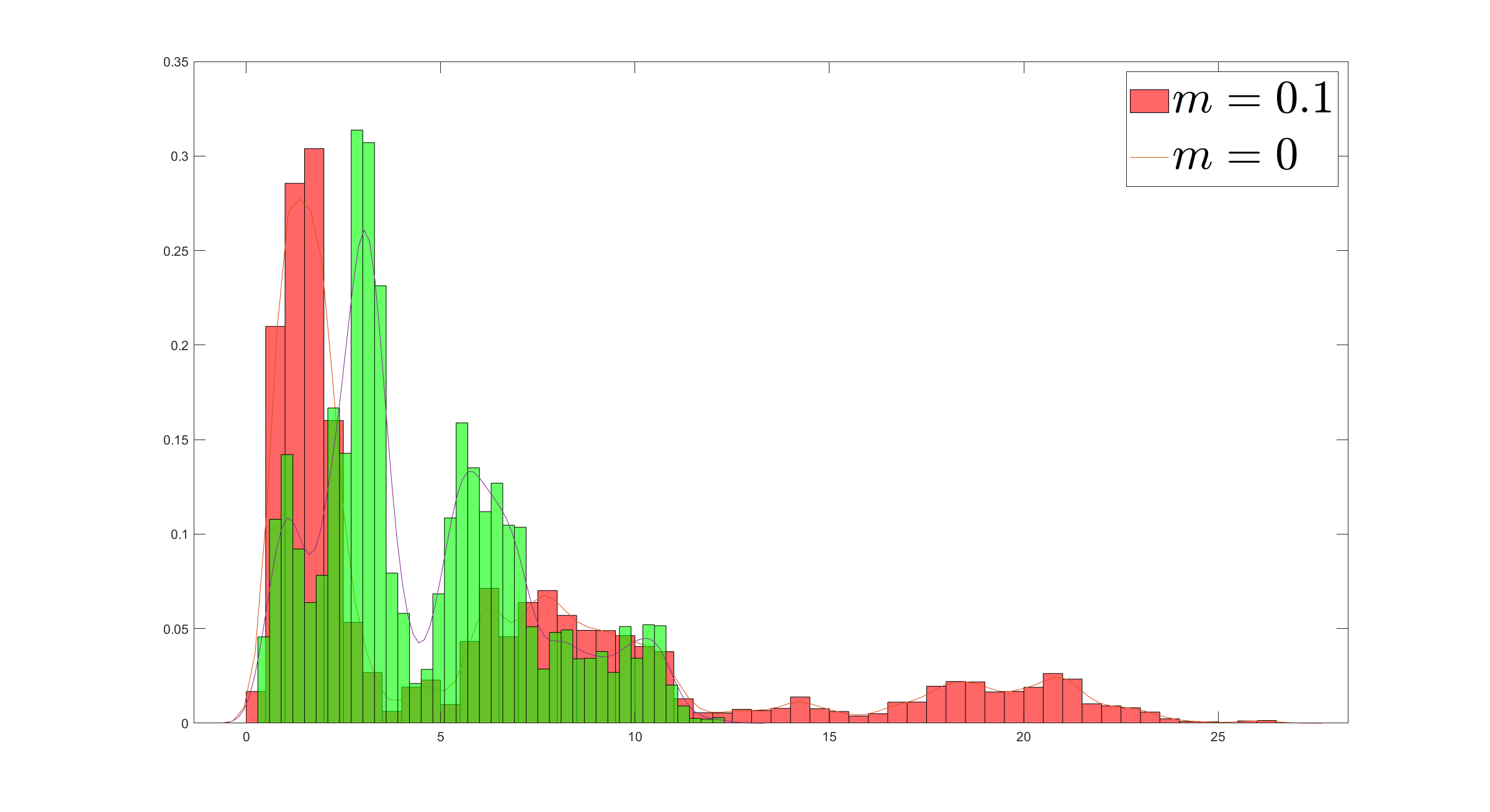}
       \caption{$m=0.1$}
       \label{fig:constdistrm01}
   \end{subfigure}
	  \begin{subfigure}{0.32\textwidth}
       \includegraphics[width=55mm,height=7cm]{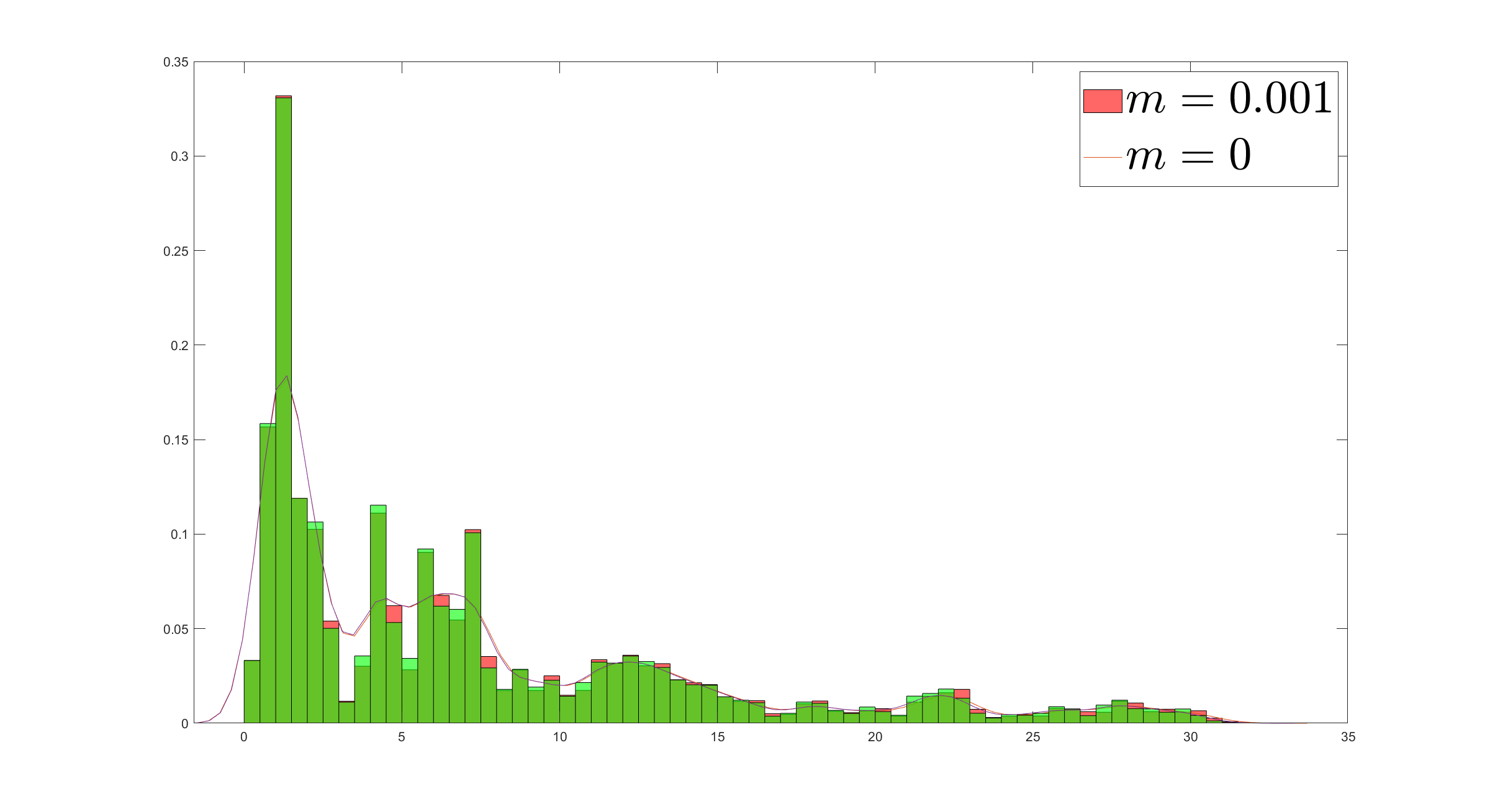}
       \caption{$m=0.001$}
       \label{fig:constdistm001}
			  \end{subfigure}
			   \caption{\small Density plot of wealth process: Exploration vs non exploration}
				\label{fig:HistogramOptimalWealth}
 \end{figure}

{
\section{Optimal portfolio with portfolio constraints and exploration for quadratic utility }\label{se:quadratiutility}
\label{seq:Optimal portfolio with portfolio constraints and exploration for quadratic utility }
We consider in this section the case with a quadratic utility function. It is worth noting that in portfolio theory the use of quadratic utility functions is very close to the classic Markowitz mean-variance portfolio and the reader is referred to e.g., \cite{duffie1991mean,bodnar2013equivalence} for discussions on connections between three quadratic optimization problems: the Markowitz mean–variance problem, the mean–variance utility function and the quadratic utility. Note that unlike mean-variance (MV) solution, the optimal solution under a quadratic utility function is not only time-consistent but, as shown below, also lies on the MV efficient frontier under mild conditions. Moreover, while it becomes extremely complex even without exploration e.g. \cite{li2002dynamic,bielecki2005continuous,li2016continuous} when a portfolio constraint is added in a MV problem, we still able to obtain closed form solutions when portfolio constraints are included in our exploratory learning setting with a quadratic utility.
 %Hence, are widely used in portfolio theory and considered as good approximations 
%
%{\bf 
%\begin{itemize}
	%\item Add here literature on quadratic utility functions and its connections with Mean-Variance problem, e.g. \cite{bodnar2013equivalence}
	%\item Add arguments related to MV problem (precommitted, equilbrium strategy,..), especially it becomes extremely complex even without exploration e.g. \cite{li2002dynamic,bielecki2005continuous,li2016continuous} when a portfolio constraint is added 
	%\item Cite and discuss the recent paper on MV with quadratic utility and RL \cite{kato2020mean}
%\end{itemize} 
%}
Note that for a quadratic utility function, the wealth process can be negative. Therefore, below we consider the amount of wealth $\theta_t$ invested in the risky asset at time $t$. To include portfolio constraints we assume that given the current wealth $X^\theta_t$, the risk investment amount $\theta_t$ at time $t$ is bounded by $a(t,X_t^{\theta})\le\theta_t \le b(t,X_t^{\theta})$, where $a$ (resp. $b$) is a deterministic {continuous} function defining the lower (resp. upper) portfolio bound. The wealth dynamics follows the following SDE
\begin{align} \label{eq_Xtnoc}
d X_t^{\theta} = & \left(r X_t^{\theta}+ \theta_t  (\mu-r) \right)  dt + \sigma \theta_t  d W_t \, , \quad X_0^{\theta} = x_0\in\bbr\,.
\end{align}
The set of admissible investment strategies is now defined by
\begin{align*}
\mathcal{D}_{[a,b]}(x_0) :=  \Bigg\{ (\theta)_{t\in[0,T]}\,&\text{is progressively measurable}\, \bigg \vert \,\; a(t,X_t^{\theta})\le\theta_t \le b(t,X_t^{\theta}),\; \\
&\text{\eqref{eq_Xtnoc} has a unique strong solution and}  \;  \E\bigg[\int_0^{T} \theta_t^2 d t\bigg]  < \infty\Bigg\}.
\end{align*}
%where $a$ (resp. $b$) is a deterministic function defining the lower (resp. upper) portfolio bound. 
Our objective is to maximize the terminal expected utility
\begin{align}
& \max_{\theta \in {\mathcal D}_{[a,b]}(x_0)}  \mathbb{E} \Big[U(X_T^{\theta})\Big], \label{eq: quadraticEU}
\end{align}
where $U(x)=Kx-\frac{1}{2}\varepsilon x^2$, with parameter $K, \varepsilon>0$.  The constant $1/\varepsilon$ reflects the agent's risk aversion and can be regarded as the risk aversion parameter in the mean–variance analysis.  We remark that the quadratic utility function $U(x)=Kx-\frac{1}{2}\varepsilon x^2$ has its global maximum at the so-called bliss point $K/\varepsilon$. Obviously, the quadratic utility function is symmetric with respect to the bliss point $K/\varepsilon$: it is increasing for $x<K/\varepsilon$ and decreasing for $x>K/\varepsilon$ and the utility values at some $x<K/\varepsilon$ and at $K/\varepsilon-x$ are equivalent. 
We consider the exploratory version of the wealth dynamics given by
\begin{align} 
d X_t^{\lambda} = \tilde{A}(t,X_t^\lambda;\lambda) dt +\tilde{B}(t,X_t^\lambda;\lambda)  d W_t \, , \quad X_0^{\lambda} = x_0,\label{eq_XtExploquadratic}
\end{align}
where
\begin{align}
\tilde{A}(t,x;\lambda) = \int_{a(t,x)}^{b(t,x)} \big(r x+ \theta  (\mu-r) \big)\lambda (\theta|t,x)   d \theta  ;\quad \tilde{B}(t,x;\lambda)=\sqrt{\int_{a(t,x)}^{b(t,x)} \sigma^2 \theta^2 \lambda (\theta|t, x) d \theta}. \label{ABtildequadratic}
\end{align}
and $\lambda(\cdot|t,x)$ is a probability density on the interval $[a(t,x),b(t,x)]$. The exploratory optimization is now stated by 
\begin{align} v(s,x;m):= \max_{\lambda\in\wt{\mathcal{H}}_{[a,b]  }}  \mathbb{E} \bigg[ U(X_T^{\lambda})-m\int_s^T\int_{a(t,X_t^\lambda)}^{b(t,X_t^\lambda)} \lambda (\theta|t,X_t^\lambda) \ln\lambda(\theta|t,X_t^\lambda) d \theta dt \big|X_s^\lambda=x\bigg].
\end{align}
where $\wt{\mathcal{H}}_{[a,b]}$ is the set of admissible feedback policies $\lambda $ that 
satisfy the following properties:

%{ADJUST the admissibility}
\begin{enumerate}
	\item For each $(t,x)\in[0,T]\times \bbr$, $\lambda (\cdot|t,x)$  is a density function on $[a(t,x),b(t,x)]$. %a.s.  and the function $\lambda (\cdot |t,x): [a(t,x),b(t,x)]\mapsto \bbr  $ is measurable; 
	\item {The mapping $ [0,T]\times \bbr\times [a(t,x), b(t,x)]\ni(t,x,\pi)\mapsto \lambda(\pi|t,x)$ is measurable}.
	%\item For each $E\in\cB(\cC)$, the process $\{\int_E\lambda_t(\pi)d\pi,\, t\in[0,T] \}$ is $\mathcal F$ progressively measurable;
	%\item For each $\lambda\in\mathcal{H}$, the exploratory SDE \eqref{eq_XtExplo} admits a unique strong solution denoted by $X^\lambda$ {which is positive} and
	%$$\mathbb{E} \Bigg[%e^{-\rho t}\bigg(\int_\cC{U}(c_tX_t^{\lambda})  \lambda_t (\pi,c) d\pi dc
	%U(X_T^{\lambda})-m \int_{0}^{T}\int_\cC \lambda (\pi|t,X_t^{\lambda})\ln\lambda (\pi|t,X_t^{\lambda}) d \pi  dt\Bigg]<\infty.$$
%\end{enumerate}
%\end{asp}
%\thaicomment{\bf Discussion on these assumptions: TO BE ADDED} 
	%The process $\{ \int_{a(t,X_t^{\lambda})}^{b(t,X_t^{\lambda})}  \lambda (\theta|t, X_t^{\lambda}) d \theta,  t\in[0,T]\}$ is $\mathcal F$ progressively measurable;
	%and $\E\bigg[\int_0^{T} \bigg(\int_{a(t,X_t^{\lambda})}^{b(t,X_t^{\lambda})} \sigma^2 \theta^2 \lambda (\theta|t, X_t^{\lambda}) d \theta\bigg) d t\bigg]  < \infty$;
	\item The exploration SDE \eqref{eq_XtExploquadratic} admits a unique strong solution denoted by $X^\lambda$ and
	$$\mathbb{E} \bigg[ U(X_T^{\lambda})-m\int_0^T\int_{a(t,X_t^\lambda)}^{b(t,X_t^\lambda)}  \lambda (\theta|t, X_t^{\lambda})\ln \lambda (\theta|t, X_t^{\lambda}) d \theta dt \bigg]<\infty.$$
\end{enumerate}
As before, the optimal value function satisfies the following HJB equation
\begin{align}
 v_t(t,x;m)+\sup_{\lambda\in\wt{\mathcal{H}}_{[a,b]}}\bigg\{\tilde{A}(t,x;\lambda)v_x(t,x;m)&+\frac{1}{2}\tilde{B}^2(t,x;\lambda)v_{xx}(t,x;m)\notag\\
&-m\int_{a(t,x)}^{b(t,x)} \lambda (\theta|t,x)  \ln\lambda(\theta|t,x) d \theta\bigg\}=0,\label{eq:constrainedHJBnocquadratic}
\end{align}
with terminal condition $v(T,x;m)=U(x)$.
%First, observe that the formula in the bracket of \eqref{eq:HJB} can be expressed as
%$$
%\int_\cC \bigg((r+\pi(\mu-r))xv_x(t,x;m)+\frac{1}{2}\sigma^2 x^2 \pi^2 v_{xx}(t,x;m) -m\ln\lambda_t\bigg)\lambda_t (\pi) d \pi.
%$$
Using again the standard argument of {DPP} we observe that under the portfolio constraint $\theta\in\mathcal D_{[a,b]}$, the optimal feedback policy now follows a truncated Gaussian distribution. 
\begin{lm}
In the exploratory constrained EU setting with quadratic utility function and $v_{xx}<0$, the optimal feedback policy $\tilde{\lambda}^{*}$ is a Gaussian distribution with mean $\tilde{\alpha}(t,x)$ and variance $\tilde{\beta}^2(t,x)$ truncated on interval $[a(t,x),b(t,x)]$, where 
\begin{equation}
\tilde{\alpha}(t,x)=-\frac{(\mu-r)v_x(t,x)}{\sigma^2 v_{xx}(t,x)};\quad \tilde{\beta}^2(t,x)=-\frac{m}{\sigma^2 v_{xx}(t,x)}.
\label{eq:alphabetatilde}
\end{equation}
The density of the optimal policy $\wt{\lambda}^{*}$ is given by
\begin{align}
 \wt{\lambda}^{*}(\theta|t, x;m)&:=\frac{1}{\wt{\beta}(t,x)}\frac{\varphi\bigg(\frac{\theta-\wt{\alpha}(t,x)}{\wt{\beta}(t,x)}\bigg)}{\Phi\bigg(\frac{b(x)-\wt{\alpha}(t,x)}{\wt{\beta}(t,x)}\bigg)-\Phi\bigg(\frac{a(x)-\wt{\alpha}(t,x)}{\wt{\beta}(t,x)}\bigg)},
\label{eq:constrainedoptpolicyquadratic}
\end{align}
where $\varphi$ and $\Phi$ are the PDF and CDF functions of the standard normal distribution, respectively.
\end{lm}
Substituting \eqref{eq:constrainedoptpolicyquadratic} back to the HJB \eqref{eq:constrainedHJBnocquadratic} we obtain the following non-linear PDE
\begin{align}
 v_t(t,x;m)+rxv_x(t,x;m)-\frac{1}{2}\frac{(\mu-r)^2v_x^2(t,x;m)}{\sigma^2 v_{xx}(t,x;m)}+\frac{m}{2}\ln\bigg(-\frac{2\pi_e m}{\sigma^2 v_{xx}(t,x;m)}\bigg)+{m\ln Z(t,x;m)}=0, \label{eq:constrainedHJBnocsimpquadratic}
\end{align} 
with terminal condition $v(T,x;m)=U(x)$, where by abusing notations, 
\begin{equation}
Z(t,x;m):={\Phi(\wt{Q}_b(t,x;m))-\Phi(\wt{Q}_a(t,x;m))}, %\frac{\phi(A(m))-\phi(B(m))}
\label{eq:}
\end{equation}
with
\begin{align}
&\wt{Q}_a(t,x;m):=\bigg(a(t,x)+\frac{(\mu-r)}{\sigma^2}\frac{ v_{x}(t,x)}{ v_{xx}(t,x)}\bigg)\sqrt{\frac{-\sigma^2  v_{xx}(t,x)}{m}};\\ &\wt{Q}_b(t,x;m):=\bigg(b(t,x)+\frac{(\mu-r)}{\sigma^2}\frac{ v_{x}(t,x)}{ v_{xx}(t,x)}\bigg)\sqrt{\frac{-\sigma^2 v_{xx}(t,x)}{m}}.
\label{eq:}
\end{align}
In the rest of this section we assume that the portfolio bounds are given by 
$$a(t,x):={-}\pi^{Merton}x+a_0(t),\quad \mbox{and}\quad b(t,x):={-}\pi^{Merton}x+b_0(t),$$
where, {as before $\pi^{Merton}= \frac{(\mu-r)}{\sigma^2}$}, $a_0 $ and $b_0$ are two time-varying continuous bounded functions with $a_0(t)<b_0(t),\, \forall t\in[0,T]$. 
%Below $\pi^{Merton}:=\frac{(\mu-r) }{\sigma^2}$. 
\begin{thm}[Quadratic utility-exploratory optimal investment under portfolio constraint]\label{Thm:constrainedquadratic}
%Assume that the portfolio bounds are given by 
%$$a(t,x):=\pi^{Merton}x+a_0(t),\quad \mbox{and}\quad b(t,x):=\pi^{Merton}x+b_0(t),$$
%where  $a_0 $ and $b_0$ are two time-varying continuous bounded functions and with $a_0(t)<b_0(t),\, \forall t\in[0,T]$ . Then, 
The optimal value function of the entropy-regularized exploratory constrained optimal investment problem with quadratic utility $U(x)=Kx-\frac{1}{2}\varepsilon x^2$ is given by
\begin{align}
\wt{v}^{[a,b]}(t,x;m)=&-\frac{1}{2}\varepsilon x^2 e^{-(\rho^2-2r)(T-t)}+Kxe^{-(\rho^2-r)(T-t)}-
 \frac{K^2}{2\varepsilon}(1-e^{-\rho^2(T-t)})\\
&+
\frac{m}{4}(\rho^2-2r)(T-t)^2+\frac{m}{2} \ln(\varepsilon^{-1}\sigma^{-2}{2\pi_e m})(T-t)+ m F(t,m),
\label{eq:quadraticvalue}
\end{align}
for $(t,x)\in[0,T]\times \bbr$, where $\rho:=\frac{(\mu-r) }{\sigma}$ is the Sharpe ratio, $F(t,m)=\int_t^T \ln (f(s,m)) ds$, 
\begin{equation}
f(t,m):=\Phi(\wt{Q}_b(t;m))-\Phi(\wt{Q}_a(t;m)),
%\Phi\bigg(\bigg(b_0(t) - \frac{K}{\varepsilon} e^{r(T-t)}\pi^{Merton}\bigg)\sqrt{\frac{\varepsilon \sigma^2e^{(\rho^2-2r)(T-t)}}{m}}\bigg)-\Phi\bigg(\bigg(a_0(t) - \frac{K}{\varepsilon} e^{r(T-t)}\pi^{Merton}\bigg)\sqrt{\frac{\varepsilon \sigma^2e^{(\rho^2-2r)(T-t)}}{m}}\bigg)\bigg),
\label{eq:ZabQuadratic}
\end{equation}
where
$$
\wt{Q}_a(t;m):=\bigg(a_0(t)-\frac{K}{\varepsilon} e^{-r(T-t)} \pi^{Merton}\bigg)\sqrt{\frac{\varepsilon\sigma^2 }{m}e^{-(\rho^2-2r)(T-t)}});
$$
$$
\wt{Q}_b(t;m):=\bigg(b_0(t)-\frac{K}{\varepsilon} e^{-r(T-t)} \pi^{Merton}\bigg)\sqrt{\frac{\varepsilon\sigma^2 }{m}e^{-(\rho^2-2r)(T-t)}}).
$$
Moreover, the optimal feedback distribution control $\wt{\lambda}^{*}(\cdot|t,x;m)$ is a Gaussian distribution with parameter $(\frac{K}{\varepsilon} e^{-r(T-t)}-x) \pi^{Merton}$ and  $\frac{m}{\varepsilon \sigma^2 }e^{(\rho^2-2r)(T-t)}$, truncated  on the interval $[a(t,x),b(t,x)]$ i.e.
\begin{equation}
\wt{\lambda}^{*}(\theta|t,x;m)=\mathcal N\left(\theta \bigg|(\frac{K}{\varepsilon} e^{-r(T-t)}-x) \pi^{Merton}, \frac{m}{\varepsilon\sigma^2 }e^{-(2r-\rho^2)(T-t)}\right)\bigg|_{[a(t,x),b(t,x)]}.
\label{eq:quadratic}
\end{equation}
%and the associated optimal wealth under $\lambda^*$ is given by the following SDE
%\begin{align} 
%d X_t^{\lambda^{*,[a,b]}} = X_t^{\lambda^{*,[a,b]}}  \bigg(r+\pi_t^{*,[a,b]} (\mu-r) \bigg) dt + \sigma \pi_t ^{*,[a,b]}  X_t^{\lambda^{*,[a,b]}}  d W_t \, , \quad X_0 = x_0 \,.
%\label{eq:constrainedwealth}
%\end{align}
%where 
%\begin{equation}
%\pi^{*,[a,b]}_t=\pi^{Merton} +\frac{\varphi\bigg((a-\pi^{Merton})\sigma m^{-1/2}\bigg)-\varphi\bigg((b-\pi^{Merton})\sigma m^{-1/2}\bigg)}{\sigma  m^{-1/2}Z_{a,b}(m)}.
%\label{eq:constrainedpi}
%\end{equation}
\end{thm}
\begin{cor}[Unconstrained exploratory quadratic portfolio]\label{cor:unconstrainedquadratic}
The optimal value function of the entropy-regularized exploratory \it{unconstrained} optimal investment problem for the quadratic utility $U(x)=Kx-\frac{1}{2}\varepsilon x^2$ is given by
\begin{align}
\wt{v}^{}(t,x;m)=&-\frac{1}{2}\varepsilon x^2 e^{-(\rho^2-2r)(T-t)}+Kxe^{-(\rho^2-r)(T-t)}-
 \frac{K^2}{2\varepsilon}(1-e^{-\rho^2(T-t)})\\
&+
\frac{m}{4}(\rho^2-2r)(T-t)^2+\frac{m}{2} \ln(\varepsilon^{-1}\sigma^{-2}{2\pi_e m})(T-t),
\label{eq:unquadraticvalue}
\end{align}
for $(t,x)\in[0,T]\times \bbr$. Moreover, the unconstrained optimal feedback distribution control is a Gaussian distribution with parameter $(\frac{K}{\varepsilon} e^{-r(T-t)}-x) \pi^{Merton}$ and  $\frac{m}{\varepsilon \sigma^2 }e^{-(2r-\rho^2)(T-t)}$.
\end{cor}
\proof This is a direct consequence of Theorem \ref{Thm:constrainedquadratic} by letting $a_0(t)\to -\infty$ and $b_0(t)\to +\infty$. \endproof
\begin{remark}Compared to the mean-variance results obtained in \cite{wang2020continuous}, the risk reward parameter $K{\epsilon}^{-1}$ plays the role of the Lagrangian multiplier $\omega$ (under their notations) in the mean-variance setting. It should be mentioned that instead of considering the true portfolio, the authors in \cite{wang2020continuous} study the mean-variance problem for the discounted portfolio, which can be obtained by assuming $r=0$ in our setting. Note that the variance of our optimal policy is different from the one obtained in  \cite{wang2020continuous} by a multiplicative factor $2\epsilon^{-1}$ since our quadratic utility is nothing else but the mean-variance utility in \cite{wang2020continuous} scaled up by $-\epsilon/2$. 
\end{remark}
Letting $m\to 0$ in Corollary \ref{cor:unconstrainedquadratic} we can obtain the unconstrained optimal portfolio without exploration for quadratic utility function.
\begin{cor}[Unconstrained quadratic portfolio]\label{cor:unconstrainedquadraticnoexploration}
For a quadratic utility $U(x)=Kx-\frac{1}{2}\varepsilon x^2$, the optimal value function of \it{unconstrained} optimal investment problem without exploration is given by
\begin{align}
\wt{v}(t,x;0)
=&-\frac{1}{2}\varepsilon x^2 e^{-(\rho^2-2r)(T-t)}+Kxe^{-(\rho^2-r)(T-t)}-
 \frac{K^2}{2\varepsilon}(1-e^{-\rho^2(T-t)})
\end{align}
and the unconstrained optimal control strategy is given by $(\frac{K}{\varepsilon} e^{-r(T-t)}-x) \pi^{Merton}$.
\end{cor}
Following similar arguments used in Proposition \ref{eq:ExploratoryTheorem2} and Lemma \ref{Le:compare} we can confirm that the exploration cost for the constrained problem is smaller than that of the unconstrained problem for a quadratic utility function.  
\begin{prop}
\label{eq:ExploratoryTheorem2quadratic}
In the constrained problem with exploration and quadratic utility function, the exploration cost is given by 
$$
\wt{L}^{[a,b]}(T,x;m)=\frac{mT}{2}+m\int_0^T \frac{\wt{Q}_a(t;m)\varphi(\wt{Q}_a(t;m))-\wt{Q}_b(t;m)\varphi((\wt{Q}_b(t;m))}{f(t,m)}d t\le \frac{mT}{2}.
$$
Moreover, $\lim_{m\to 0} \wt{L}^{[a,b]}(T,x;m)=0$
\end{prop}
%\proof Lemma \ref{Le:compare}. \endproof
%Intuitively, $ \psi(t,x;m)$ measures the exploration effect. It is easy to see that  $ \psi(t,x;m)\to 0$ as $m\to 0$, hence, $v(t,x;m)\to v(t,x;0)=\ln x+(r+\frac{1}{2}\frac{(\mu-r)^2}{\sigma^2})$ which is the optimal value function in the absence of exploration. Clearly, this is equivalent to $k(t)\to 1$ and $l(t)\to (r+\frac{1}{2}\frac{(\mu-r)^2}{\sigma^2}) $. 
Similarly to Theorems \ref{Thm-updateconstrained}-\ref{Thm-convergenceconstrained} for logarithmic utility, a policy improvement theorem can be shown for 
quadratic utility functions. 
}
{
Below we show that the optimal solution of the exploratory quadratic utility problem is mean-variance efficient (see e.g. \cite{luenberger1997investment,kato2020mean,duffie1991mean} for similar discussions in the case without exploration). 
\begin{prop}\label{Pro: MVequivalence}
Assume that the agent's initial wealth is smaller than the discounted reward level $\frac{K}{\epsilon}$, i.e. $x_0\le \frac{K}{\epsilon} e^{-rT}$. Then, the exploratory unconstrained optimal portfolio belongs to the mean-variance frontier.
\end{prop}
\begin{remark}Note that since the exploration parameter only effects the diffusion term, we can see that Proposition \ref{Pro: MVequivalence} also holds for the unconstrained case without exploration. Moreover, it can be shown that Proposition \ref{Pro: MVequivalence} still holds true for the constrained case with exploration. Indeed, recall the optimal constrained exploratory solution whose dynamics is given by 
$$
d X^{\wt{\lambda}^*}_t=  \tilde{A}(t,X^{\wt{\lambda}^*}_t;\wt{\lambda}^*) dt+ \wt{B}(t,X^{\wt{\lambda}^*}_t;\wt{\lambda}^*) dW_t,$$ where $\tilde{A}$, $\wt{B}$ are given by \eqref{eq:constrainedpiquadratic}. Note that the second term of $\wt{\theta}_t^{*,[a,b]}$ in \eqref{eq:constrainedpiquadratic} is negative because
$$
\frac{\varphi\big(\wt{Q}_a(t;m)\big)-\varphi\big(\wt{Q}_b(t;m)\big)}{f(t,m)}<0.
$$ Using the comparison principle of ordinary differential equations we can conclude that 
$\E[X^{\wt{\lambda}^*}_T]\le \E[X^{0,\wt{\lambda}^*}_T]\le  \frac{K}{\epsilon}$, which implies that the optimal solution of the expected quadratic utility problem with both portfolio constraints and exploration lies on the (constrained) mean-variance frontier.
\end{remark}
}
{
\subsection{Implementation}
We consider the case without constraint.
Recall the value function
\begin{align}
\wt{v}^{}(t,x;m)=&-\frac{1}{2}\varepsilon x^2 e^{-(\rho^2-2r)(T-t)}+Kxe^{-(\rho^2-r)(T-t)}-
 \frac{K^2}{2\varepsilon}(1-e^{-\rho^2(T-t)})\\
&+
\frac{m}{4}(\rho^2-2r)(T-t)^2+\frac{m}{2} \ln(\varepsilon^{-1}\sigma^{-2}{2\pi_e m})(T-t),
\label{eq:unquadraticvalue1}
\end{align}
for $(t,x)\in[0,T]\times \bbr_+$. 
We choose to parametrize the value function as 
\begin{align*}
J^\theta (t,x;m)=-\frac{1}{2}\epsilon(xe^{r(T-t)}-K/\epsilon)^2e^{-\theta_3(T-t)}
+K^2/\epsilon e^{-\theta_3(T-t)}
-\frac{1}{2}K^2/\epsilon
+\theta_2(T-t)^2 +\theta_1(T-t).
\end{align*}
That is,
$$
\left\{
\begin{array}{ll}
&\theta_1=m/2\ln(2\pi_e m\epsilon^{-1}\sigma^{-2})\\
&\theta_2=m/4(\rho^2-2r)\\
&\theta_3=\rho^2.
\end{array}
\right.
$$
From here the test functions are chosen as
\begin{equation}
\left\{
\begin{array}{ll}
\frac{\partial J^\theta}{\partial\theta_1}=(T-t),\\
\frac{\partial J^\theta}{\partial\theta_2}=(T-t)^2,\\
\frac{\partial J^\theta}{\partial\theta_3}=\frac{1}{2}\epsilon(xe^{r(T-t)}-K/\epsilon)^2e^{-\theta_3(T-t)}(T-t)
-K^2/\epsilon e^{-\theta_3(T-t)}(T-t).
\end{array}
\right.
\end{equation}
Also, recall  the unconstrained optimal feedback distribution control is a Gaussian distribution with parameter $(\frac{K}{\varepsilon} e^{-r(T-t)}-x) \pi^{Merton}$ and  $\frac{m}{\varepsilon \sigma^2 }e^{-(2r-\rho^2)(T-t)}$.
For the policy distribution,
we parametrize it as follows
\begin{align*}
L(\phi_1,\phi_2,\phi_3)=
\mathcal N\bigg((\frac{K}{\varepsilon} e^{-r(T-t)}-x) \phi_1, e^{\phi_2+\phi_3(T-t)}\bigg).
\end{align*}
From here, it can be seen that
the entropy is given by
\begin{align*}
H(\phi_1,\phi_2,\phi_3)= 0.5 (\ln(2\pi_e) + 1 +\phi_2 + \phi_3 (T - t));
\end{align*}
 The log-likelihood, $l(\phi_1,\phi_2,\phi_3):=\ln L(\phi_1,\phi_2,\phi_3)$ is given by
\begin{equation*}
l(\phi_1,\phi_2,\phi_3)(y)=-\frac{1}{2}\ln (2\pi)-\frac{1}{2}(\phi_2+\phi_3(T-t))
-\frac{1}{2}\bigg(y - \phi_1(e^{-r(T-t)}K/\epsilon -x)\bigg)^2 e^{-\phi_2-\phi_3(T-t)}.
\label{eq:}
\end{equation*}
Also, the derivatives of the log-likelihood are given by
\begin{align*}
&\frac{\partial l}{\partial \phi_1}=
\bigg(y- \phi_1(e^{-r(T-t)}K/\epsilon -x)\bigg) e^{-\phi_2-\phi_3(T-t)}(e^{-r(T-t)}K/\epsilon -x)\\
%&\red{(a - \phi_1(K/\epsilonexp(-r(T-t))-x))(K/\epsilonexp(-r(T-t))-x)exp(-\phi_2-\phi_3(T-t))}\\
&\frac{\partial l}{\partial \phi_2}=-\frac{1}{2}+\frac{1}{2}\bigg(y - \phi_1(e^{-r(T-t)}K/\epsilon -x)\bigg)^2 e^{-\phi_2-\phi_3(T-t)}\\
%&\red{\frac{\partial l}{\partial \phi_2}=-0.5+0.5.*(a - \phi_1(K/\epsilon.*exp(-r.*(T-t))-x))^2.*exp(-\phi_2-\phi_3(T-t))}\\
&\frac{\partial l}{\partial \phi_3}=-\frac{1}{2}(T-t)+\frac{1}{2}\bigg(y - \phi_1(e^{-r(T-t)}K/\epsilon -x)\bigg)^2 e^{-\phi_2-\phi_3(T-t)}(T-t).%\\
%&\frac{\partial l}{\partial \phi_3}=-0.5.*(T-t)+ 0.5.*(a - \phi_1(K/\epsilon.* exp(-r(T-t))-x))^2 exp(-\phi_2-\phi_3(T-t))(T-t)
\end{align*}
\subsection{Numerical example}
We illustrate
the results for quadratic utility function.
Specifically, we consider the scenario 
without constraints with $r=0.02, \mu=0.05, \sigma=0.30, T=1, x_0=0.5$ and $\epsilon=1$. We remark that the condition $x_0\le \frac{K}{\epsilon} e^{-rT}$ is fulfilled. 
First, we run the algorithm
using $1000$ iterations, the learning rates which are used to update $\theta$ and $\phi$ are chosen to be $\eta_\theta=0.01$,
$\eta_\phi=0.01$, respectively with the decay rate $l(i)=1/i^{0.51}$.
Moreover, we choose the exploration rate $m=0.01$.
Figure \ref{fig: quadratic-theta-convergence}
plots the evolution
of coefficients of $J^{\theta}$. In this case,
the true coefficients are $(\theta_1,\theta_2,\theta_3)=(-0.001797,-0.000075,0.010000)$.
\begin{figure}[H] 	% <---
   \begin{subfigure}{0.32\textwidth}
       \includegraphics[width=55mm,height=7cm]{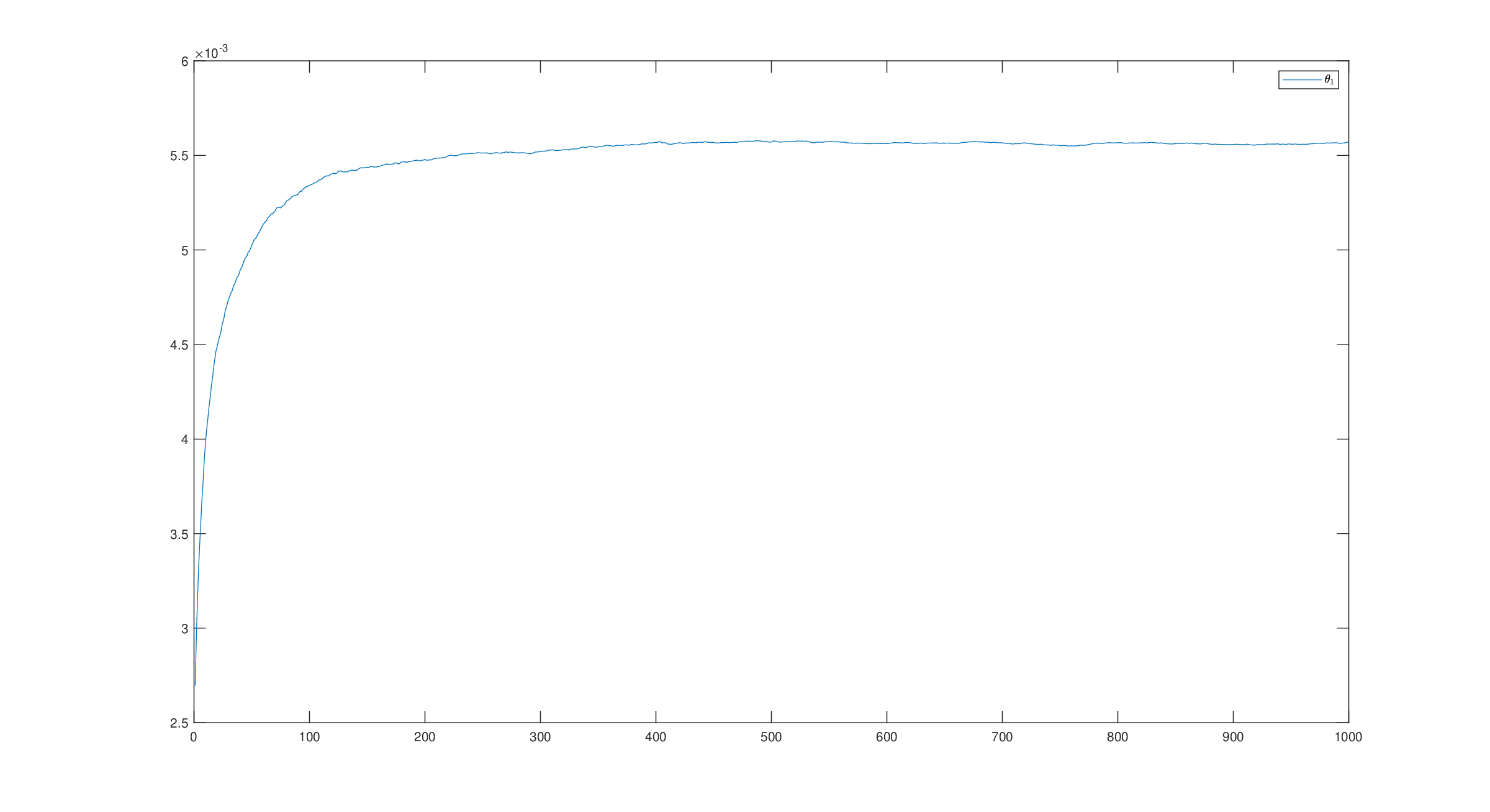}
       \caption{$\theta_1$}
       \label{fig:theta1}
   \end{subfigure}
  \begin{subfigure}{0.32\textwidth}
       \includegraphics[width=55mm,height=7cm]{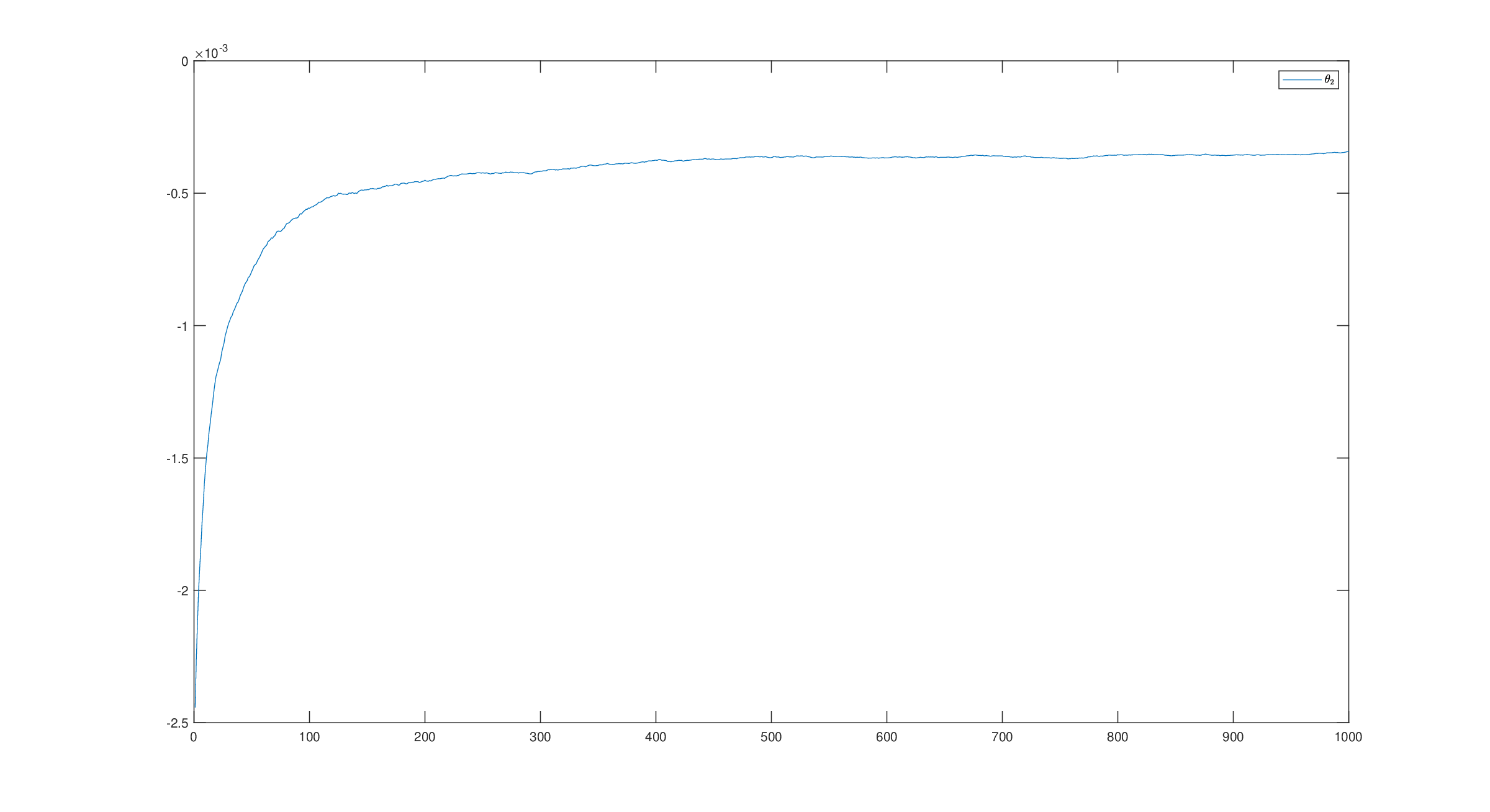}
       \caption{$\theta_2$}
       \label{fig:theta2}
   \end{subfigure}
	  \begin{subfigure}{0.32\textwidth}
       \includegraphics[width=55mm,height=7cm]{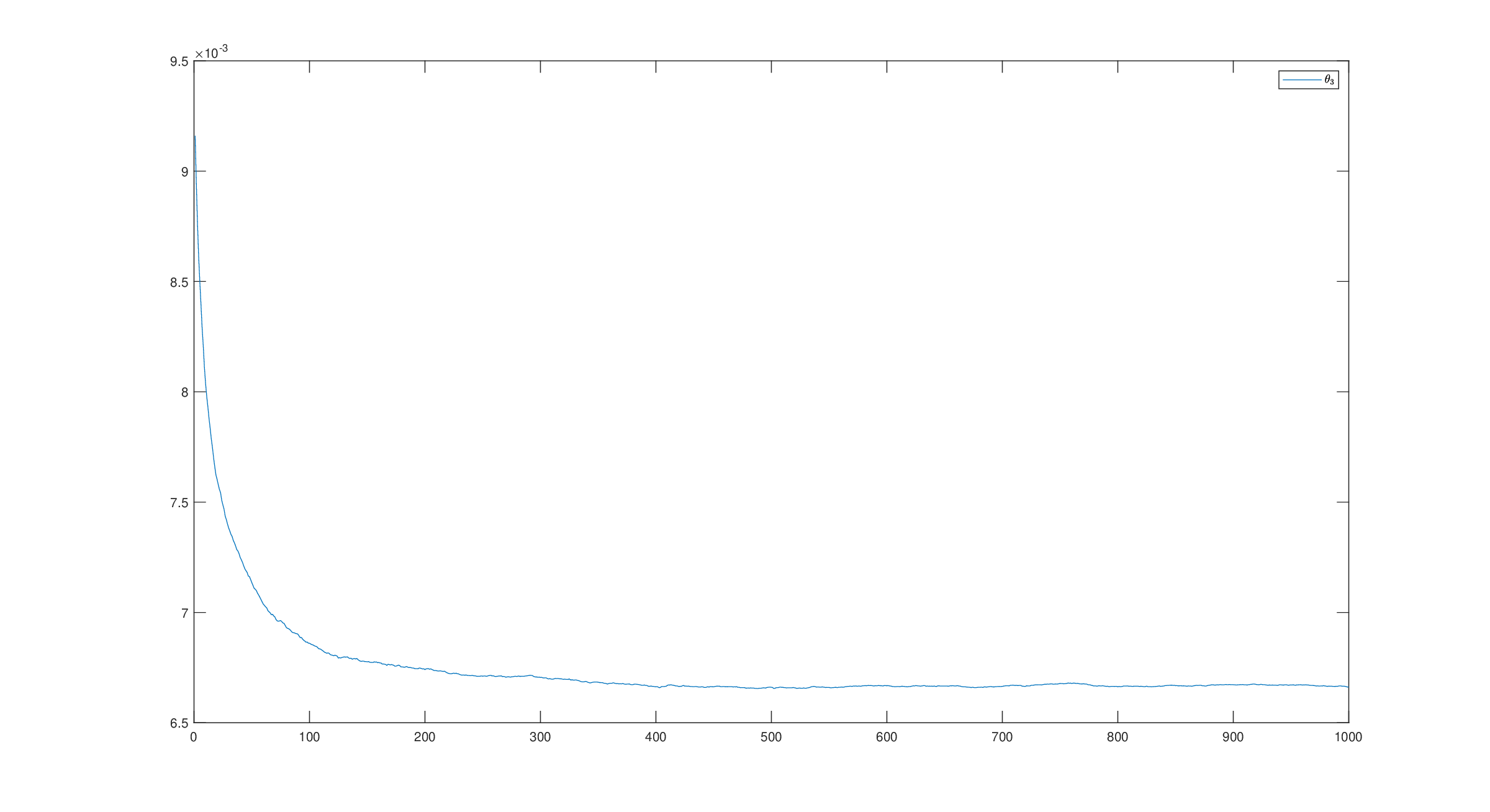}
       \caption{$\theta_3$}
       \label{fig:theta3}
			  \end{subfigure}
			   \caption{Convergence of $\theta_1$ (Left), $\theta_2$ (Center) and $\theta_3$ (Right)}
				\label{fig: quadratic-theta-convergence}
 \end{figure}
In this case, it is evident from Figure \ref{fig: quadratic-theta-convergence}
that $(\theta_1,\theta_2,\theta_3)$ converges to its true counterpart.
Next, recall from Corollary \ref{cor:unconstrainedquadratic}
the optimal feedback distribution control is a Gaussian distribution with parameter $(\frac{K}{\varepsilon} e^{-r(T-t)}-x) \pi^{Merton}$ and  $\frac{m}{\varepsilon \sigma^2 }e^{-(2r-\rho^2)(T-t)}$.
To have a concrete comparison,
we choose $x=0.5, t=0.5$; we then run
the algorithm to 
obtain estimates for $\mu,\sigma$.
Using the estimates for $\mu,\sigma$,
in Figure \ref{fig: quadratic-theta-convergence} 
we plot the density of the true optimal control
versus the density of the learned
optimal control for various exploration rates
$m=1, m=0.1, m=0.01$.
	\begin{figure}[H] 	% <---
   \begin{subfigure}{0.32\textwidth}
       \includegraphics[width=55mm,height=7cm]{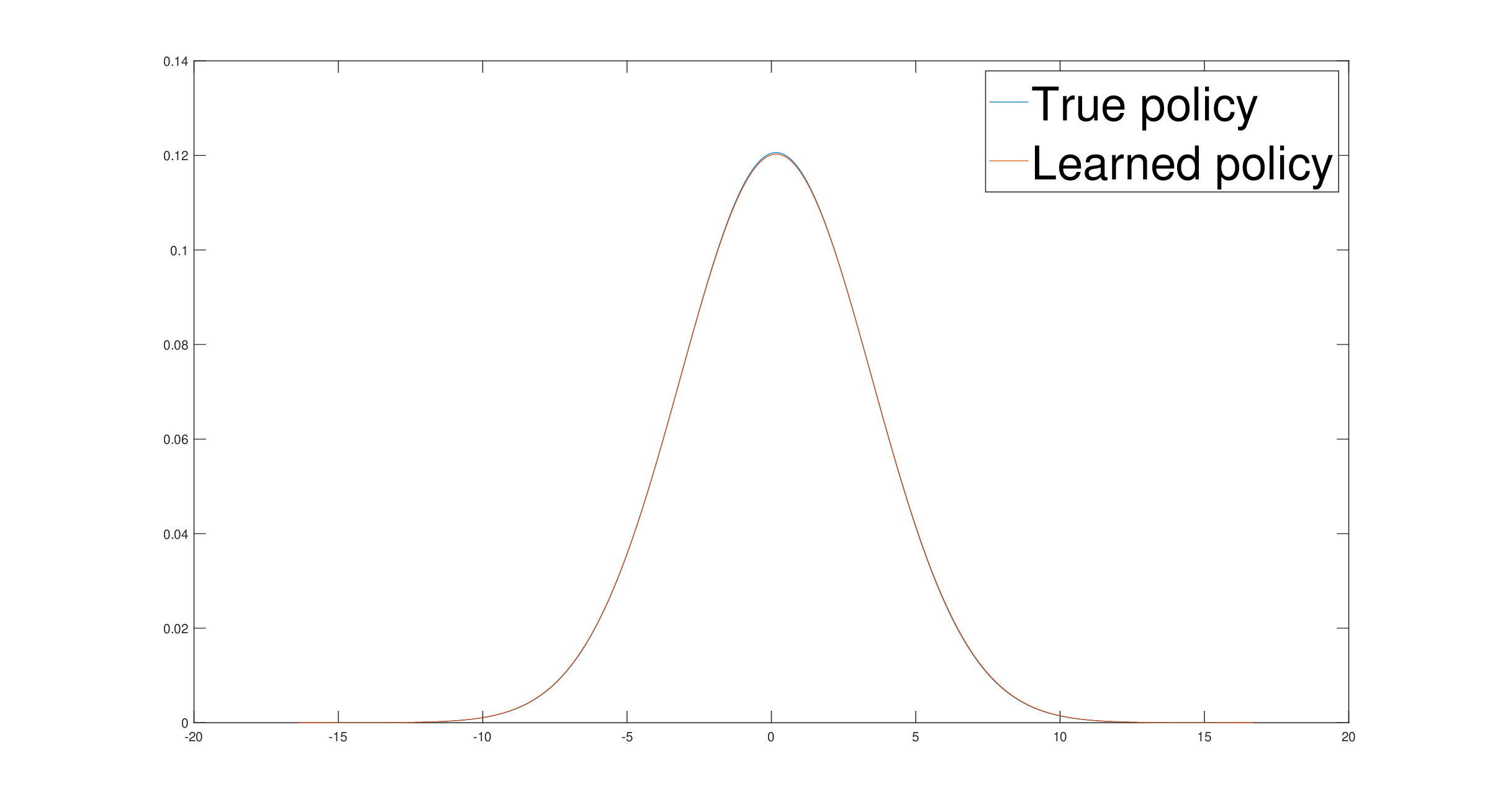}
       \caption{$m=1$}
       \label{fig:m1}
   \end{subfigure}
  \begin{subfigure}{0.32\textwidth}
       \includegraphics[width=55mm,height=7cm]{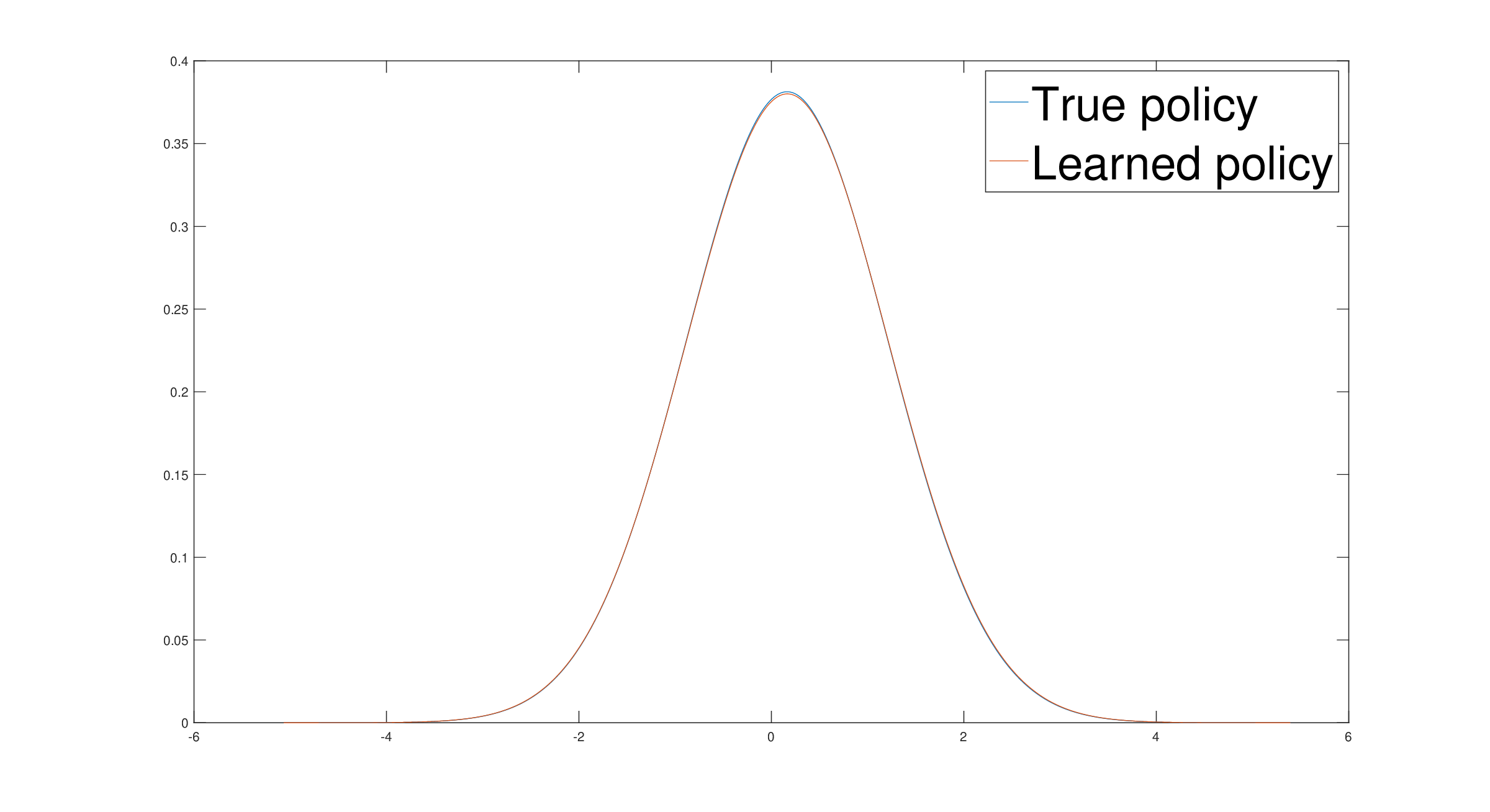}
       \caption{$m=0.1$}
       \label{fig:m2}
   \end{subfigure}
	  \begin{subfigure}{0.32\textwidth}
       \includegraphics[width=55mm,height=7cm]{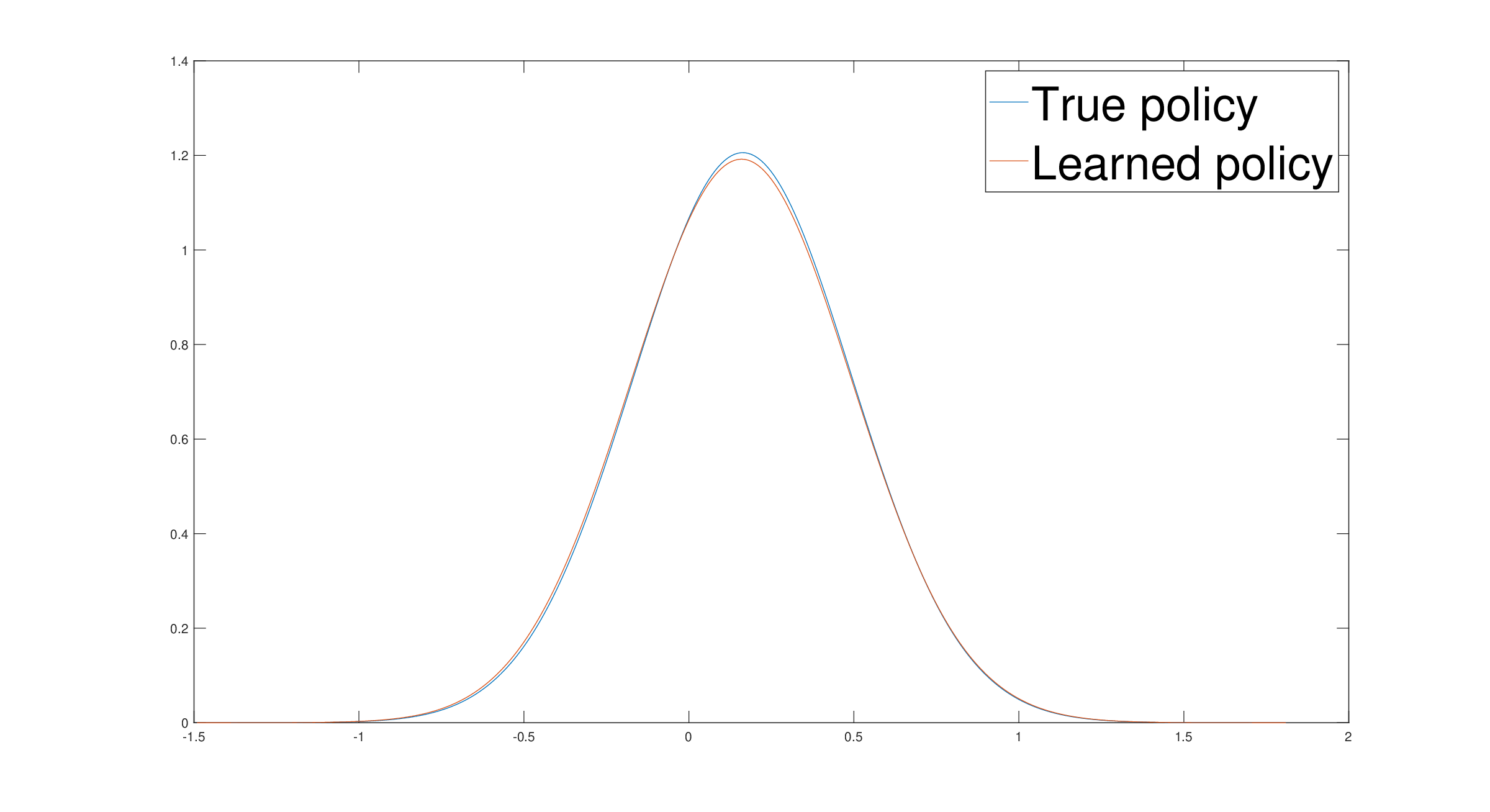}
       \caption{$m=0.01$}
       \label{fig:m3}
			  \end{subfigure}
			   \caption{Quadratic utility: True policy versus learned policy}
				\label{fig: quadratic-theta-convergence}
 \end{figure}			
 It is clear from Figure \ref{fig: quadratic-theta-convergence}
that the learned policies are matched
extremely well
to the true policies for different exploration rates.
This once again confirms
the theoretical findings presented in this section. We remark that compared to the existing literature on exploratory MV problem studied in e.g. \cite{wang2020continuous,dai2020learning,jia2022a,jia2022policy} which has to deal with an additional Lagrange multiplier, our numerical implementation for the quadratic utility is much simpler thanks to our closed form solutions. As mentioned above, as the classical MV problem is time-inconsistent problem, the optimal value and strategy might be only significant when computing as the initial time $t=0$. In our case, the optimal investment strategy and value function can be learned at any time in $t\in[0,T$]. 
Finally, we remark that the above learning procedure can be extended to the constrained problem by slightly adopting the above learning implementation as it is done for logarithm utility. Since the paper is rather lengthy we decided not to present this implementation. 
}			

{
\section{An extension to random coefficient models}\label{Sec: factor}
This section shows that our model can be extended beyond the classical Black-Scholes model. In particular, we assume that the drift and the volatility of the risky asset $S$ are driven by a state process $y$ as follows:
\begin{equation}
dS_t=S_t( \mu(y_t) dt +\sigma(y_t) d W^S_t),
\label{eq:}
\end{equation}
and
\begin{equation}
d y_t =\mu_Y(y_t) d t+ \sigma_Y(y_t) d W_t^Y,
\label{eq:yfactor}
\end{equation}
where $W^S,W^y$ are two independent Brownian motions. We assume that $\mu_Y$, $\sigma_Y$ are globally Lipschitz and linearly bounder functions so that there exists a unique strong solution $y$ to the factor SDE \eqref{eq:yfactor}. It is worth mentioning that due to the additional randomness $y$, the market is incomplete. Such a factor model has been well studied in the literature. % {\bf \large cite Some papers}. 
Now, for each $m\in\bbr$, the exploratory optimal value function is defined by
\begin{align}
&v(t,x,y;m):= \max_{\lambda \in \mathcal{H} }  \mathbb{E} \Bigg[ U(X_T^{\lambda})-m\int_{t}^{T}\int_\bbr \lambda(\pi|t,X_t^{\lambda})\ln\lambda(\pi|t,X_t^{\lambda}) d \pi   ds \bigg|X^\lambda_t=x, y_t=y\Bigg]\notag,
\end{align}
where as before, $X^\lambda$ is the exploratory wealth process and $\mathcal{H} $ is the set of admissible distributions.
Hence, the optimal value function $v$ satisfies the following HJB equation
\begin{align}
&v_t(t,x,y;m)+\mu_Y v_y(t,x,y;m)+\frac{1}{2} \sigma_Y^2(y)v_{yy}(t,x,y;m)\notag\\
+&\sup_{\lambda}\bigg\{\int_\bbr
\bigg((r+\pi(\mu(y)-r))xv_x(t,x,y;m)+\frac{1}{2}\sigma^2 (y) x^2 \pi^2 v_{xx}(t,x,y;m)
-m\ln\lambda (\pi|t,x,y)\bigg)\lambda (\pi|t,x,y)  d \pi \bigg\},\label{eq:svHJB}
\end{align}
with terminal condition $v(T,x,y;m)=\ln(x)$. Following the same steps in Section 3, the optimal distribution $\lambda^*$ is given by
\begin{align}
\lambda^*(\pi|t,x,y;m)&:=\frac{\exp\{\frac{1}{m} 
\big((r+\pi(\mu(y)-r))xv_x(t,x,y;m)+\frac{1}{2}\sigma^2(y) x^2 \pi^2 v_{xx}(t,x,y;m)\big)\}}
{\int_\mathbb R \exp\bigg\{\frac{1}{m} \big((r+\pi(\mu(y)-r))xv_x(t,x,y;m)+\frac{1}{2}\sigma^2 (y) x^2 \pi^2 v_{xx}(t,x,y;m)\big)  d \pi\bigg\}},
\label{eq:lambdaoptSV}
\end{align}
which is Gaussian with mean $\alpha$ and variance $\beta^2$ ({assuming that $v_{xx}<0$}) defined by
\begin{equation}
\alpha=-\frac{(\mu(y)-r)xv_x}{\sigma^2(y) x^2 v_{xx}};\quad \beta^2=-\frac{m}{\sigma^2(y) x^2 v_{xx}}.
\label{eq:alphabeta}
\end{equation}
As before, using the basic properties of Gaussian laws we obtain the following PDE
\begin{align}
 v_t(t,x,y;m)&+\mu_Y(y) v_y(t,x,y;m)+\frac{1}{2} \sigma_Y^2(y)v_{yy}(t,x,y;m)+rxv_x(t,x,y;m)\notag\\
&-\frac{1}{2}\frac{(\mu(y)-r)^2v_x^2(t,x;m)}{\sigma^2 (y) v_{xx}(t,x,y;m)}+\frac{m}{2}\ln\bigg(-\frac{2\pi_e m}{\sigma^2(y) x^2 v_{xx}(t,x,y;m)}\bigg)=0,
\label{eq:SVPDE}
\end{align} 
with terminal condition $v(T,x,y;m)=\ln(x).$ Compared to the previous sections, we have to deal with a two-dimension PDE. To solve \eqref{eq:SVPDE}, we seek for an ansatz of the form $v(t,x,y;m)=\ln x+ f(t,y;m)$, where $f$ is a smooth function that depends only on $t$ and $y$. For this ansatz we obtain the following PDE for $f$
\begin{align}
 f_t(t,y;m)&+\mu_Y (y)f_y(t,y;m)+\frac{1}{2} \sigma_Y^2(y)f_{yy}(t,y;m)+ h(y)=0, \quad f(T,y;m)=0\label{eq:sVHJBnocsimp}
\end{align}
where $h(y):=\bigg(r+\frac{1}{2}\frac{(\mu(y)-r)^2}{\sigma^2 (y) }\bigg)+\frac{m}{2}\ln\bigg(\frac{2\pi_e m}{\sigma^2(y) }\bigg)$. 
\begin{thm}\label{Thm:SVthm}
Assume that $\mu_Y$ and $\sigma_Y$ are bounded differentiable functions with bounded derivatives and 
\begin{equation}
\sup_{y\in \bbr} \max\{\sigma_Y(y),\sigma^{-1}_Y(y), \sigma^{-1}(y)\}<\infty\quad \mbox{and} \quad \sup_{y\in\bbr}\bigg| \frac{(\mu(y)-r)^2}{\sigma^2 (y) }\bigg|<\infty.
\label{eq:}
\end{equation}
Then, the $v(t,x,y;m)= \ln x+ f(t,y;m)$, where $f$ solves the PDE \eqref{eq:sVHJBnocsimp}, is the optimal value function of the exploratory problem. Moreover, the optimal distribution control $\lambda^*$ is Gaussian with mean $\frac{(\mu(y)-r) }{\sigma^2(y)}$ and variance $\frac{m}{\sigma^2(y) }$
%\end{equation}
and the exploratory wealth process is given by 
\begin{align} 
d X_t^{\lambda^*} = X_t^{\lambda^*} \bigg(r+\frac{(\mu (y_t)-r)^2 }{\sigma^2(y_t)}\bigg) dt + \sqrt{\bigg(m +\frac{(\mu(y_t)-r)^2 }{\sigma^2(y_t)}\bigg)} X_t^{\lambda^*}  d W_t^S \, , \quad X_0^{\lambda^*}= x_0 .
\end{align}
\end{thm}
\proof 
Changing variable $g(t,y;m):=f(T-t,y;m),$ we can rewrite the PDE \eqref{eq:sVHJBnocsimp} as
\begin{align}
 -g_t(t,y;m)&+\mu_Y (y)g_y(t,y;m)+\frac{1}{2} \sigma_Y^2(y)g_{yy}(t,y;m)= h(y), \quad g(0,y;m)=0.\label{eq:sVHJBnocsimp2}
\end{align}
By assumption,  $h$ is continuous bounded and the Cauchy problem \eqref{eq:sVHJBnocsimp2}  satisfies all conditions in Theorems 12 and 17 (Chapter 1) in \cite{friedman2008partial}. Therefore, there exists unique solution $g(t,y;m)$ in  $[0,T]\times \bbr$. It is now straightforward to see that $\ln(x)+ f(t,y;m)$ with $f(t,y;m)=g(T-t,y;m)$ satisfies \eqref{eq:SVPDE}. Note that by Feynman-Kac's theorem, the function $f(t,y;m)$ can be represented as
\begin{equation}
f(t,y;m)=\E^Q \bigg[\int_t^T h(\wt{y}_s) d s\bigg| \wt{y}_t=y\bigg],
\label{eq:}
\end{equation}
where \begin{equation}
d \wt{y}_t =\mu_Y(y_t) d t+\sigma_Y(\wt{y}_t) d W^Q_t,
\label{eq:}
\end{equation} and $Q$ is a probability measure and $W^Q$ is a Brownian motion under $Q$.
By following the same steps as in the proof of Theorem \ref{Thm:logx} we can show that \eqref{eq:lambdaoptSV} is admissible and hence is an optimal policy. The corresponding optimal exploratory wealth process is given by
\begin{align} 
d X_t^{\lambda*} = X_t^{\lambda^*} \bigg(r+\frac{(\mu (y_t)-r)^2 }{\sigma^2(y_t)}\bigg) dt + \sqrt{\bigg(m +\frac{(\mu(y_t)-r)^2 }{\sigma^2(y_t)}\bigg)} X_t^{\lambda^*}  d W_t^S \, , \quad X_0 = x_0 .
\end{align} 
\endproof
\begin{rem}
\begin{itemize}
	\item It is obvious that $f(t,y;m)=(r+\frac{1}{2}\frac{(\mu-r)^2}{\sigma^2}) (T-t)+ \frac{m}{2}\ln(\sigma^{-2}{2\pi_e m}) (T-t)$ when $\sigma_Y=\mu_Y=0$. In this case, we are back to the classical Black-Scholes model in Section 3.
	\item Theorem \ref{Thm:SVthm} can be extended to the case where the factor process $y$ is correlated with the risky asset $S$. However, the exploratory setting should be adjusted to appropriately capture the correlation, see e.g. \cite{dai2023learning}.
	\item An extension of Theorem \ref{Thm:SVthm} to the case where the strategy is bounded in a given interval $[a,b]$ can be done by adapting the discussion in Section \ref{se:constrainedLog}.  
\end{itemize}
\end{rem}
}	
\section{Conclusion}\label{Conclusion}

We study an exploration version of the continuous-time expected utility maximization problem with reinforcement learning. We show that the optimal feedback policy of the exploratory problem with exploration is Gaussian. However, when the risky investment ratio is restricted on a given portfolio interval, the constrained optimal exploration policy now follows a truncated Gaussian distribution. For logarithm {and quadratic} utility functions, the solution to the exploratory problem can be obtained in closed form  which converges to the classical expected utility counterpart when the exploration weight goes to zero. For interpretable RL algorithms, a policy improvement theorem  is provided. Finally, we devise an implementable reinforcement learning algorithm by casting the optimal problem in a martingale framework. Our work can be extended to various directions. For example, it would be interesting to consider both consumption and investment with a general utility functions in more general market settings. We foresee that the $q$-learning framework developed in  \cite{jia2022q}
might be useful in such a setting.
It would be also interesting to extend
the work to a higher dimension case.
We leave these exciting research problems for future studies.\\[1cm]

{\raggedleft{\textbf{\large{Acknowledgements}}}}\\

{Thai Nguyen acknowledges the support of the Natural Sciences and Engineering Research Council of Canada [RGPIN-2021-02594]. %We would like to thank the participants of the 26th International Congress on Insurance: Mathematics and Economics for fruitful discussions and comments that helped us improve earlier versions of this manuscript. 
}
\bibliographystyle{apa}%
%{elsarticle-num-names} 
\bibliography{RL}

\begin{thebibliography}{61}
\expandafter\ifx\csname natexlab\endcsname\relax\def\natexlab#1{#1}\fi
\providecommand{\url}[1]{\texttt{#1}}
\providecommand{\href}[2]{#2}
\providecommand{\path}[1]{#1}
\providecommand{\DOIprefix}{doi:}
\providecommand{\ArXivprefix}{arXiv:}
\providecommand{\URLprefix}{URL: }
\providecommand{\Pubmedprefix}{pmid:}
\providecommand{\doi}[1]{\href{http://dx.doi.org/#1}{\path{#1}}}
\providecommand{\Pubmed}[1]{\href{pmid:#1}{\path{#1}}}
\providecommand{\bibinfo}[2]{#2}
\ifx\xfnm\relax \def\xfnm[#1]{\unskip,\space#1}\fi
%Type = Article
\bibitem[{Silver et~al.(2016)Silver, Huang, Maddison, Guez, Sifre, Van
  Den~Driessche, Schrittwieser, Antonoglou, Panneershelvam, Lanctot
  et~al.}]{silver2016mastering}
\bibinfo{author}{D.~Silver}, \bibinfo{author}{A.~Huang}, \bibinfo{author}{C.~J.
  Maddison}, \bibinfo{author}{A.~Guez}, \bibinfo{author}{L.~Sifre},
  \bibinfo{author}{G.~Van Den~Driessche}, \bibinfo{author}{J.~Schrittwieser},
  \bibinfo{author}{I.~Antonoglou}, \bibinfo{author}{V.~Panneershelvam},
  \bibinfo{author}{M.~Lanctot}, et~al.,
\newblock \bibinfo{title}{Mastering the game of go with deep neural networks
  and tree search},
\newblock \bibinfo{journal}{nature} \bibinfo{volume}{529}
  (\bibinfo{year}{2016}) \bibinfo{pages}{484--489}.
%Type = Article
\bibitem[{Silver et~al.(2017)Silver, Schrittwieser, Simonyan, Antonoglou,
  Huang, Guez, Hubert, Baker, Lai, Bolton et~al.}]{silver2017mastering}
\bibinfo{author}{D.~Silver}, \bibinfo{author}{J.~Schrittwieser},
  \bibinfo{author}{K.~Simonyan}, \bibinfo{author}{I.~Antonoglou},
  \bibinfo{author}{A.~Huang}, \bibinfo{author}{A.~Guez},
  \bibinfo{author}{T.~Hubert}, \bibinfo{author}{L.~Baker},
  \bibinfo{author}{M.~Lai}, \bibinfo{author}{A.~Bolton}, et~al.,
\newblock \bibinfo{title}{Mastering the game of go without human knowledge},
\newblock \bibinfo{journal}{nature} \bibinfo{volume}{550}
  (\bibinfo{year}{2017}) \bibinfo{pages}{354--359}.
%Type = Book
\bibitem[{Bertsekas(2019)}]{bertsekas2019reinforcement}
\bibinfo{author}{D.~Bertsekas}, \bibinfo{title}{Reinforcement learning and
  optimal control}, \bibinfo{publisher}{Athena Scientific},
  \bibinfo{year}{2019}.
%Type = Inproceedings
\bibitem[{Williams et~al.(2017)Williams, Wagener, Goldfain, Drews, Rehg, Boots,
  and Theodorou}]{williams2017information}
\bibinfo{author}{G.~Williams}, \bibinfo{author}{N.~Wagener},
  \bibinfo{author}{B.~Goldfain}, \bibinfo{author}{P.~Drews},
  \bibinfo{author}{J.~M. Rehg}, \bibinfo{author}{B.~Boots},
  \bibinfo{author}{E.~A. Theodorou},
\newblock \bibinfo{title}{Information theoretic mpc for model-based
  reinforcement learning},
\newblock in: \bibinfo{booktitle}{2017 IEEE International Conference on
  Robotics and Automation (ICRA)}, \bibinfo{organization}{IEEE},
  \bibinfo{year}{2017}, pp. \bibinfo{pages}{1714--1721}.
%Type = Article
\bibitem[{Kaelbling et~al.(1996)Kaelbling, Littman, and
  Moore}]{kaelbling1996reinforcement}
\bibinfo{author}{L.~P. Kaelbling}, \bibinfo{author}{M.~L. Littman},
  \bibinfo{author}{A.~W. Moore},
\newblock \bibinfo{title}{Reinforcement learning: A survey},
\newblock \bibinfo{journal}{Journal of artificial intelligence research}
  \bibinfo{volume}{4} (\bibinfo{year}{1996}) \bibinfo{pages}{237--285}.
%Type = Inproceedings
\bibitem[{Schneckenreither and
  Haeussler(2019)}]{schneckenreither2019reinforcement}
\bibinfo{author}{M.~Schneckenreither}, \bibinfo{author}{S.~Haeussler},
\newblock \bibinfo{title}{Reinforcement learning methods for operations
  research applications: The order release problem},
\newblock in: \bibinfo{booktitle}{Machine Learning, Optimization, and Data
  Science: 4th International Conference, LOD 2018, Volterra, Italy, September
  13-16, 2018, Revised Selected Papers 4}, \bibinfo{organization}{Springer},
  \bibinfo{year}{2019}, pp. \bibinfo{pages}{545--559}.
%Type = Article
\bibitem[{Bertsimas and Thiele(2006)}]{bertsimas2006robust}
\bibinfo{author}{D.~Bertsimas}, \bibinfo{author}{A.~Thiele},
\newblock \bibinfo{title}{A robust optimization approach to inventory theory},
\newblock \bibinfo{journal}{Operations research} \bibinfo{volume}{54}
  (\bibinfo{year}{2006}) \bibinfo{pages}{150--168}.
%Type = Inproceedings
\bibitem[{Nevmyvaka et~al.(2006)Nevmyvaka, Feng, and
  Kearns}]{nevmyvaka2006reinforcement}
\bibinfo{author}{Y.~Nevmyvaka}, \bibinfo{author}{Y.~Feng},
  \bibinfo{author}{M.~Kearns},
\newblock \bibinfo{title}{Reinforcement learning for optimized trade
  execution},
\newblock in: \bibinfo{booktitle}{Proceedings of the 23rd international
  conference on Machine learning}, \bibinfo{year}{2006}, pp.
  \bibinfo{pages}{673--680}.
%Type = Article
\bibitem[{Schnaubelt(2022)}]{schnaubelt2022deep}
\bibinfo{author}{M.~Schnaubelt},
\newblock \bibinfo{title}{Deep reinforcement learning for the optimal placement
  of cryptocurrency limit orders},
\newblock \bibinfo{journal}{European Journal of Operational Research}
  \bibinfo{volume}{296} (\bibinfo{year}{2022}) \bibinfo{pages}{993--1006}.
%Type = Inproceedings
\bibitem[{Hendricks and Wilcox(2014)}]{hendricks2014reinforcement}
\bibinfo{author}{D.~Hendricks}, \bibinfo{author}{D.~Wilcox},
\newblock \bibinfo{title}{A reinforcement learning extension to the
  almgren-chriss framework for optimal trade execution},
\newblock in: \bibinfo{booktitle}{2014 IEEE Conference on Computational
  Intelligence for Financial Engineering \& Economics (CIFEr)},
  \bibinfo{organization}{IEEE}, \bibinfo{year}{2014}, pp.
  \bibinfo{pages}{457--464}.
%Type = Article
\bibitem[{Moody et~al.(1998)Moody, Wu, Liao, and
  Saffell}]{moody1998performance}
\bibinfo{author}{J.~Moody}, \bibinfo{author}{L.~Wu}, \bibinfo{author}{Y.~Liao},
  \bibinfo{author}{M.~Saffell},
\newblock \bibinfo{title}{Performance functions and reinforcement learning for
  trading systems and portfolios},
\newblock \bibinfo{journal}{Journal of Forecasting} \bibinfo{volume}{17}
  (\bibinfo{year}{1998}) \bibinfo{pages}{441--470}.
%Type = Article
\bibitem[{Wang and Zhou(2020)}]{wang2020continuous}
\bibinfo{author}{H.~Wang}, \bibinfo{author}{X.~Y. Zhou},
\newblock \bibinfo{title}{Continuous-time mean--variance portfolio selection: A
  reinforcement learning framework},
\newblock \bibinfo{journal}{Mathematical Finance} \bibinfo{volume}{30}
  (\bibinfo{year}{2020}) \bibinfo{pages}{1273--1308}.
%Type = Article
\bibitem[{Wang(2019)}]{wang2019large}
\bibinfo{author}{H.~Wang},
\newblock \bibinfo{title}{Large scale continuous-time mean-variance portfolio
  allocation via reinforcement learning},
\newblock \bibinfo{journal}{arXiv preprint arXiv:1907.11718}
  (\bibinfo{year}{2019}).
%Type = Article
\bibitem[{Dai et~al.(2020)Dai, Dong, and Jia}]{dai2020learning}
\bibinfo{author}{M.~Dai}, \bibinfo{author}{Y.~Dong}, \bibinfo{author}{Y.~Jia},
\newblock \bibinfo{title}{Learning equilibrium mean-variance strategy},
\newblock \bibinfo{journal}{Available at SSRN 3770818}  (\bibinfo{year}{2020}).
%Type = Article
\bibitem[{Jia and Zhou(2022)}]{jia2022policy}
\bibinfo{author}{Y.~Jia}, \bibinfo{author}{X.~Y. Zhou},
\newblock \bibinfo{title}{Policy gradient and actor-critic learning in
  continuous time and space: Theory and algorithms},
\newblock \bibinfo{journal}{Journal of Machine Learning Research}
  \bibinfo{volume}{23} (\bibinfo{year}{2022}) \bibinfo{pages}{1--55}.
%Type = Article
\bibitem[{Gosavi(2009)}]{gosavi2009reinforcement}
\bibinfo{author}{A.~Gosavi},
\newblock \bibinfo{title}{Reinforcement learning: A tutorial survey and recent
  advances},
\newblock \bibinfo{journal}{INFORMS Journal on Computing} \bibinfo{volume}{21}
  (\bibinfo{year}{2009}) \bibinfo{pages}{178--192}.
%Type = Article
\bibitem[{Hambly et~al.(2021)Hambly, Xu, and Yang}]{hambly2021recent}
\bibinfo{author}{B.~Hambly}, \bibinfo{author}{R.~Xu},
  \bibinfo{author}{H.~Yang},
\newblock \bibinfo{title}{Recent advances in reinforcement learning in
  finance},
\newblock \bibinfo{journal}{arXiv preprint arXiv:2112.04553}
  (\bibinfo{year}{2021}).
%Type = Article
\bibitem[{Jaimungal(2022)}]{jaimungal2022reinforcement}
\bibinfo{author}{S.~Jaimungal},
\newblock \bibinfo{title}{Reinforcement learning and stochastic optimisation},
\newblock \bibinfo{journal}{Finance and Stochastics} \bibinfo{volume}{26}
  (\bibinfo{year}{2022}) \bibinfo{pages}{103--129}.
%Type = Article
\bibitem[{Charpentier et~al.(2021)Charpentier, Elie, and
  Remlinger}]{charpentier2021reinforcement}
\bibinfo{author}{A.~Charpentier}, \bibinfo{author}{R.~Elie},
  \bibinfo{author}{C.~Remlinger},
\newblock \bibinfo{title}{Reinforcement learning in economics and finance},
\newblock \bibinfo{journal}{Computational Economics}  (\bibinfo{year}{2021})
  \bibinfo{pages}{1--38}.
%Type = Article
\bibitem[{Doya(2000)}]{doya2000reinforcement}
\bibinfo{author}{K.~Doya},
\newblock \bibinfo{title}{Reinforcement learning in continuous time and space},
\newblock \bibinfo{journal}{Neural computation} \bibinfo{volume}{12}
  (\bibinfo{year}{2000}) \bibinfo{pages}{219--245}.
%Type = Article
\bibitem[{Fr{\'e}maux et~al.(2013)Fr{\'e}maux, Sprekeler, and
  Gerstner}]{fremaux2013reinforcement}
\bibinfo{author}{N.~Fr{\'e}maux}, \bibinfo{author}{H.~Sprekeler},
  \bibinfo{author}{W.~Gerstner},
\newblock \bibinfo{title}{Reinforcement learning using a continuous time
  actor-critic framework with spiking neurons},
\newblock \bibinfo{journal}{PLoS computational biology} \bibinfo{volume}{9}
  (\bibinfo{year}{2013}) \bibinfo{pages}{e1003024}.
%Type = Article
\bibitem[{Lee and Sutton(2021)}]{lee2021policy}
\bibinfo{author}{J.~Lee}, \bibinfo{author}{R.~S. Sutton},
\newblock \bibinfo{title}{Policy iterations for reinforcement learning problems
  in continuous time and space—fundamental theory and methods},
\newblock \bibinfo{journal}{Automatica} \bibinfo{volume}{126}
  (\bibinfo{year}{2021}) \bibinfo{pages}{109421}.
%Type = Book
\bibitem[{Sutton and Barto(2018)}]{sutton2018reinforcement}
\bibinfo{author}{R.~S. Sutton}, \bibinfo{author}{A.~G. Barto},
  \bibinfo{title}{Reinforcement learning: An introduction},
  \bibinfo{publisher}{MIT press}, \bibinfo{year}{2018}.
%Type = Article
\bibitem[{Liu et~al.(2020)Liu, Chen, and Jiang}]{liu2020dynamic}
\bibinfo{author}{Y.~Liu}, \bibinfo{author}{Y.~Chen},
  \bibinfo{author}{T.~Jiang},
\newblock \bibinfo{title}{Dynamic selective maintenance optimization for
  multi-state systems over a finite horizon: A deep reinforcement learning
  approach},
\newblock \bibinfo{journal}{European Journal of Operational Research}
  \bibinfo{volume}{283} (\bibinfo{year}{2020}) \bibinfo{pages}{166--181}.
%Type = Article
\bibitem[{Lee and Lee(2021)}]{lee2021multi}
\bibinfo{author}{H.-R. Lee}, \bibinfo{author}{T.~Lee},
\newblock \bibinfo{title}{Multi-agent reinforcement learning algorithm to solve
  a partially-observable multi-agent problem in disaster response},
\newblock \bibinfo{journal}{European Journal of Operational Research}
  \bibinfo{volume}{291} (\bibinfo{year}{2021}) \bibinfo{pages}{296--308}.
%Type = Article
\bibitem[{Wang et~al.(2020)Wang, Zariphopoulou, and
  Zhou}]{wang2020reinforcement}
\bibinfo{author}{H.~Wang}, \bibinfo{author}{T.~Zariphopoulou},
  \bibinfo{author}{X.~Y. Zhou},
\newblock \bibinfo{title}{Reinforcement learning in continuous time and space:
  A stochastic control approach.},
\newblock \bibinfo{journal}{Journal of Machine Learning Research}
  \bibinfo{volume}{21} (\bibinfo{year}{2020}) \bibinfo{pages}{1--34}.
%Type = Article
\bibitem[{Jia and Zhou(2022)}]{jia2022a}
\bibinfo{author}{Y.~Jia}, \bibinfo{author}{X.~Y. Zhou},
\newblock \bibinfo{title}{Policy evaluation and temporal-difference learning in
  continuous time and space: A martingale approach},
\newblock \bibinfo{journal}{Journal of Machine Learning Research}
  \bibinfo{volume}{23} (\bibinfo{year}{2022}) \bibinfo{pages}{1--55}.
%Type = Article
\bibitem[{Barnard(1993)}]{barnard1993temporal}
\bibinfo{author}{E.~Barnard},
\newblock \bibinfo{title}{Temporal-difference methods and markov models},
\newblock \bibinfo{journal}{IEEE Transactions on Systems, Man, and Cybernetics}
  \bibinfo{volume}{23} (\bibinfo{year}{1993}) \bibinfo{pages}{357--365}.
%Type = Incollection
\bibitem[{Baird(1995)}]{baird1995residual}
\bibinfo{author}{L.~Baird},
\newblock \bibinfo{title}{Residual algorithms: Reinforcement learning with
  function approximation},
\newblock in: \bibinfo{booktitle}{Machine Learning Proceedings 1995},
  \bibinfo{publisher}{Elsevier}, \bibinfo{year}{1995}, pp.
  \bibinfo{pages}{30--37}.
%Type = Article
\bibitem[{Guo et~al.(2022)Guo, Hu, Xu, and Zhang}]{guo2022general}
\bibinfo{author}{X.~Guo}, \bibinfo{author}{A.~Hu}, \bibinfo{author}{R.~Xu},
  \bibinfo{author}{J.~Zhang},
\newblock \bibinfo{title}{A general framework for learning mean-field games},
\newblock \bibinfo{journal}{Mathematics of Operations Research}
  (\bibinfo{year}{2022}).
%Type = Article
\bibitem[{Mou et~al.(2021)Mou, Zhang, and Zhou}]{mou2021robust}
\bibinfo{author}{C.~Mou}, \bibinfo{author}{W.~Zhang},
  \bibinfo{author}{C.~Zhou},
\newblock \bibinfo{title}{Robust exploratory mean-variance problem with drift
  uncertainty},
\newblock \bibinfo{journal}{arXiv preprint arXiv:2108.04100}
  (\bibinfo{year}{2021}).
%Type = Article
\bibitem[{Tang et~al.(2022)Tang, Zhang, and Zhou}]{tang2022exploratorySIAM}
\bibinfo{author}{W.~Tang}, \bibinfo{author}{Y.~P. Zhang},
  \bibinfo{author}{X.~Y. Zhou},
\newblock \bibinfo{title}{Exploratory hjb equations and their convergence},
\newblock \bibinfo{journal}{SIAM Journal on Control and Optimization}
  \bibinfo{volume}{60} (\bibinfo{year}{2022}) \bibinfo{pages}{3191--3216}.
%Type = Article
\bibitem[{Huang et~al.(2022)Huang, Wang, and Zhou}]{huang2022convergence}
\bibinfo{author}{Y.-J. Huang}, \bibinfo{author}{Z.~Wang},
  \bibinfo{author}{Z.~Zhou},
\newblock \bibinfo{title}{Convergence of policy improvement for
  entropy-regularized stochastic control problems},
\newblock \bibinfo{journal}{arXiv preprint arXiv:2209.07059}
  (\bibinfo{year}{2022}).
%Type = Article
\bibitem[{Cuoco(1997)}]{cuoco1997optimal}
\bibinfo{author}{D.~Cuoco},
\newblock \bibinfo{title}{Optimal consumption and equilibrium prices with
  portfolio constraints and stochastic income},
\newblock \bibinfo{journal}{Journal of Economic Theory} \bibinfo{volume}{72}
  (\bibinfo{year}{1997}) \bibinfo{pages}{33--73}.
%Type = Article
\bibitem[{Dai et~al.(2023)Dai, Dong, and Jia}]{dai2023learning}
\bibinfo{author}{M.~Dai}, \bibinfo{author}{Y.~Dong}, \bibinfo{author}{Y.~Jia},
\newblock \bibinfo{title}{Learning equilibrium mean-variance strategy},
\newblock \bibinfo{journal}{Mathematical Finance} \bibinfo{volume}{33}
  (\bibinfo{year}{2023}) \bibinfo{pages}{1166--1212}.
%Type = Article
\bibitem[{Dai et~al.(2021)Dai, Jin, Kou, and Xu}]{dai2021dynamic}
\bibinfo{author}{M.~Dai}, \bibinfo{author}{H.~Jin}, \bibinfo{author}{S.~Kou},
  \bibinfo{author}{Y.~Xu},
\newblock \bibinfo{title}{A dynamic mean-variance analysis for log returns},
\newblock \bibinfo{journal}{Management Science} \bibinfo{volume}{67}
  (\bibinfo{year}{2021}) \bibinfo{pages}{1093--1108}.
%Type = Book
\bibitem[{Fleming and Soner(2006)}]{fleming2006controlled}
\bibinfo{author}{W.~H. Fleming}, \bibinfo{author}{H.~M. Soner},
  \bibinfo{title}{Controlled Markov processes and viscosity solutions},
  volume~\bibinfo{volume}{25}, \bibinfo{publisher}{Springer Science \& Business
  Media}, \bibinfo{year}{2006}.
%Type = Book
\bibitem[{Pham(2009)}]{pham2009continuous}
\bibinfo{author}{H.~Pham}, \bibinfo{title}{Continuous-time stochastic control
  and optimization with financial applications}, volume~\bibinfo{volume}{61},
  \bibinfo{publisher}{Springer Science \& Business Media},
  \bibinfo{year}{2009}.
%Type = Article
\bibitem[{Chen and Vellekoop(2017)}]{chen2017optimal}
\bibinfo{author}{A.~Chen}, \bibinfo{author}{M.~Vellekoop},
\newblock \bibinfo{title}{Optimal investment and consumption when allowing
  terminal debt},
\newblock \bibinfo{journal}{European Journal of Operational Research}
  \bibinfo{volume}{258} (\bibinfo{year}{2017}) \bibinfo{pages}{385--397}.
%Type = Article
\bibitem[{Kamma and Pelsser(2022)}]{kamma2022near}
\bibinfo{author}{T.~Kamma}, \bibinfo{author}{A.~Pelsser},
\newblock \bibinfo{title}{Near-optimal asset allocation in financial markets
  with trading constraints},
\newblock \bibinfo{journal}{European Journal of Operational Research}
  \bibinfo{volume}{297} (\bibinfo{year}{2022}) \bibinfo{pages}{766--781}.
%Type = Book
\bibitem[{Karatzas and Shreve(1998)}]{karatzas1998methods}
\bibinfo{author}{I.~Karatzas}, \bibinfo{author}{S.~E. Shreve},
  \bibinfo{title}{Methods of mathematical finance},
  volume~\bibinfo{volume}{39}, \bibinfo{publisher}{Springer},
  \bibinfo{year}{1998}.
%Type = Book
\bibitem[{Luenberger(1998)}]{luenberger1997investment}
\bibinfo{author}{D.~G. Luenberger}, \bibinfo{title}{Investment science},
  \bibinfo{publisher}{Oxford university press}, \bibinfo{year}{1998}.
%Type = Book
\bibitem[{Fleming(1976)}]{fleming1976generalized}
\bibinfo{author}{W.~H. Fleming}, \bibinfo{title}{Generalized solutions in
  optimal stochastic control}, \bibinfo{publisher}{Brown Univ.},
  \bibinfo{year}{1976}.
%Type = Article
\bibitem[{Fleming and Nisio(1984)}]{fleming1984stochastic}
\bibinfo{author}{W.~H. Fleming}, \bibinfo{author}{M.~Nisio},
\newblock \bibinfo{title}{On stochastic relaxed control for partially observed
  diffusions},
\newblock \bibinfo{journal}{Nagoya Mathematical Journal} \bibinfo{volume}{93}
  (\bibinfo{year}{1984}) \bibinfo{pages}{71--108}.
%Type = Article
\bibitem[{Nicole~el et~al.(1987)Nicole~el, Du~Huu, and
  Monique}]{nicole1987compactification}
\bibinfo{author}{K.~Nicole~el}, \bibinfo{author}{N.~Du~Huu},
  \bibinfo{author}{J.-P. Monique},
\newblock \bibinfo{title}{Compactification methods in the control of degenerate
  diffusions: existence of an optimal control},
\newblock \bibinfo{journal}{Stochastics: an international journal of
  probability and stochastic processes} \bibinfo{volume}{20}
  (\bibinfo{year}{1987}) \bibinfo{pages}{169--219}.
%Type = Inproceedings
\bibitem[{Donsker and Varadhan(2006)}]{donsker2006large}
\bibinfo{author}{M.~Donsker}, \bibinfo{author}{S.~Varadhan},
\newblock \bibinfo{title}{Large deviations for markov processes and the
  asymptotic evaluation of certain markov process expectations for large
  times},
\newblock in: \bibinfo{booktitle}{Probabilistic Methods in Differential
  Equations: Proceedings of the Conference Held at the University of Victoria,
  August 19--20, 1974}, \bibinfo{organization}{Springer}, \bibinfo{year}{2006},
  pp. \bibinfo{pages}{82--88}.
%Type = Incollection
\bibitem[{Rubinstein(1977)}]{rubinstein1977strong}
\bibinfo{author}{M.~Rubinstein},
\newblock \bibinfo{title}{The strong case for the generalized logarithmic
  utility model as the premier model of financial markets},
\newblock in: \bibinfo{booktitle}{Financial Dec Making Under Uncertainty},
  \bibinfo{publisher}{Elsevier}, \bibinfo{year}{1977}, pp.
  \bibinfo{pages}{11--62}.
%Type = Article
\bibitem[{Pulley(1983)}]{pulley1983mean}
\bibinfo{author}{L.~B. Pulley},
\newblock \bibinfo{title}{Mean-variance approximations to expected logarithmic
  utility},
\newblock \bibinfo{journal}{Operations Research} \bibinfo{volume}{31}
  (\bibinfo{year}{1983}) \bibinfo{pages}{685--696}.
%Type = Article
\bibitem[{Gerrard et~al.(2023)Gerrard, Kyriakou, Nielsen, and
  Vodi{\v{c}}ka}]{gerrard2023optimal}
\bibinfo{author}{R.~Gerrard}, \bibinfo{author}{I.~Kyriakou},
  \bibinfo{author}{J.~P. Nielsen}, \bibinfo{author}{P.~Vodi{\v{c}}ka},
\newblock \bibinfo{title}{On optimal constrained investment strategies for
  long-term savers in stochastic environments and probability hedging},
\newblock \bibinfo{journal}{European Journal of Operational Research}
  \bibinfo{volume}{307} (\bibinfo{year}{2023}) \bibinfo{pages}{948--962}.
%Type = Incollection
\bibitem[{Merton(1975)}]{merton1975optimum}
\bibinfo{author}{R.~C. Merton},
\newblock \bibinfo{title}{Optimum consumption and portfolio rules in a
  continuous-time model},
\newblock in: \bibinfo{booktitle}{Stochastic optimization models in finance},
  \bibinfo{publisher}{Elsevier}, \bibinfo{year}{1975}, pp.
  \bibinfo{pages}{621--661}.
%Type = Article
\bibitem[{El~Karoui and Jeanblanc-Picqu{\'e}(1998)}]{el1998optimization}
\bibinfo{author}{N.~El~Karoui}, \bibinfo{author}{M.~Jeanblanc-Picqu{\'e}},
\newblock \bibinfo{title}{Optimization of consumption with labor income},
\newblock \bibinfo{journal}{Finance and Stochastics} \bibinfo{volume}{2}
  (\bibinfo{year}{1998}) \bibinfo{pages}{409--440}.
%Type = Book
\bibitem[{Friedman(2008)}]{friedman2008partial}
\bibinfo{author}{A.~Friedman}, \bibinfo{title}{Partial differential equations
  of parabolic type}, \bibinfo{publisher}{Courier Dover Publications},
  \bibinfo{year}{2008}.
%Type = Book
\bibitem[{Kotz et~al.(2004)Kotz, Balakrishnan, and
  Johnson}]{kotz2004continuous}
\bibinfo{author}{S.~Kotz}, \bibinfo{author}{N.~Balakrishnan},
  \bibinfo{author}{N.~L. Johnson}, \bibinfo{title}{Continuous multivariate
  distributions, Volume 1: Models and applications},
  volume~\bibinfo{volume}{1}, \bibinfo{publisher}{John Wiley \& Sons},
  \bibinfo{year}{2004}.
%Type = Article
\bibitem[{Csisz{\'a}r(1975)}]{csiszar1975divergence}
\bibinfo{author}{I.~Csisz{\'a}r},
\newblock \bibinfo{title}{I-divergence geometry of probability distributions
  and minimization problems},
\newblock \bibinfo{journal}{The annals of probability}  (\bibinfo{year}{1975})
  \bibinfo{pages}{146--158}.
%Type = Article
\bibitem[{Duffie and Richardson(1991)}]{duffie1991mean}
\bibinfo{author}{D.~Duffie}, \bibinfo{author}{H.~R. Richardson},
\newblock \bibinfo{title}{Mean-variance hedging in continuous time},
\newblock \bibinfo{journal}{The Annals of Applied Probability}
  (\bibinfo{year}{1991}) \bibinfo{pages}{1--15}.
%Type = Article
\bibitem[{Bodnar et~al.(2013)Bodnar, Parolya, and
  Schmid}]{bodnar2013equivalence}
\bibinfo{author}{T.~Bodnar}, \bibinfo{author}{N.~Parolya},
  \bibinfo{author}{W.~Schmid},
\newblock \bibinfo{title}{On the equivalence of quadratic optimization problems
  commonly used in portfolio theory},
\newblock \bibinfo{journal}{European Journal of Operational Research}
  \bibinfo{volume}{229} (\bibinfo{year}{2013}) \bibinfo{pages}{637--644}.
%Type = Article
\bibitem[{Li et~al.(2002)Li, Zhou, and Lim}]{li2002dynamic}
\bibinfo{author}{X.~Li}, \bibinfo{author}{X.~Y. Zhou}, \bibinfo{author}{A.~E.
  Lim},
\newblock \bibinfo{title}{Dynamic mean-variance portfolio selection with
  no-shorting constraints},
\newblock \bibinfo{journal}{SIAM Journal on Control and Optimization}
  \bibinfo{volume}{40} (\bibinfo{year}{2002}) \bibinfo{pages}{1540--1555}.
%Type = Article
\bibitem[{Bielecki et~al.(2005)Bielecki, Jin, Pliska, and
  Zhou}]{bielecki2005continuous}
\bibinfo{author}{T.~R. Bielecki}, \bibinfo{author}{H.~Jin},
  \bibinfo{author}{S.~R. Pliska}, \bibinfo{author}{X.~Y. Zhou},
\newblock \bibinfo{title}{Continuous-time mean-variance portfolio selection
  with bankruptcy prohibition},
\newblock \bibinfo{journal}{Mathematical Finance: An International Journal of
  Mathematics, Statistics and Financial Economics} \bibinfo{volume}{15}
  (\bibinfo{year}{2005}) \bibinfo{pages}{213--244}.
%Type = Article
\bibitem[{Li and Xu(2016)}]{li2016continuous}
\bibinfo{author}{X.~Li}, \bibinfo{author}{Z.~Q. Xu},
\newblock \bibinfo{title}{Continuous-time markowitz’s model with constraints
  on wealth and portfolio},
\newblock \bibinfo{journal}{Operations research letters} \bibinfo{volume}{44}
  (\bibinfo{year}{2016}) \bibinfo{pages}{729--736}.
%Type = Article
\bibitem[{Kato et~al.(2020)Kato, Nakagawa, Abe, and Morimura}]{kato2020mean}
\bibinfo{author}{M.~Kato}, \bibinfo{author}{K.~Nakagawa},
  \bibinfo{author}{K.~Abe}, \bibinfo{author}{T.~Morimura},
\newblock \bibinfo{title}{Mean-variance efficient reinforcement learning by
  expected quadratic utility maximization},
\newblock \bibinfo{journal}{arXiv preprint arXiv:2010.01404}
  (\bibinfo{year}{2020}).
%Type = Article
\bibitem[{Jia and Zhou(2023)}]{jia2022q}
\bibinfo{author}{Y.~Jia}, \bibinfo{author}{X.~Y. Zhou},
\newblock \bibinfo{title}{{Q}-learning in continuous time},
\newblock \bibinfo{journal}{Journal of Machine Learning Research}
  (\bibinfo{year}{2023}).

\end{thebibliography}
{
\appendix
\section{Technical proofs of Section \ref{se:Log}}
\subsection{Proof of Theorem \ref{Thm:logx}}
%\begin{thm}\label{Thm:logx}
 %The optimal value function of the entropy-regularized exploratory optimal investment problem with logarithmic utility $U(x)=\ln x$ is given by
%\begin{align}
%v(t,x;m)=&\ln x+
 %\bigg(r+\frac{1}{2}\frac{(\mu-r)^2}{\sigma^2} \bigg)(T-t)+\frac{m}{2} \ln(\sigma^{-2}{2\pi_e m})(T-t)
%\label{eq:valueopt}
%\end{align}
%for $(t,x)\in[0,T]\times \bbr_+$. Moreover, the optimal feedback distribution control $\lambda^*(\pi):=\lambda^*(\pi; x,m)$ {(which is independent of $t$)} is a Gaussian with mean $\frac{(\mu-r) }{\sigma^2}$ and variance $\frac{m}{\sigma^2 }$, i.e.
%\begin{equation}
%\lambda^*(\pi)=\mathcal N\left(\pi\bigg|\frac{(\mu-r) }{\sigma^2}, \frac{m}{\sigma^2 }\right)
%\label{eq:lambdaopt}
%\end{equation}
%and the associated optimal wealth under $\lambda^*$ is given by the following SDE
%\begin{align} 
%d X_t^{\lambda*} = X_t^{\lambda^*} \bigg(r+\frac{(\mu-r)^2 }{\sigma^2}\bigg) dt + \sqrt{\bigg(m +\frac{(\mu-r)^2 }{\sigma^2}\bigg)} X_t^{\lambda^*}  d W_t \, , \quad X_0 = x_0 \,.\label{eq:Wealthoptlamb}
%\end{align}
%\end{thm}
\proof Recall that $U(x)=\ln (x)$. We find the solution of the HJB \eqref{eq:HJBnocsimp} under the form $v(t,x;m)=k(t) \ln x+l(t)$, where $k, l:[0,T]\to \bbr$ are smooth functions satisfying the terminal condition $k(T)=1$, $l(T)=0$. 
Direct calculation shows that \eqref{eq:HJBnocsimp} boils down to 
\begin{align}
k'(t)\ln x +\bigg(r+\frac{1}{2}\frac{(\mu-r)^2}{\sigma^2}\bigg) k(t)-\frac{m}{2}\ln k(t)+l'(t)+c_m=0, \forall (t,x)\in [0,T)\times \bbr_+,
\end{align}
where $c_m:=\frac{m}{2} \ln(\sigma^{-2}{2\pi_e m})$ and $\pi_e=3.14...$ is the Archimedes' constant.  
It follows that $k'(t) =0$ and $$(r+\frac{1}{2}\frac{(\mu-r)^2}{\sigma^2}) k(t)-\frac{m}{2}\ln k(t)+l'(t)+c_m=0.$$ Hence, $k(t)=1$ and
$$
l(t)=- (r+\frac{1}{2}\frac{(\mu-r)^2}{\sigma^2})t-c_m t+C,
$$
where the constant $C$ is chosen such that the terminal condition $l(T)=0$ is fulfilled. Direction calculation leads to 
$$
l(t)= (r+\frac{1}{2}\frac{(\mu-r)^2}{\sigma^2})(T-t)+c_m (T-t)$$
and the solution of PDE \eqref{eq:HJBnocsimp} is given by \eqref{eq:valueopt}.
%\begin{align}
%v(t,x;m)=& \ln x+
 %(r+\frac{1}{2}\frac{(\mu-r)^2}{\sigma^2})(T-t)+c_m (T-t).
%\label{eq:}
%\end{align}
%\red{Thai: Can we compute $\mathbb E[X_t^{\lambda*}]$ and $Var(X_t^{\lambda*}$?.
%We will report these two quantities in the table. If not computed in closed form, we have to simulate it instead???}
Let us first show that $v(t,x;m)$ in \eqref{eq:valueopt} is the optimal value function. Indeed, it follows that the optimal distribution $\lambda^*$ is now given by a Gaussian distribution with mean $\alpha=\frac{(\mu-r) }{\sigma^2}$ (independent of the exploration parameter $m$) and variance 
$\beta^2=\frac{m}{\sigma^2}. $
The law of optimal feedback Gaussian control allows to determine the exploration wealth drift and volatility from \eqref{eq:hatAB} as
\begin{align*}
\hat{A}{(t, x;\lambda^*)}=x\bigg(r+\frac{(\mu-r)^2 }{\sigma^2}\bigg); \quad
\hat{B}^2{(t, x;\lambda^*)}=x^2 \bigg(m+\frac{(\mu-r)^2 }{\sigma^2}\bigg).
\end{align*}
Hence, the exploration wealth dynamics is given by \eqref{eq:Wealthoptlamb}.
%\begin{align} \label{eq_XtExplo00}
%d X_t^{\lambda^*} = X_t^{\lambda} \bigg(r+\frac{(\mu-r)^2 }{\sigma^2}\bigg) dt + \sqrt{\bigg(\frac{m}{\sigma^2 } +\frac{(\mu-r)^2 }{\sigma^4}\bigg)}\sigma  X_t^{\lambda}  d W_t \, , \quad X_0 = x_0 \,.
%\end{align}
Now, it is clear that the SDE \eqref{eq:Wealthoptlamb} admits solution given by
	\begin{align} \label{eq_XtExplo0}
X_t^{\lambda^*} = x_0 \exp\bigg\{ \bigg(r+\frac{1 }{2}\frac{(\mu-r)^2 }{\sigma^2}-\frac{1 }{2}m \bigg) t + \sqrt{m +\frac{(\mu-r)^2 }{\sigma^2}} W_t\bigg\}.
\end{align}
Moreover, it can be seen directly from \eqref{eq_XtExplo0} that 
	$$
	%\E[ \ln [X_T^{\lambda*}]=%\ln x_0+	\E[ \bigg(r+\frac{(\mu-r)^2 }{\sigma^2}\bigg)-\frac{1 }{2}\bigg(\frac{m}{(m(T-t)+1)} +\frac{(\mu-r)^2 }{\sigma^2}\bigg)) T+ \sqrt{\bigg(\frac{m}{ (m(T-t)+1)} +\frac{(\mu-r)^2 }{\sigma^2}\bigg)} W_t]=\ln x_0+
	\E[ \ln [X_T^{\lambda*}]]=\ln x_0+	 \bigg(r+\frac{(\mu-r)^2 }{2\sigma^2}\bigg)T-\frac{m T }{2}<\infty,
	$$
	which means that $\lambda^*$  is admissible. \endproof

\subsection{Proof of Theorem \ref{Thm-update}}
\proof
Recall first that {for any $\lambda$ admissible (not necessarily Gaussian), the corresponding value function $v^{{\lambda}}(s,x;m)$ defined by \eqref{eq:valuefunction}}  solves the following average PDE
$$
v_t^{{\lambda}}(t,x;m)+\int_\cC \bigg((r+\pi(\mu-r))xv_x^{{\lambda}}(t,x;m)+\frac{1}{2}\sigma^2 x^2 \pi^2 v_{xx}^{{\lambda}}(t,x;m) -m\ln\lambda (\pi|t,x) \bigg)\lambda (\pi|t,x)  d \pi=0
$$
 with terminal condition $v^{{\lambda}}(T,x;m)=U(x)$. It follows that
\begin{align*}
v_t^{{\lambda}}(t,x;m)+\sup_{{\hat{\lambda}}}\int_\cC \bigg((r+\pi(\mu-r))xv_x^{{\lambda}}(t,x;m)+\frac{1}{2}\sigma^2 x^2 \pi^2 v_{xx}^{{\lambda}}(t,x;m) -m\ln{\hat{\lambda }}(\pi|t,x)\bigg){\hat{\lambda}} (\pi|t,x) d \pi\ge 0.
\end{align*}
Note that the above supremum is attained at the updated policy $\wt{\lambda}$ defined by \eqref{eq:pischeme}. In other words,
\begin{align*}
v_t^{{\lambda}}(t,x;m)+\int_\cC \bigg((r+\pi(\mu-r))xv_x^{{\lambda}}(t,x;m)+\frac{1}{2}\sigma^2 x^2 \pi^2 v_{xx}^{{\lambda}}(t,x;m) -m\ln\wt{\lambda}(\pi|t,x)  \bigg)\wt{\lambda}(\pi|t,x)  d \pi\ge 0,%\label{eq: aver}
\end{align*}
{which implies that
\begin{align}
v_t^{{\lambda}}(t,x;m)+\int_\cC \bigg((r+\pi(\mu-r))xv_x^{{\lambda}}(t,x;m)&+\frac{1}{2}\sigma^2 x^2 \pi^2 v_{xx}^{{\lambda}}(t,x;m) \bigg)\wt{\lambda} (\pi|t,x)  d \pi\notag\\
&\ge m \int_\cC \ln\wt{\lambda}(\pi|t,x) \wt{\lambda}(\pi|t,x)   d \pi\ge 0.\label{eq: aver}
\end{align}
}
Now, for the policy $\wt{\lambda}$, the corresponding value function is given by
\begin{align}\label{eq:wlambdatile}
v^{\wt{\lambda}}(t,x;m):=\mathbb{E} \Bigg[U(X_T^{\wt{\lambda}})-m\int_t^T \int_\cC \wt{\lambda}(\pi|s,X^{\wt{\lambda}}_s) \ln\wt{\lambda}(\pi|s,X^{\wt{\lambda}}_s)  d \pi ds|X^{\wt{\lambda}}_t=x\Bigg].
\end{align}
{Let $Y_t:=v^\lambda(t,X_t^{\wt{\lambda}};m)$.} By Itô's formula we obtain
 \begin{align*}
v^{{\lambda}}(s,X_s^{\wt{\lambda}};m)&=v^{{\lambda}}(t,x;m)+
\int_t^s \hat{B}{(u,  X_u^{\wt{\lambda}};\wt{\lambda} )} 
 v^{{\lambda}}_{x}(u,X_u^{\wt{\lambda}};m) d W_u\\
&+
\int_t^s  \bigg( v^{{\lambda}}_t(u,X_u^{\wt{\lambda}};m)+\hat{A}{(u, X_u^{\wt{\lambda}};\wt{\lambda})}v^{{\lambda}}_x(u,X_u^{\wt{\lambda}};m)+\frac{1}{2} \hat{B}^2{(u,  X_u^{\wt{\lambda}};\wt{\lambda}) }
 v^{{\lambda}}_{xx}(u,X_u^{\wt{\lambda}};m)\bigg) du.
\end{align*}
Let $\{\tau_n\}_{n\ge 1}$ be a localization sequence of the above stochastic integral, i.e.  
$\tau_n:=\min\{s\ge t: \E[\int_t^{s\wedge \tau_n} \hat{B}^2{(u,  X_u^{\wt{\lambda}};\wt{\lambda}) }  (v^{{\lambda}}_{x}(u,X_u^{\wt{\lambda}};m))^2 d u]\ge n\}\wedge T$. 
Clearly, $\lim_{n\to \infty} \tau_n=T$. By \eqref{eq: aver}  we obtain {for $t\le s\le T$},
 \begin{align*}
&v^{{\lambda}}(t,x;m)
=\E\bigg[v^{{\lambda}}({s\wedge \tau_n},X_{s\wedge \tau_n}^{\wt{\lambda}};m)\notag\\&-
\int_t^{s\wedge \tau_n}  \bigg( v^{{\lambda}}_t(u,X_u^{\wt{\lambda}};m)+\hat{A}{(u, X_u^{\wt{\lambda}};\wt{\lambda})}v^{{\lambda}}_x(u,X_u^{\wt{\lambda}};m)+\frac{1}{2} \hat{B}^2{(u,  X_u^{\wt{\lambda}};\wt{\lambda}) }
 v^{{\lambda}}_{xx}(u,X_u^{\wt{\lambda}};m)\bigg) du\bigg\vert X_{t}^{\wt{\lambda}}=x \bigg]\\
&\leq \E\bigg[v^{{\lambda}}({s\wedge \tau_n},X_{s\wedge \tau_n}^{\wt{\lambda}};m)\notag-m\int_t^{s\wedge \tau_n} \int_\cC \ln\wt{\lambda}(\pi|u,X^{\wt{\lambda}}_u)\wt{\lambda}(\pi|u,X^{\wt{\lambda}}_u)d \pi du\bigg\vert X_{t}^{\wt{\lambda}}=x \bigg].
\end{align*}
%by \eqref{eq: aver}. 
{By} taking $s=T$ , sending $n\to\infty$ {and using \eqref{eq:wlambdatile}} we obtain that for $(t,x)\in[0,T)\times \bbr_+$
$$v^{{\lambda}}(t,x;m) \leq \E\bigg[U(X_{T}^{\wt{\lambda}})\notag-m\int_t^{T} \int_\cC \ln\wt{\lambda}(\pi|u,X^{\wt{\lambda}}_u)\wt{\lambda}(\pi|u,X^{\wt{\lambda}}_u) d \pi du\bigg\vert X_{t}^{\wt{\lambda}}=x \bigg]=v^{\wt{\lambda}}(t,x;m).$$
\endproof
\subsection{Proof of Theorem \ref{Th:policyupdatelog}} \label{Ap: update}
\proof  Observe first that $v^{{\lambda^0}}(s,x;m)$ solves the following average PDE
$$
v_t^{{\lambda^0}}(t,x;m)+\int_\cC \bigg((r+\pi(\mu-r))xv_x^{{\lambda^0}}(t,x;m)+\frac{1}{2}\sigma^2 x^2 \pi^2 v_{xx}^{{\lambda^0}}(t,x;m) -m\ln\lambda_t^0\bigg)\lambda_t^0 (\pi) d \pi=0.
$$
Solving the above PDE with terminal condition $v^{{\lambda^0}}(T,x;m)=\ln x$ and  ${\lambda}^0(\pi| t,x; m)= \mathcal N(\pi|
 a, b^2)$ with $a,b >0$, we obtain  $v_t^{{\lambda^0}}(t,x;m)=\ln x+ h(t)$, where $h$ is a continuous function depending only on $t$ and $h(T)=0$. Therefore, the value function $v^{{\lambda^0}}$ satisfies hypothesis in Theorem \ref{Thm-update} and its conclusions apply. In particular, the next Gaussian policy \begin{equation}\label{eq:pischeme0}
{\lambda}^1(\pi| t,x; m)= \mathcal N\bigg(\pi\bigg|
-\frac{(\mu-r)xv_x^{{\lambda^0}}}{\sigma^2 x^2 v_{xx}^{{\lambda^0}}},-\frac{m}{\sigma^2 x^2 v^{{\lambda^0}}_{xx}}\bigg)=\mathcal N\bigg(\pi\bigg|
\frac{(\mu-r)}{\sigma^2 },\frac{m}{\sigma^2}\bigg)
\end{equation}
is admissible and 
\begin{equation}
v^{{\lambda}^1}(t,x; m)\ge v^{\lambda^0}(t,x; m), \quad  (t,x)\in[0,T)\times \bbr_+,
\label{eq:}
\end{equation}
where $v^{{\lambda}^1}(t,x; m)$ be the value function corresponding to this new  policy ${\lambda}^1(\pi| t,x; m)$. Again, $v^{{\lambda^1}}(s,x;m)$ solves the following average PDE
$$
v_t^{{\lambda^1}}(t,x;m)+\int_\cC \bigg((r+\pi(\mu-r))xv_x^{{\lambda^1}}(t,x;m)+\frac{1}{2}\sigma^2 x^2 \pi^2 v_{xx}^{{\lambda^1}}(t,x;m) -m\ln\lambda^1(\pi|t,x;m)\bigg)\lambda^1 (\pi|t,x;m) d \pi=0.
$$
Direct computations related to the distribution of ${\lambda}^1$ lead to the following PDE
\begin{align*}
 v_t^{{\lambda^1}}(t,x;m)&+\bigg(r+\frac{(\mu-r)^2}{\sigma^2}\bigg) x v_x^{{\lambda^1}}(t,x;m) +\frac{1}{2}\bigg(\frac{(\mu-r)^2}{\sigma^2}+m \bigg)x^2v_{xx}^{{\lambda^1}}(t,x;m)+\frac{m}{2}\bigg(1+\ln\bigg(\frac{2\pi_e m}{\sigma^2}\bigg)\bigg)=0.
\end{align*}
It is hence straightforward to see that $v^{\lambda^*}( t,x, m)=v(t,x;m)$ given by \eqref{eq:valueopt} solves the above PDE and the proof is complete. \endproof
\subsection{Proof of Theorem \ref{Thm:martingale}}
\proof The proof is similar to that in \cite{jia2022a,jia2022policy} and can be done as follows. First, by the Markov property of the process $X^\lambda_t$, it can be observed directly that
\begin{align}
Y_s&=J(t,X^{\lambda}_s;{\lambda})-m\int_t^s \int_\cC \lambda (\pi|u,X^{\lambda}_u)  \ln\lambda (\pi|u,X^{\lambda}_u)  d \pi du\notag\\&=
\E\bigg[U(X_T^\lambda) -m\int_s^T \int_\cC \lambda (\pi|u,X^{\lambda}_u)  \ln\lambda (\pi|u,X^{\lambda}_u)  d \pi du\bigg|X_s^\lambda\bigg]-m\int_t^s \int_\cC \lambda (\pi|u,X^{\lambda}_u)  \ln\lambda (\pi|u,X^{\lambda}_u)  d \pi du\notag\\
&=
\E\bigg[U(X_T^\lambda) -m\int_t^T \int_\cC \lambda (\pi|u,X^{\lambda}_u)  \ln\lambda (\pi|u,X^{\lambda}_u)  d \pi du\bigg|X_s^\lambda\bigg]=\E\big[Y_T|X_s^\lambda\big]\notag
.
\end{align} 
\endproof
\section{Technical Proofs of Section  \ref{se:constrainedLog}}
\subsection{Proof of Theorem \ref{Thm:constrainedlogx}}

\proof Similar to the unconstrained problem, we find the solution of the HJB \eqref{eq:constrainedHJBnocsimp} under the form $v^{[a,b]}(t,x;m)=k(t) \ln x+l(t)$, where $k, l:[0,T]\to \bbr+$ are smooth functions satisfying the terminal condition $k(T)=1$, $l(T)=0$. 
Direct calculation shows that
$$
A(m)=(a-\pi^{Merton})\sigma m^{-1/2};\quad B(m)=(b-\pi^{Merton})\sigma m^{-1/2},
$$
which are independent of $(t,x)$. Hence, $Z(m)=Z_{a,b}(m)$ is also independent of $t,x$. Now, \eqref{eq:constrainedHJBnocsimp} boils down to
\begin{align}
k'(t)\ln x +(r+\frac{1}{2}\frac{(\mu-r)^2}{\sigma^2}) k(l)-\frac{m}{2}\ln k(t)+l'(t)+c_m{+m\ln Z_{a,b}(m)}=0, 
\end{align}
for $(t,x)\in [0,T)\times \bbr_+$, where $c_m:=\frac{m}{2} \ln(\sigma^{-2}{2\pi_e m})$. 
It follows that $k(t)=1$ and
$$
l(t)=- (r+\frac{1}{2}\frac{(\mu-r)^2}{\sigma^2})t-c_m t+C-m t \ln Z_{a,b}(m),
$$
where the constant $C$ is chosen such that the terminal condition $l(T)=0$ is fulfilled. Direction calculation leads to 
$$
l(t)= (r+\frac{1}{2}\frac{(\mu-r)^2}{\sigma^2})(T-t)+c_m (T-t){+m\ln Z_{a,b}(m)} (T-t)$$
and the solution of PDE \eqref{eq:constrainedHJBnocsimp} is given by \eqref{eq:constrainedvalueopt}. 
%\begin{align}
%v^{[a,b]}(t,x;m)=& \ln x+
 %(r+\frac{1}{2}\frac{(\mu-r)^2}{\sigma^2})(T-t)+c_m (T-t)\thaicomment{+m\ln Z_{a,b}(m)} (T-t).
%\label{eq:}
%\end{align}
%\red{Thai: Can we compute $\mathbb E[X_t^{\lambda*}]$ and $Var(X_t^{\lambda*}$?.
%We will report these two quantities in the table. If not computed in closed form, we have to simulate it instead???}
It is straightforward to verify that $v^{[a,b]}$ defined by \eqref{eq:constrainedvalueopt} solves \eqref{eq:constrainedHJBnocsimp}. Indeed, it follows that the optimal distribution $\lambda^{*,[a,b]}$ is now given by a Gaussian distribution with mean $\alpha=\frac{(\mu-r) }{\sigma^2}$ (independent of exploration) and variance 
$\beta^2=\frac{m}{\sigma^2} $ truncated on the interval $[a,b]$. The density of $\lambda^{*,[a,b]}$ is given by \eqref{eq:constrainedpi}
%\begin{align}
 %\lambda^{*,[a,b]}_t(\pi;m)&:=\sigma m^{-1/2}\frac{\phi\big((\pi-\pi^{Merton})\sigma m^{-1/2}\big)}{\Phi((b-\pi^{Merton})\sigma m^{-1/2})-\Phi((a-\pi^{Merton})\sigma m^{-1/2})}
%\label{eq:constrainedoptpolicy}
%\end{align}
which is independent of the wealth level. This truncated Gaussian control allows to determine the exploration wealth drift and volatility as
\begin{align}
\hat{A}(t, x;\lambda^*)=x\bigg(r+\pi_t^{*,[a,b]} (\mu-r)\bigg); \quad
{\hat{B}^2(t,x;\lambda^*)=x^2 \sigma^2 (q_t^{*[a,b]})^2,}
\end{align}
where $\pi_t^{*,[a,b]}$ {and $(q_t^{*[a,b]})^2$ are} given by \eqref{eq:constrainedpi}. This leads to the exploration wealth dynamics given by \eqref{eq:constrainedwealth}.
%$$
%d X_t^{\lambda^{*,[a,b]}} = X_t^{\lambda^{*,[a,b]}}  \bigg(r+\pi_t^{*,[a,b]} (\mu-r) \bigg) dt + \sigma \pi_t ^{*,[a,b]}  X_t^{\lambda^{*,[a,b]}}  d W_t \, , \quad X_0 = x_0 \,.$$
Now, it is clear that the above SDE \eqref{eq:constrainedwealth} admits a unique solution $X^{\lambda^{*,[a,b]}}$ and 	%\E[ \ln [X_T^{\lambda*}]=%\ln x_0+	\E[ \bigg(r+\frac{(\mu-r)^2 }{\sigma^2}\bigg)-\frac{1 }{2}\bigg(\frac{m}{(m(T-t)+1)} +\frac{(\mu-r)^2 }{\sigma^2}\bigg)) T+ \sqrt{\bigg(\frac{m}{ (m(T-t)+1)} +\frac{(\mu-r)^2 }{\sigma^2}\bigg)} W_t]=\ln x_0+
		$\E[ \ln [X_T^{\lambda^{*,[a,b]}}]]<\infty,
	$
	which means that ${\lambda^{*,[a,b]}}$  is admissible.\endproof
\subsection{Proof of Lemma \ref{Le:compare}}
\proof Observe first that the function $x\varphi(x)$ is increasing in $(-1,1)$ and takes negative (resp. positive) values when $x<0$ (resp. $x>0$). Therefore, $A(m)\varphi(A(m))-B(m)\varphi(B(m))\le 0$
if
\begin{itemize}	
\item $A(m)< 0< B(m)$, which implies that $a<\pi^{Merton}<b$;
	\item $0\le A(m)\le B(m)\le 1$, which implies that $\pi^{Merton}\le a<b\le \pi^{Merton}+\sqrt{m}\sigma^{-1}$;
	\item $-1\le A(m)\le B(m)\le 0$,  which implies that $\pi^{Merton}-\sqrt{m}\sigma^{-1}\le a<b<\pi^{Merton}$.
\end{itemize}
For each of these cases, by compararing Propositions \ref{eq:ExploratoryTheorem2} and \ref{eq:ExploratoryTheorem} we obtain
$$
L^{[a,b]}(T,x;m)=\frac{mT}{2}+mT \frac{A(m)\varphi(A(m))-B(m)\varphi(B(m))}{2Z_{a,b}(m)}
\leq \frac{mT}{2}=L(T,x;m)
$$ 
and Lemma \ref{Le:compare} is proved. \endproof
{
\section{Technical proofs for Section 6}
\subsection{Proof of Theorem \ref{Thm:constrainedquadratic}}
\proof We find the solution of the HJB \eqref{eq:constrainedoptpolicyquadratic} under the form $\wt{v}^{[a,b]}(t,x;m)=k(t) x^2+l(t) x+ q(t)$, where $k, l, q:[0,T]\to \bbr$ are smooth functions satisfying the terminal condition $k(T)=-\frac{\varepsilon}{2}$, $l(T)=K$ and $q(T)=0$. Differenting this ansatz, plugging into the PDE \eqref{eq:constrainedoptpolicyquadratic} and studying the coefficients of $x^2$ and $x$ we obtain
$$
k(t)=-\frac{1}{2}\varepsilon e^{-(\rho^2-2r)(T-t)}, \quad l(t)=Ke^{-(\rho^2-r)(T-t)}.
$$
Similarly, we obtain the following ODE when considering the coefficient of $x^0$:
\begin{align}
q'(t)-\frac{1}{2}\rho^2\frac{l^2(t)}{2 k(t)}-\frac{m}{2}\ln (-2k(t))+ \frac{m}{2} \ln(\sigma^{-2}{2\pi_e m})+m\ln (Z(t,x;m))=0, \quad q(T)=0.
\end{align}
Direct calculation shows that 
\begin{align}
&\wt{Q}_a(t,x;m)=\bigg(a_0(t)-\frac{K}{\varepsilon} e^{-r(T-t)} \pi^{Merton}\bigg)\sqrt{\frac{\varepsilon\sigma^2 }{m}e^{-(\rho^2-2r)(T-t)}})=\wt{Q}_a(t;m);\\ &\wt{Q}_b(t,x;m)=\bigg(b_0(t)-\frac{K}{\varepsilon} e^{-r(T-t)} \pi^{Merton}\bigg)\sqrt{\frac{\varepsilon\sigma^2 }{m}e^{-(\rho^2-2r)(T-t)}}\bigg)=\wt{Q}_b(t;m),
\label{eq:}
\end{align}
which are independent of $x$. Hence, $Z(t,x;m)=f(t,m)$ given in \eqref{eq:ZabQuadratic} which is also independent of $x$.
Therefore,  
$$
q(t)=-\frac{K^2}{2\varepsilon}(1-e^{-\rho^2(T-t)})+\frac{m}{4}(\rho^2-2r)(T-t)^2+\frac{m}{2} \ln(\varepsilon^{-1}\sigma^{-2}{2\pi_e m})(T-t)-m \int_t^T  \ln(f(s,m)) ds.
$$
Now, it is straightforward to verify that $\wt{v}^{[a,b]}$ defined by \eqref{eq:quadraticvalue} solves \eqref{eq:constrainedoptpolicyquadratic}. 
Next, it follows that the optimal feedback distribution control $\wt{\lambda}^{*}$ is a Gaussian distribution with parameter $(\frac{K}{\varepsilon} e^{r(T-t)}-x) \pi^{Merton}$ and  $\frac{m}{\varepsilon \sigma^2 }e^{(2r-\rho^2)(T-t)}$, truncated  on the interval $[a(t,x),b(t,x)]$ i.e.
\begin{equation}
\wt{\lambda}^{*}(\theta|t,x;m)=\mathcal N\left(\theta \bigg|(\frac{K}{\varepsilon} e^{-r(T-t)}-x) \pi^{Merton}, \frac{m}{\varepsilon\sigma^2 }e^{-(2r-\rho^2)(T-t)}\right)\bigg|_{[a(t,x),b(t,x)]}.
\label{eq:quadratic}
\end{equation}
This explicit truncated Gaussian policy allows to determine the exploration wealth drift and volatility as
\begin{align*}
&\tilde{A}(t,x;\wt{\lambda}^*) = r x+(\mu-r) \wt{\theta}_t^{*,[a,b]},\quad \wt{B}(t,x;\wt{\lambda}^*):=\sigma\wt{q}_t^{*,[a,b]},
%\bigg((\frac{K}{\varepsilon} e^{r(T-t)}-x) \pi^{Merton}   +\frac{\varphi\big(\wt{Q}_a(t;m)\big)-\varphi\big(\wt{Q}_b(t;m)\big)}{f(t,m)\sqrt{\frac{m}{\varepsilon\sigma^2 }e^{(2r-\rho^2)(T-t)}}}\bigg),
\end{align*}
%and $\wt{B}(\pi,x,\wt{\lambda}^*):=\sigma\wt{q}_t^{*,[a,b]},$
%\begin{align}
%\wt{B}^2(\pi,x,\wt{\lambda}^*):=(\wt{q}_t^{*,[a,b]})^2 \sigma^2.
%\end{align}
where 
\begin{equation}
\wt{\theta}_t^{*,[a,b]}= \big(\frac{K}{\varepsilon} e^{-r(T-t)}-x\big) \pi^{Merton}  +\sqrt{\frac{m} {\varepsilon\sigma^2 e^{-(\rho^2-2r)(T-t)}}}\frac{\varphi\big(\wt{Q}_a(t;m)\big)-\varphi\big(\wt{Q}_b(t;m)\big)}{f(t,m)}
\label{eq:constrainedpiquadratic}
\end{equation}
and
\begin{align} 
(\wt{q}_t^{*,[a,b]})^2=  (\wt{\theta}_t ^{*,[a,b]})^2+
 \frac{m} {\varepsilon\sigma^2 e^{-(\rho^2-2r)(T-t)}}&\Bigg(1+
\frac{\wt{Q}_a(t;m) \varphi(\wt{Q}_a(t;m))-\wt{Q}_b(t;m)\varphi(\wt{Q}_b(t;m)\bigg)}{f(t,m)}\notag\\
&-
\bigg(\frac{\varphi(\wt{Q}_a(t;m))-\varphi((\wt{Q}_b(t;m))}{f(t,m)}\bigg)^2\Bigg),
\end{align}
which guarantees that the SDE \eqref{eq_XtExploquadratic} admits strong solution. 
%$$
%d X_t^{\lambda^{*,[a,b]}} = X_t^{\lambda^{*,[a,b]}}  \bigg(r+\pi_t^{*,[a,b]} (\mu-r) \bigg) dt + \sigma \pi_t ^{*,[a,b]}  X_t^{\lambda^{*,[a,b]}}  d W_t \, , \quad X_0 = x_0 \,.$$
Finally, it is straightforward to verify the integrability conditions to conclude that $\wt{\lambda}^{*}$ is admissible.\endproof
\subsection{Proof of Proposition \ref{Pro: MVequivalence}}
%\begin{prop}\label{Pro: MVequivalence}
%Assume that the agent's initial wealth is smaller that the discounted reward level $\frac{K}{\epsilon}$, i.e. $x_0\le \frac{K}{\epsilon} e^{-rT}$. Then, the exploratory optimal portfolio belongs to the mean-variance frontier.
%\end{prop}
\proof First, from the quadratic utility form it is easy to see that for any admissible terminal portfolio $X_T$,
\begin{equation}
\E[K X_T-\frac{1}{2}\epsilon X_T^2]=-\frac{1}{2}\epsilon\bigg(\E[X_T]-\frac{K}{\epsilon}\bigg)^2+\frac{K^2}{2\epsilon}-\frac{1}{2}Var[X_T].
\label{eq:quafacto}
\end{equation}
Observe that when $Var[X_T]=\Gamma$ is fixed, the right hand side of \eqref{eq:quafacto} is increasingly monotone with respect to $\E[X_T]$ as long as $\E[X_T]\le  \frac{K}{\epsilon}$. 
Therefore, among policies $\theta$ with a fixed variance $Var[X_T]=\Gamma$ and mean $\E[X_T]\le  \frac{K}{\epsilon}$, the policy that maximizes the quadratic expected utility will provide the highest mean $\E[X_T]$ of the right hand-side of \eqref{eq:quafacto}. In other words, to show that the {\it unconstrained} exploratory optimal terminal wealth of the quadratic expected utility maximization problem $X^{0,\wt{\lambda}^*}$ lies in the mean-variance efficient frontier, it suffices to show that $\E[X^{0,\wt{\lambda}^*}_T]\le  \frac{K}{\epsilon}$.
Indeed, from \eqref{eq:constrainedpiquadratic}, taking $a_0(t)=-\infty$ and $b_0(t)=+\infty$, the unconstrained optimal wealth dynamics with exploration is given by \begin{align*}
 dX^{0,\wt{\lambda}^*}_t&=\bigg(rX^{\wt{\lambda}^*}_t+ (\mu-r) \big(\frac{K}{\varepsilon} e^{-r(T-t)}-X^{0,\wt{\lambda}^*}_t\big) \pi^{Merton}\bigg) dt \\
&+\sigma\sqrt{\big(\frac{K}{\varepsilon} e^{-r(T-t)}-X^{0,\wt{\lambda}^*}_t\big)^2 (\pi^{Merton})^2 +\frac{m} {\varepsilon\sigma^2 e^{-(\rho^2-2r)(T-t)}}}dW_t,
%\bigg((\frac{K}{\varepsilon} e^{r(T-t)}-x) \pi^{Merton}   +\frac{\varphi\big(\wt{Q}_a(t;m)\big)-\varphi\big(\wt{Q}_b(t;m)\big)}{f(t,m)\sqrt{\frac{m}{\varepsilon\sigma^2 }e^{(2r-\rho^2)(T-t)}}}\bigg),
\end{align*}
%and $\wt{B}(\pi,x,\wt{\lambda}^*):=\sigma\wt{q}_t^{*,[a,b]},$
%\begin{align}
%\wt{B}^2(\pi,x,\wt{\lambda}^*):=(\wt{q}_t^{*,[a,b]})^2 \sigma^2.
%\end{align}
which implies that $\E[X^{0,\wt{\lambda}^*}_t]=x_0+ \int_0^t \big(\frac{K}{\varepsilon} e^{-r(T-s)}-\E[X^{0,\wt{\lambda}^*}_s]\big) \rho^2 ds$. Put $\psi(t):=\E[X^{0,\wt{\lambda}^*}_t]$ we obtain the following ODE 
$$\psi'(t)=\rho^2\frac{K}{\epsilon}e^{-r(T-t)}+(r-\rho^2)\psi(t).$$
Solving this ODE with initial condition $\psi(0)=x_0$ we obtain
\begin{equation}
\psi(t)=\E[X^{0,\wt{\lambda}^*}_t]=\frac{K}{\epsilon}e^{-r(T-t)}+x_0e^{(r-\rho^2)t}-\frac{K}{\epsilon}e^{-r(T-t)}e^{-\rho^2 t}
\label{eq:}
\end{equation}
and $\E[X^{0,\wt{\lambda}^*}_T]=\frac{K}{\epsilon}+(x_0e^{rT}-\frac{K}{\epsilon})e^{-\rho^2 T}$. It follows directly from the condition $x_0\le \frac{K}{\epsilon} e^{-rT}$ that $\E[X^{0,\wt{\lambda}^*}_T]\le  \frac{K}{\epsilon}$. 
\endproof
}

\end{document}